\documentclass[12pt]{article}
\pdfoutput=1

\usepackage{latexsym}
\usepackage{amssymb,amsfonts,amsmath}
\usepackage{bbm}
\usepackage{amssymb}
\usepackage[colorlinks=true,
linkcolor = blue]{hyperref}
\usepackage{amsfonts}
\usepackage{dsfont}
\usepackage{slashed, tensor}
\usepackage{booktabs}
\usepackage{graphicx}
\usepackage{mathrsfs}

\parskip=\medskipamount

\newcommand{\be}{\begin{equation}}
    \newcommand{\ee}{\end{equation}}

\def\cN{{\cal N}}
\def\dfrac{\displaystyle \frac }

\def\W{\mathcal{ W}}
\def\N{\mathcal{ N}}
\def\bW{\mathcal{ \bar W}}
\def\a{{\alpha}}
\def\da{{\dot\alpha}}
\def\b{{\beta}}
\def\db{{\dot\beta}}
\def\g{{\gamma}}
\def\dg{{\dot\gamma}}
\def\btheta{{\bar \theta}}

\def\hp{\hat +}
\def\hm{\hat -}
\def\dD{{\bar D}}
\def\dF{{\bar F}}
\def\V{{\mathcal{V}}}

\newcommand{\bea}{\begin{eqnarray}}
\newcommand{\eea}{\end{eqnarray}}
\newcommand{\ba}{\begin{array}}
    \newcommand{\ea}{\end{array}}
\newcommand{\nn}{\nonumber \\}
\newcommand{\lb}{\label}

\numberwithin{equation}{section}

\title{}
\author{}
\date{}

\begin{document}

\begin{center}
{\Large \bf New bi-harmonic superspace formulation \\
\vspace{0.2cm}

of $4D, \mathcal{ N}=4$ SYM theory}
\end{center}

\begin{center}

{\bf I.L. Buchbinder${}^a$\footnote{joseph@tspu.edu.ru}, E.A.
Ivanov${}^{b,c}$\footnote{eivanov@theor.jinr.ru}, V.A.
Ivanovskiy${}^c$\footnote{ivanovskiy.va@phystech.edu}} \vspace{5mm}

\footnotesize{ ${}^{a}${\it Center of Theoretical Physics,
Tomsk State Pedagogical University, 634061, Tomsk, Russia, \\
\it National Research Tomsk State University, 634050, Tomsk, Russia}\\
${}^{b}$ {\it Bogoliubov Laboratory of Theoretical Physics,
JINR, 141980 Dubna, Moscow region, Russia}\\
${}^{c}$ {\it Moscow Institute of Physics and Technology, 141700
Dolgoprudny, Moscow Region, Russia} }


\end{center}

    \begin{abstract}
We develop a novel bi-harmonic $\N=4$ superspace formulation of the $\mathcal
{N} = 4 $ supersymmetric Yang-Mills theory (SYM) in four dimensions.
In this approach, the $ \mathcal{N} = 4 $ SYM superfield constraints
are solved in terms of on-shell $ \mathcal {N} = 2 $ harmonic superfields.
Such an approach provides a convenient tool of constructing the
manifestly $ \mathcal{N} = 4$ supersymmetric invariants and further rewriting them in $ \mathcal {N} = 2 $ harmonic  superspace.
In particular, we present $\N=4$ superfield form of the leading term in the $\mathcal{N} = 4 $ SYM effective action which was known
previously in $\N=2$ superspace formulation.

    \end{abstract}
\vspace{0.3cm}

\begin{centerline}
{\it Dedicated to S.James Gates Jr.  on the occasion of his
70th birthday}
\end{centerline}

 \setcounter{footnote}{0}
    \section{Introduction}
The $\mathcal{ N}=4$ supersymmetric Yang-Mills (SYM)  theory in four-dimensional Minkowski space
exhibits many remarkable properties on both the classical and the quantum
levels. Apparently, it is the most symmetric field theoretical
model in physics known to date. The model is gauge invariant, has the maximal
extended rigid supersymmetry (that is, the maximal spin in the relevant gauge
supermultiplet is equal to one) and possesses $R$-symmetry $SU(4)\sim SO(6)$,
as well as the whole $SU(2,2|4)$ superconformal symmetry. As its most remarkable property,  it is a finite
quantum field theory free from any anomalies. $\mathcal{ N}=4$
SYM theory  bears a close relation  with string/brane theory and is a
key object of the modern AdS/CFT activity. Nowadays, various (classical and quantum) aspects of this theory remain a subject of intensive study.
Of particular interest is working out the relevant superspace approaches highlighting one or another side of the rich symmetry structure
of this theory.

$\mathcal{ N}=4$ SYM theory was originally deduced in a component
formulation via dimensional reduction from the
ten-dimensional $\mathcal{ N}=1$ SYM theory \cite{litlink 0} and proceeding from the dual spinor model \cite{Glio}\footnote{See also ref. [20] in \cite{Glio}. The necessary ingredients
for constructing the component action are contained in \cite{Fayet} where, in particular, the term ``hypermultiplet'' was introduced.}.
Subsequently, it was further developed and used by many authors
under diverse angles (see, e.g., the reviews \cite{litlink
Sohn,litlink 6,litlink Kov,litlink HP,litlink BBIKO, litlink
23,litlink 22,BuIv2} and books \cite{litlink 1001,litlink W,litlink
GIOS,litlink BK,litlink John}). The on-shell field content of the
theory amounts to the vector field, six real scalar fields and four
Weyl spinor fields, all being in the adjoint representation of the
gauge group. These fields can be combined into $\N=1$ superfields
encompassing the vector multiplet and three chiral multiplets (see
e.g. \cite{litlink 1001,litlink BK}) or into $\N=2$ harmonic
superfields \cite{litlink 10,litlink GIOS} encompassing the relevant
vector multiplet and a hypermultiplet. In the component description,
all four supersymmetries of the theory under consideration are
on-shell and hidden. In the $\N=1$ superfield description, one out
of four supersymmetries is manifest and the other three are still
hidden and on-shell. When $\N=2$ harmonic superfields are used, two
supersymmetries are manifest and the other two are hidden. The
superfield formulation in terms of unconstrained $\N=4$ superfields,
with all four supersymmetries being manifest and off-shell, is  as
yet unknown\footnote{There is still $\N=3$ superfield formulation
manifesting 3 out of 4 underlying supersymmetries \cite{N3HSS}. It
is based upon a somewhat complicated  techniques of $\N=3$ harmonic
superspace which by now have not sill been enough developed,
especially in the quantum domain (see, however, a recent ref.
\cite{Sieg}).}.

There are at least two large domains of tasks where the manifestly
supersymmetric formulations are of crucial importance. The first one is related to quantum calculations in supersymmetric field theories. The manifest
supersymmetry provides tools to keep these calculations under an efficient control and to write down
all admissible contributions to the quantum effective action of the given theory up to some numerical
coefficients. The second circle of problems is associated with the analysis of the low-energy dynamics
in the string/brane theory. It is desirable to explicitly know the admissible supersymmetric invariants, in four or higher
dimensions, describing the low-energy string/brane interactions. In
particular, the low-energy D3-brane interactions are described in
terms of $\N=4$ SYM theory, so it is useful to have a list of all possible
$\N=4$ invariant functionals containing the vector fields.
In all cases, we need to be aware of a general technique of
constructing supersymmetric invariants. In $\N=4$ SYM theory, this
problem is most complicated just because no manifestly
$\cN=4$ supersymmetric formulation of this theory is available by now.

Taking into account the lack of such a general off-shell description, for setting up $\N=4$ supersymmetric invariants, especially
in the context of  quantum effective actions, there were worked out the approaches employing various harmonic superspaces
with the lesser number of manifest supersymmetries  (see the review \cite{litlink 22}
and reference therein). These approaches allow one to construct {\it on-shell}
$\N =4$ supersymmetric invariants, which display the manifest
$\mathcal{ N}=2$ supersymmetry and an additional hidden $\mathcal{
N}=2$ supersymmetry. In some special cases only the hidden supersymmetry
proves to have an on-shell closure, while for checking manifest supersymmetry the equations
of motions are not needed.

In this paper we propose a new method to
construct the on-shell $\N=4$ superinvariants. As the starting point, we
rewrite the standard superspace constraints of $ \mathcal {N} = 4$
SYM theory \cite{Son} in $ \mathcal {N} = 4 $ bi-harmonic
superspace involving two independent sets of harmonic $SU(2)$ variables. This form of the $ 4D, \mathcal {N} = 4$ SYM constraints still
preserves the manifest $\N=4$ supersymmetry. Our crucial observation is that these constraints can be solved explicitly in terms of few $
4D, \mathcal {N} = 2$ harmonic superfields subjected to some on-shell constraints. As a result, proceeding from  the manifestly $\N=4$ supersymmetric
invariants we can express, in a simple way,  these invariants in terms of $
\mathcal {N} = 2 $ superfields\footnote{Another type of the bi-harmonic approach to $\N=4$ SYM was worked out in \cite{BeSam}. As distinct from the one used here, it does not
allow a direct passing to $\N=2$ superfields. An interpretation of the on-shell $\N=4$ SYM constraints in the harmonic superspace with
the $\frac{SU(4)}{[SU(2)\times SU(2)'\times U(1)]}$ harmonic part \cite{IKaVO} was suggested in \cite{HoWe} for finding out the restrictions imposed
by superconformal symmetry on some correlation functions of the $\N=4$ SYM superfield strengths.}.
Our approach is $4D, \mathcal{N}=4$ counterpart of
a similar method worked out in \cite {litlink 9} for $ \mathcal {N} = (1,1) $ SYM theory in 6 dimensions.

Effective actions play an important role in quantum field theory. The
low-energy effective action of $\N =4$ SYM theory is not an exception.
According to \cite{litlink John}, it can be matched with the effective action
of a D3-brane  propagating in the  $AdS_5$ background. The
bi-harmonic superspace approach can be used to find the low-energy
effective action. Using our method, we reconstruct not only the known result for the leading term in the effective action
\cite{litlink 8,litlink BIP}, but also present some higher-order supersymmetric invariants which can hopefully be identified with the next parts of
the derivative expansion of the effective action.

The paper is organized as follows. In section \ref{s2.0} we recall
the basics of  $\mathcal{ N}=2$  harmonic superspace approach,  including the
formulation of $ \mathcal {N} = 4 $ SYM theory in this
superspace. Also the expression for the leading low-energy
effective action is presented, and the method for its calculation
is briefly outlined. Sections \ref{3.0} and \ref{s3} are devoted to the definition of basics of bi-harmonic
$\mathcal{ N}=4$ superspace and its implications in $\mathcal{ N}=4$ SYM theory. The constraints of
$\mathcal{ N}=4$ SYM theory are rewritten in bi-harmonic superspace in section \ref{3.0} and then
are solved in section \ref{s3}. Different $\mathcal{ N}=2$ superfields
which specify this solution are matched with the objects defined in section
\ref{s2.0}. In section \ref {s6.0} it is shown how the
bi-harmonic superspace approach allows one to construct $\mathcal{ N}=4$ supersymmetric
invariants in terms of $\N=2$ harmonic superfields. In addition,
we show that the low-energy effective action given earlier in $\N=2$ harmonic superspace
can be rewritten, in a rather simple form, in terms of the bi-harmonic superspace quantities.
The main results of our work are summarized in Conclusions. Appendices A and B collect some technical details.

    \section{$4D,\ \mathcal{ N}=4$  SYM theory in harmonic superspace \label{s2.0}}
    The bi-harmonic superspace we are going to deal with is an extension of $\mathcal{ N}=2$ harmonic superspace.
     Therefore, we start by giving here some basic facts about $\mathcal{ N}=2$  harmonic superspace and superfields.
    For more details, see ref. \cite{litlink 10,litlink GIOS}.
    \subsection{Brief survey of ${\cal N}=2$ harmonic superspace}
    The standard $4D, \,\mathcal{ N}=2$ superspace amounts to the set of coordinates
    \begin{equation}
        z^M=(x^m, \theta^{\alpha}_i, \bar\theta^{\dot{\alpha}i} ),
    \end{equation}
    where $x^m$, $m=0,1,2,3\,$,   are the Minkowski space coordinates and $\theta^{\alpha}_i$, $\bar\theta^{\dot{\alpha}i}$,
    $i=1,2 $, $\alpha,\dot\alpha=1,2$, are spinor Grassmann coordinates.

    In order to pass to harmonic superspace we add to these coordinates the harmonics  $u^{\pm i}$ ($u^{-}_i=(u^{+i})^*$, $u^{+i}u^{-}_i=1$)
    which represent the ``harmonic sphere'' $SU (2)_R/U (1)$, with $SU (2)_R$ being the $R$-symmetry group realized on the doublet indices $i,\ k$.
    The  $4D,\, \mathcal{ N} = 2$  harmonic  superspace in the central basis is
    defined as the enlarged coordinate set
    \begin{equation}
        Z=(z, u)=(x^m, \theta^{\alpha}_i, \bar\theta^{\dot{\alpha}i}, u^{\pm i}).
    \end{equation}
    In the analytic basis it is parametrized by the coordinates
    \begin{eqnarray}
        &Z_{\rm an}=(x^m_{\rm an}, \theta^{\pm}_{\alpha}, \bar\theta^{\pm}_{\dot{\alpha}}, u^{\pm i}), \label{HSSa} \\
        &x^m_{\rm an}=x^m-2i\theta^{(i}\sigma^m\bar\theta^{j)}u^+_iu^-_j, \quad\theta^{\pm}_\alpha=u^{\pm}_i\theta^{i}_\alpha, \qquad\bar\theta^{\pm}_{\dot\alpha}=u^{\pm}_i\bar\theta^{i}_{\dot\alpha}.
    \end{eqnarray}
    The crucial feature of the analytic basis is that it manifests the existence of subspace  involving only half of the original Grassmann coordinates
    \begin{equation}
        \zeta=(x^m_{\rm an}, \theta^{+}_{\alpha}, \bar\theta^{+}_{\dot{\alpha}}, u^{\pm i}),
        \label{11111}
    \end{equation}
    such that it closed  under $4D$, $\mathcal{ N}=2$ supersymmetry transformations.
    The set (\ref{11111}) parametrizes what is called the ``harmonic analytic superspace''.

    The important ingredients of the harmonic superspace approach are the spinor and harmonic derivatives.
    In the analytic basis, they are defined as
    \begin{eqnarray}
            &&D^{+}_\alpha=\frac{\partial}{\partial \theta^{-\alpha}}, \qquad
            \bar D^{+}_{\dot\alpha}=\frac{\partial}{\partial\bar \theta^{-\dot\alpha}},\nn
            &&D^{-}_\alpha=-\frac{\partial}{\partial \theta^{+\alpha}}+2i\bar \theta^{-\dot\alpha}\partial_{\alpha\dot \alpha}, \qquad
            \bar D^{-}_{\dot\alpha}=-\frac{\partial}{\partial \btheta^{+\dot\alpha}}+2i \theta^{-\alpha}\partial_{\alpha\dot \alpha},\nn
            &&D^{++}=u^{+i}\frac{\partial}{\partial u^{-i}}-2i\theta^{+\alpha}\bar\theta^{+\dot\alpha}\partial_{\alpha\dot \alpha}
            +\theta^{+\alpha}\frac{\partial}{\partial \theta^{-\alpha}}+\bar\theta^{+\dot\alpha}\frac{\partial}{\partial \bar\theta^{-\dot\alpha}},\nn
            &&D^{--}=u^{-i}\frac{\partial}{\partial u^{+i}} -2i\theta^{-\alpha}\bar\theta^{-\dot\alpha}\partial_{\alpha\dot \alpha}+\theta^{-\alpha}
            \frac{\partial}{\partial \theta^{+\alpha}}+\bar\theta^{-\dot\alpha}\frac{\partial}{\partial \bar\theta^{+\dot\alpha}}.
            \label{DeriVat}
    \end{eqnarray}
    They are related to the spinor derivatives in the central basis,
    \bea
     D^i_\a=\frac{\partial}{\partial\theta^{\a}_i}+i\btheta^{\da i}\partial_{\a\da}, \qquad \dD_{\da i}=-\frac{\partial}{\partial\btheta^{\da i}}
    -i\theta^{\a}_i\partial_{\a\da}\,,\label{N2SpinCB}
    \eea
as
\bea
D^{\pm }_\a =D^i_\a u^\pm_i\,, \qquad \dD_{\da}^\pm =\dD_{\da}^i u^\pm_i\,.
\eea

    The harmonic derivatives $D^{\pm\pm}$, together with the harmonic $U(1)$ charge operator
    $$
    D^0 = u^{+i}\frac{\partial}{\partial u^{+i}} - u^{-i}\frac{\partial}{\partial u^{-i}} +
    \theta^{+\alpha}\frac{\partial}{\partial \theta^{+\alpha}}+\bar\theta^{+\dot\alpha}\frac{\partial}{\partial \bar\theta^{+\dot\alpha}} -
    \theta^{-\alpha}
    \frac{\partial}{\partial \theta^{-\alpha}} -\bar\theta^{-\dot\alpha}\frac{\partial}{\partial \bar\theta^{-\dot\alpha}}\,,
    $$
    form an $SU(2)$ algebra,
    \begin{equation}
        [D^{++}, D^{--}] = D^0\,, \quad [ D^0, D^{\pm\pm}] = \pm 2 D^{\pm\pm}\,.
    \end{equation}
In the central basis, the harmonic derivatives are simply
\be
D^{\pm\pm} = \partial^{\pm\pm} = u^{\pm i}\frac{\partial}{\partial u^{\mp i}}\,, \quad D^0 = \partial^0 = u^{+i}\frac{\partial}{\partial u^{+i}} - u^{-i}\frac{\partial}{\partial u^{-i}}\,.
\ee
Of course, the (super)algebra of the spinor and harmonic derivatives does not depend on the choice of basis in $\mathcal{N}=2$ harmonic superspace.

    The harmonic superfields (as well as the harmonic projections of the spinor covariant derivatives) carry a definite
    integer harmonic $U(1)$ charges, $D^0 \Phi^{q}(Z) = q \Phi^{q}(Z), \,[D^0, D^\pm_{\alpha, \dot\alpha}] = \pm D^\pm_{\alpha, \dot\alpha} $.
    The harmonic $U(1)$ charge  is assumed to be strictly preserved in any superfield action defined on the superspaces (\ref{HSSa}) or (\ref{11111}). Just
    due to this requirement, all invariants are guaranteed to depend just on two parameters of the ``harmonic sphere'' $SU (2)_R/U (1)$.

        In addition, we will use the identities
    \begin{eqnarray}
    &&(\theta^{\pm})^2=\theta^{\pm\alpha}\theta^{\pm}_\alpha, \qquad (\bar \theta^{\pm})^2=\bar \theta^{\pm}_{\dot\alpha}\bar \theta^{\pm\dot\alpha},\nn
    &&(D^{\pm})^2=D^{\pm\alpha}D^{\pm}_\alpha, \qquad (\bar D^{\pm})^2=\bar D^{\pm}_{\dot\alpha}\bar D^{\pm\dot\alpha},\nn
    &&(D^{+})^4=\frac{1}{16}(D^{+})^2(\bar D^{+})^2, \qquad (D^{-})^4=\frac{1}{16}(D^{-})^2(\bar D^{-})^2,
    \end{eqnarray}
    and the following definition of the integration measures over the total harmonic superspace and its analytic subspace
    \begin{equation}
     dud^{12}z=d^4x du(D^{+})^4(D^{-4}), \qquad d\zeta^{-4}=d^4x_{\rm an}du(D^{-4}).\label{measHSS}
    \end{equation}

    The ``shortness'' of the spinor derivatives $D^{+}_\alpha,\bar D^{+}_{\dot\alpha}$ in the analytic basis (\ref{DeriVat}) reflects the existence
    of the analytic harmonic subspace (\ref{11111}) in the general harmonic superspace (\ref{HSSa}): one can define an analytic ${\cal N}=2$
    superfield by imposing the proper covariant ``Grassmann analyticity'' constraints on a general harmonic superfield, viz., $D^{+}_\alpha \Phi^{q}(Z) =
    \bar D^{+}_{\dot\alpha}\Phi^{q}(Z) = 0 \;\Rightarrow \;\Phi^{q}(Z) = \varphi^{q}(\zeta)\,.$ The harmonic derivative $D^{++}$ commutes with these spinor derivatives and
    so possesses a unique property of preserving the Grassmann harmonic analyticity: $D^{++}\Phi^{q}(Z)$ is an analytic superfield if $ \Phi^{q}(Z)$ is.

    \subsection{$\mathcal{ N}=4$ SYM action \label{s2.2}}
    When formulated in $\mathcal{ N}=2$ harmonic superspace, $\mathcal{ N}=4$
    vector gauge  multiplet can be viewed as a ``direct sum'' of gauge $\mathcal{ N}=2$
    superfield  $\V^{++}$ and the hypermultiplet superfield $q^{+}_A=(q^+, -\tilde
    q^{+})$. Both these superfields are analytic,
    \begin{equation}
   D^{+}_\alpha {\cal V}^{++} = \bar D^{+}_{\dot\alpha}{\cal V}^{++}  = 0\,, \quad D^{+}_\alpha q^{+}_A = \bar D^{+}_{\dot\alpha}q^{+}_A  = 0\,, \label{AnalitVq}
     \end{equation}
    and belong to the same adjoint representation of the gauge group.

    $\mathcal{ N}=2$ gauge
    multiplet  $\V^{++}$ is described dy the classical action \cite{litlink 9}
    \begin{equation}
    S^{
        \mathcal{N}=2}_{\text{SYM}}=\frac{1}{2}\sum\limits_{n=2}^{\infty}\text{tr}\frac{(-i)^n}{n}\int
    d^{12}z du_1\dots du_n\frac{\V^{++}(z,u_1)\dots
        \V^{++}(z,u_n)}{(u^+_1u^+_2)\dots(u^+_nu^+_1)},
    \label{0001}
    \end{equation}
    where $d^{12}z = d^4x d^8\theta$  and the harmonic distributions $1/(u^+_1u^+_2), \cdots$
    are defined in \cite{litlink GIOS}.

    This action yields the following equations of motion
    \begin{equation}
    (D^{+})^2\W=0, \qquad (\bar{D}^{+})^2\bW=0\,,
    \label{equations}
    \end{equation}
    where $D^+_\alpha$ and $\bar{D}^{+}_{\dot\alpha}$ were defined in (\ref{DeriVat}) and $\W$,
    $\bar \W$ are the chiral and antichiral gauge superfield strengths ,
    \begin{equation}
    \W=-\frac{1}{4}(\bar D^{+})^2\V^{--}, \qquad
    \bW=-\frac{1}{4}(D^{+})^2\V^{--}, \label{DefWW}
    \end{equation}
    with $\V^{--}$ being a non-analytic harmonic gauge connection related to $\V^{++}$ by the harmonic flatness
    condition
    \begin{eqnarray}
    &&D^{++}\V^{--}-D^{--}\V^{++}+i[\V^{++},\V^{--}]=0\, \Longleftrightarrow \, [\nabla^{++},\, \nabla^{--}] = 0\,, \label{0004} \\
    && \nabla^{\pm\pm} := D^{\pm\pm} + i [\V^{\pm\pm},\,\,.].\label{0004a}
    \end{eqnarray}
Note that in the considered ``$\lambda$'' frame, in which the gauge group is represented by transformations with the manifestly analytic gauge parameters,
the spinor derivatives $D^{+}_{\alpha}$ and $\bar{D}^{+}_{\dot\alpha}$ require no gauge connection terms as they are gauge-covariant on their own right. The gauge-covariant
derivatives  $\nabla^{-}_{\alpha}$ and $\bar\nabla^{-}_{\dot\alpha}$ are defined as
 \begin{equation}
\nabla^{-}_{\alpha} := [\nabla^{--}, D^+_\alpha]\,, \qquad \bar\nabla^{-}_{\dot\alpha} := [\nabla^{--}, \bar D^+_{\dot\alpha}]\,.
  \end{equation}
Using these definitions and the relation (\ref{0004}), one can check that
\begin{equation}
\bar\nabla^{-}_{\dot\alpha}{\cal W} = 0\,, \qquad \nabla^{-}_{\alpha}\bar{\cal W} = 0\,,
\end{equation}
while the rest of (anti)chirality conditions, $\bar
D^+_{\dot\alpha}{\cal W} =  D^+_\alpha\bar{\cal W} = 0\,,$  directly
follows from the definition (\ref{DefWW}). It also follows from the
definition (\ref{DefWW}) that the superfield strengths ${\cal W},\bar{\cal W}$ satisfy the reality condition
\begin{equation}
(D^{+})^2\W =(\bar{D}^{+})^2\bW\,,  \label{RealWW}
\end{equation}
as well as the conditions of the covariant harmonic independence
\begin{equation}
\nabla^{\pm\pm}\W = \nabla^{\pm\pm}\bW = 0\,. \label{HarmIndep}
\end{equation}
Let us point out that both the (anti)chirality of the superfield strengths and the constraints (\ref{RealWW}), (\ref{HarmIndep}) hold off shell, as the consequences
of the definition (\ref{DefWW}), the flatness condition (\ref{0004})  and the analyticity of the gauge connection ${\cal V}^{++}$, eq. (\ref{AnalitVq}).
Using (\ref{RealWW}), one can cast the equations of motion  (\ref{equations}) into an equivalent form
\begin{equation}
F^{++} = 0\,, \quad F^{++} := \frac{1}{16}\left(D^{+}\right)^{2}\left(\bar D^{+}\right)^{2}\V^{--}\,. \label{eqsViaV--}
\end{equation}
The superfield $F^{++}$ is analytic and satisfies the off-shell constraint $\nabla^{++} F^{++} = 0\,.$

    The classical action for the hypermultiplet in the adjoint representation  reads \cite{litlink 10,litlink GIOS}
    \begin{equation}
    S_q=\frac{1}{2}\text{tr}\int d\zeta^{-4}q^{+}_A\nabla^{++}q^{+A}=
    \frac{1}{2}\text{tr}\int d\zeta^{-4}q^{+}_A\left(D^{++}q^{+A}+i[\V^{++},q^{+A}]\right).
    \label{0005}
    \end{equation}

    The action of $\mathcal{ N}=4$ SYM theory in ${\cal N}=2$ harmonic superspace is the sum of the
    actions  (\ref{0001}) and (\ref{0005}),
    \begin{equation}
    S^{ \mathcal{N}=4}_{\text{SYM}}=S^{ \mathcal{N}=2}_{\text{SYM}}+S_q.
    \label{2.40}
    \end{equation}
    This action yields the equations of motion
        \begin{eqnarray}
    \nabla^{++}q^{+}_A=0, \qquad F^{++} = -i[q^{+A},q^{+}_A]\,, \label{EqMoVq}
    \end{eqnarray}
where the second equation is just the modification of (\ref{eqsViaV--}) by the hypermultiplet source term.

    The total action (\ref{2.40}) is manifestly $\mathcal{N}=2$ supersymmetric by construction.
    Also, it is invariant under the hidden  $\mathcal{ N}=2$ supersymmetry transformations which
    complement the manifest $\mathcal {N} = 2 $ supersymmetry to the full $\mathcal {N} = 4 $
    supersymmetry
    \begin{eqnarray}
    \delta \V^{++}=\left[\epsilon^{A\alpha}\theta^{ +}_{\alpha}-\bar\epsilon^A_{\dot\alpha}\bar\theta^{ +\dot\alpha}\right]q^{+}_{A}, \quad
    \delta q^{+}_A=-\frac{1}{32}(D^{+})^2(\bar D^{+})^2\left[\epsilon^{\alpha}_A\theta^{-}_\alpha \V^{--}+\bar\epsilon^{}_{A\dot \alpha}\bar\theta^{-\dot\alpha} \V^{--}\right],
    \label{00.5}
    \end{eqnarray}
    with $\bar\epsilon^{}_{A\dot \alpha}$ and $\epsilon^{\alpha}_A$ as new anticommuting parameters.
    The algebra of these transformations is closed modulo terms proportional to the classical equations
    of motion. Only the manifest $\mathcal{ N}=2$ supersymmetry in (\ref{2.40}) is off-shell closed.

    We also note that the actions (\ref{0001}), (\ref{0005}) and hence their sum (\ref{2.40}) are
    manifestly invariant under the automorphism group $SU(2)_R\times SU(2)_{PG} \times U(1)_R\,$. The group $SU(2)_{PG}$ acts
    on the doublet indices of the hypermultiplet superfield $q^+_A$ and commutes with manifest ${\cal N}=2$ supersymmetry (but forms a semi-direct product with
    the hidden supersymmetry (\ref{00.5})), while $U(1)_R$ acts as a phase transformation of the Grassmann variables
    and covariant  spinor derivatives. It forms a semi-direct product with both types of supersymmetry, like the $R$-symmetry group $SU(2)_R$.

    \subsection{The leading low-energy effective action in $ \mathcal {N} = 4 $ SYM theory}
    The $ \mathcal {N} = 4 $ supersymmetric leading low-energy effective
    action is the exact contribution to the quantum effective action of
    $ \mathcal {N} = 4 $ SYM theory in the Coulomb phase (see, e.g., reviews \cite{litlink 23, litlink 22}
    and references therein). From a formal point of view, such
    an action is some on-shell $ \mathcal {N} = 4 $ supersymmetric
    invariant constructed out of the abelian $ \mathcal {N} = 2$
    superfields $\V^{++}$ and $q^{+}_A$ belonging to the Cartan subalgebra of the gauge group. All other components of these superfields,
    being ``heavy'', in the Coulomb phase can be integrated out in the relevant functional integral and so do not contribute to the effective action.
    As we will see, such invariants can be written easily enough in terms of bi-harmonic
    superfields. However, before doing this, we will briefly remind how
    such an $\mathcal{ N}=4$ invariant is written through $\mathcal{ N}=2$
    harmonic superfields, limiting ourselves, for simplicity, to the gauge group $SU(2)$.

    Construction of the leading low-energy effective action in $ \mathcal {N} = 4 $ SYM
    theory begins, as a starting point, from $ \mathcal {N} = 2 $ invariant low-energy effective action ${\cal S}_{eff}$
    written through the non-holomorphic effective potential $\mathcal{H}( W, {\bar W})$ in the form:

\begin{equation}
    {\cal S}_{eff} = \int d^{12}zdu\,\mathcal{H}( W, {\bar W}), \qquad
    \mathcal{H}( \W, {\bW})=c\ \text{ln}\left(\frac{
        \W}{\Lambda}\right)\text{ln}\left(\frac{{\bW}}{\Lambda}\right),
    \label{2.46}
    \end{equation}
    where $\Lambda$ is an arbitrary scale\footnote{In fact, the action does not depend on $\Lambda$ in
    virtue of the (anti)chirality of $(\bW)\W$.}, the  $\W, {\bW}$ satisfy the equations of motion
    (\ref{equations}) and $c$ is some constant.
    The non-holomorphic effective potential was studied and the constant $c$ was calculated in many papers by various methods
    (see the reviews \cite{litlink BBIKO, litlink 23, litlink 22} and references therein).

    The complete leading low-energy $\mathcal{N}=4$ SYM effective action is an extension of
    the effective action (\ref{2.46}) by some hypermultiplet-dependent terms, such that the result
    is invariant under the hidden  $\mathcal{N}=2$ supersymmetry transformations (\ref{00.5}).
    It was computed in a closed form in \cite{litlink 8,litlink BIP} and reads
    \begin{equation}
    \Gamma=\int d^{12}zdu\ \left[c\ \text{ln}\left(\frac{ \W}{\Lambda}\right)
    \text{ln}\left(\frac{{\bW}}{\Lambda}\right)
    +\mathcal{L}\left(-2\frac{q^{+A}q^{-}_{A}}{ \W{\bW}}
    \right)\right],\label{2.50}
    \end{equation}
    with
    \begin{equation}
    \mathcal{L}(Z)=c\sum\limits_{n=1}^{\infty}\frac{Z^n}{n^2(n+1)}=c\left[(Z-1)\frac{\ln (1-Z)}{Z}
    +\text{Li}_2(Z)-1\right],\label{2.500}
    \end{equation}
    where $\text{Li}_2(Z)$ is the Euler dilogarithm function. The part
    dependent on the hypermultiplet $ q^{+A} $ is fixed, up to the numerical coefficient $c$, by the requirement that the effective
    action $\Gamma $ be invariant under both manifest $ \mathcal {N} =
    2$ supersymmetry and hidden on-shell $ \mathcal {N} = 2$
    supersymmetry. As a result, the effective action (\ref{2.50}) is invariant of
    the total $\mathcal {N} = 4$ supersymmetry and depends on all fields of the abelian $ \mathcal {N} = 4$
    vector multiplet. The coefficient $c$ should be the same in both the gauge field sector and the
    hypermultiplet sector of the low-energy effective action due to ${\cal N}=4$
    supersymmetry. This was confirmed  in \cite{litlink BIP} by the direct quantum
    supergraph calculation. The precise value of $c$ will be of no interest for our
    further consideration.

    The expressions (\ref{2.46}) and (\ref{2.50}) will be used in what follows in order
    to demonstrate the power of $\mathcal {N} = 4$ bi-harmonic superspace method which automatically and in a rather simple way
    yields the on-shell $\mathcal {N} = 4$ invariant (\ref{2.50}), (\ref{2.500}).

    \section{$\mathcal{N}=4$ bi-harmonic superspace and superfields \label{3.0}}
    This and subsequent sections deal with the construction and the applications of
    an extended $ \mathcal {N} = 4 $ bi-harmonic superspace and the relevant bi-harmonic
    superfields.
    \subsection{Bi-harmonic superspace}
        The standard $\mathcal{ N}=4$ superspace involves the coordinates
    \begin{equation}
    z^M=(x^m, \theta^{\alpha}_I, \bar\theta^{\dot{\alpha}I} ), \label{4SS}
    \end{equation}
    where $x^m$, $m=0,1,2,3$ are the Minkowski space coordinates, while $\theta^{\alpha}_I$ and
    $\bar\theta^{\dot{\alpha}I}$, $I=1,\dots,4,$ $\alpha,\dot\alpha=1,2\,$ are anticommuting Grassmann coordinates. They transform
    in the fundamental representation of the $\mathcal{N} =4$ $R$-symmetry group $U(4)$ acting on the index $I$.

    The  spinor derivatives in the central basis are defined as
    \begin{eqnarray}
    D^I_\a=\frac{\partial}{\partial\theta^{\a}_I}+i\btheta^{\da I}\partial_{\a\da}, \qquad \dD_{\da I}=-\frac{\partial}{\partial\btheta^{\da I}}
    -i\theta^{\a}_I\partial_{\a\da}\,.\label{N4Spin}
    \end{eqnarray}
    Like in the $\N=2$ case, passing to the bi-harmonic extension of (\ref{4SS}) allows one to make manifest Grassmann analyticity with respect
    to some set of spinorial coordinates.

        In order to introduce  $SU(2)$ harmonics we reduce the $R$-symmetry group $SU(4)$  to $SU(2)\times SU(2)\times U(1)$ in the following way:
        we substitute the index $ I $ by two indices $ i, A = {1,2} $ according to the rule
    \begin{eqnarray}
    &I=1\Leftrightarrow i=1,\qquad I=2 \Leftrightarrow i=2,\nn
    &I=3\Leftrightarrow A=1,\qquad I=4\Leftrightarrow A=2\,.
    \label{rule}
    \end{eqnarray}
     The first $SU(2)$ acts on the indices $i, j$ and coincides with $SU(2)_R$, while
     the second $SU(2)$ acts on the indices $A, B$ and will be identified with $SU(2)_{PG}$ of section \ref{s2.0}.
     The indices $i$ or $A$ are raised and lowered according to the ordinary $SU(2)$ rules, using
     the antisymmetric tensors $e^{ij}$, $e^{AB}$ and $e_{ij}$, $e_{AB}$. The extra $U(1)$ will be identified with $U(1)_R$ of the previous section.
     It  transforms $\theta^{\alpha\,i}$ and $\hat\theta^{\alpha\, A}$ by the mutually conjugated phase factors\footnote{The alternative bi-harmonic superspace of  \cite{BeSam}
     corresponds to the principal embedding of $SO(4) \sim SU(2) \times SU(2)$ in $SU(4)$, such that the Grassmann variables are organized into a complex 4-vector of $SO(4) \subset SU(4)$
     while the harmonics are still associated with the left and right  $SU(2)$ factors. In our case we deal with the diagonal embedding of $SU(2) \times SU(2)$ in $SU(4)$.}.

     As the next step we introduce two sets of the harmonic variables   $u^{\pm}_i$ and $v^{\hat \pm}_A$, which parametrize these two $SU(2)$ groups.
    Respectively, the full set of the $\N=4$ superspace coordinates is extended to
    \begin{equation}
    \hat Z=\Big(x^{m}, \theta^{\alpha i}, \bar\theta^{\dot\alpha i},
    \hat \theta^{\alpha A}, \bar{\hat\theta}^{\dot\alpha A}, u^\pm_i, v^{\hat\pm}_A\Big).
    \end{equation}

    The analytic basis of this bi-harmonic superspace is defined as the set of the coordinates
    \begin{equation}
    \hat Z_{\rm an}=\Big(x_{\rm an}^{m}, \theta^{\pm}_{\a}, \bar\theta^{\pm}_{\da},
    \theta^{\hat \pm}_{\a}, \bar\theta^{\hat \pm}_{\da}, u^\pm_i, v^{\hat\pm}_A\Big),
    \end{equation}
    where\footnote{The action of the standard generalized conjugation $\widetilde {}$
        on different objects in the central basis of the bi-harmonic superspace is given by the following rules
        \begin{equation*}
        \widetilde{ f^{iA}}=\overline{f^{iA}}=\bar f_{iA},\qquad
        \widetilde{\theta_{\alpha i}}=\bar \theta^i_{\dot{\alpha}}, \qquad
        \widetilde{\theta_{\alpha A}}=\bar \theta^A_{\dot{\alpha}},\qquad
        \widetilde{u^{\pm}_i}=u^{\pm i}, \qquad
        \widetilde{v^{\hat \pm}_A}=v^{\hat\pm A}.
        \end{equation*}
        In the analytic basis this operation acts as follows
        \begin{equation*}
        \widetilde{\theta^{\pm}_{\alpha }}=\bar \theta^{\pm}_{\dot{\alpha}},\qquad
        \widetilde{\bar \theta^\pm_{\dot\alpha}}=- \theta^\pm_{{\alpha}},\qquad
        \widetilde{\theta^{\hat \pm}_{\alpha}}=\bar \theta^{\hat \pm}_{\dot{\alpha}},\qquad
        \widetilde{\bar \theta^{\hat\pm}_{\dot\alpha}}=- \theta^{\hat\pm}_{{\alpha}}.
        \end{equation*}}
    \begin{eqnarray}
    &\theta^{\pm \alpha}=\theta^{\alpha i}u^{\pm}_i, \qquad \theta^{\hat \pm \alpha}=\hat\theta^{\alpha A}v^{\hat \pm}_A,\nn
    &\bar{ \theta}^{\pm \dot\alpha}=\bar{\theta}^{\dot\alpha i}u^{\pm}_i,  \qquad  \bar{\theta}^{\hat \pm \dot\alpha}=\bar{\hat\theta}^{\dot\alpha A}v^{\hat \pm}_A, \nn
    &x^m_{\text{an}}=x^m-2i\theta^{(i}\sigma^m\bar\theta^{j)}u^+_iu^-_j-2i\hat\theta^{(A}\sigma^m\bar{\hat\theta}^{B)}v^{\hat +}_A v^{\hat-}_B.
    \end{eqnarray}

    Also, we define the spinor and harmonic derivatives in the central basis like in section \ref{s2.0},
    \begin{eqnarray}
    &D^{\pm}_\alpha= D^i_\alpha u^{\pm}_i, \qquad
    D^{\hat\pm}_\alpha=\hat D^A_\alpha v^{\hat\pm}_A, \qquad
    \bar D^{\pm}_{\dot\alpha}=\bar D^i_{\dot\alpha} u^{\pm}_i, \qquad
    \bar D^{\hat\pm}_{\dot\alpha}=\bar{\hat D}^A_{\dot\alpha} v^{\hat\pm}_A, \nn
    &\partial^{\pm\pm}=u^{\pm i}\dfrac{\partial}{\partial u^{\mp i}}\,, \qquad \partial^{\hat\pm \hat\pm}=v^{\hat\pm A}\dfrac{\partial}{\partial v^{\hat\mp A}}\,,
    \end{eqnarray}
    where $D^i_\alpha$, $\hat D^A_\alpha$ and their c.c.  are the usual spinor derivatives with respect to $\theta^i_\a$, $\hat\theta^A_\a$ and their c.c..
    They are obtained from (\ref{N4Spin}) by splitting the $SU(4)$ index $I$ according to the rule (\ref{rule}).

     Spinor derivatives in the analytical basis look the same as in the previous section. The difference is that there are now two types of derivatives,
    with ``hat'' and without ``hat''. For example, $D^{+}_\a$ and $D^{\hp}_\a$ read
    \begin{eqnarray}
    D^{+}_\a=\frac{\partial}{\partial \theta^{-\alpha}}, \qquad D^{\hp}_\a=\frac{\partial}{\partial \theta^{\hm\alpha}}.
    \end{eqnarray}

    The same is true for harmonic derivatives. For example, $D^{++}$ and $D^{\hp\hp}$ read
    \begin{eqnarray}
         &&D^{++}=u^{+i}\frac{\partial}{\partial u^{-i}}-2i\theta^{+\alpha}\bar\theta^{+\dot\alpha}\partial_{\alpha\dot \alpha}
        +\theta^{+\alpha}\frac{\partial}{\partial \theta^{-\alpha}}+\bar\theta^{+\dot\alpha}\frac{\partial}{\partial \bar\theta^{-\dot\alpha}},\nn
        &&D^{\hp\hp}=v^{\hp A}\frac{\partial}{\partial v^{\hm A}}-2i\theta^{\hp\alpha}\bar\theta^{\hp\dot\alpha}\partial_{\alpha\dot \alpha}
        +\theta^{\hp\alpha}\frac{\partial}{\partial \theta^{\hm\alpha}}+\bar\theta^{\hp\dot\alpha}\frac{\partial}{\partial \bar\theta^{\hm\dot\alpha}}.
    \end{eqnarray}

    Since $D^{+}_\a,\ \dD^{+}_\da,\ D^{\hp}_\a,\ \dD^{\hp}_\da$ mutually anticommute and all are ``short'' in the analytic basis,
    there are three different types of  analytic subspaces in $\N=4$ bi-harmonic superspace,  in contrast to $\N=2$ harmonic superspace and, correspondingly, three different types of the
    Grassmann analyticity. These are the ``half-analytic'' subspace corresponding to nullifying $D^{+}_\alpha, \bar D^{+}_{\dot\alpha}$ on the appropriate superfields,
    the ``half-analytic'' subspace with nullifying $D^{\hp}_\alpha, \bar D^{\hp}_{\dot\alpha},$
    and the full analytic subspace, with four independent Grassmann-analyticity constraints.  Respectively, they amount to the following sets of coordinates
       \begin{eqnarray}
    &&\zeta_I = \Big(x^{m}_{\text{an}}, \theta^{+\alpha}, \bar\theta^{+\dot\alpha },
     \theta^{\hat \pm\alpha }, \bar{\theta}^{\hat\pm\dot\alpha }, u^\pm_i, v^{\hat\pm}_A\Big),\nn
       &&\zeta_{II} = \Big(x^{m}_{\text{an}}, \theta^{\pm\alpha}, \bar\theta^{\pm\dot\alpha },
     \theta^{\hp\alpha }, \bar{\theta}^{\hp\dot\alpha }, u^\pm_i, v^{\hat\pm}_A\Big),\nn
       &&\zeta_A = \Big(x^{m}_{\text{an}}, \theta^{+\alpha}, \bar\theta^{+\dot\alpha },
     \theta^{\hp\alpha }, \bar{\theta}^{\hp\dot\alpha }, u^\pm_i, v^{\hat\pm}_A\Big). \label{BiAnal}
    \end{eqnarray}
    All these subspaces  are closed under $4D,\, \N=4$ supersymmetry transformations.
    \subsection{Bi-harmonic superfields of $\mathcal{N}=4$ SYM theory
        \label{s1}}
    We start with the gauge-covariant  derivatives in the standard $\mathcal{N}=4$ superspace
    \begin{equation}
        \nabla_{\alpha }^{I}=D_{\alpha }^{I}+i\mathcal{A}_{\alpha }^{I}, \qquad \bar \nabla_{\dot \alpha I}=\bar D_{\dot\alpha I}+i\mathcal{\bar A}_{\dot\alpha I},\qquad \nabla_{\a \db }= \partial_{\alpha \db}+i\mathcal{V}_{\alpha \db},
    \end{equation}
    where $\mathcal{A}_{\alpha }^{I}$, $\mathcal{\bar A}_{\dot\alpha I}$ and $\mathcal{V}_{\alpha \db}$ are spinor and vector superfield gauge connections.
    In $\mathcal{N}=4$ SYM theory these derivatives satisfy the constraints \cite{Son}
    \begin{eqnarray}
    &\{\nabla_{\alpha }^{I}, \nabla_{\beta }^J\}
    = -2i\epsilon_{\alpha\beta} W^{IJ},\nn
    &\{\bar \nabla_{\da I}, \bar\nabla_{\db J}\}
     = 2i\epsilon_{\da\db} \bar W_{IJ},\nn
    &\{\nabla_{\alpha}^I, \bar\nabla_{\dot\beta J}\} = -2i\delta^I_J\nabla_{\alpha\dot\beta}\,.
    \label{1.0}
    \end{eqnarray}
    Here $W^{IJ}=-W^{JI}$ is a real $\mathcal{N}=4$  superfield strength.
    The reality condition reads
    \begin{equation}
    \overline{W^{IJ}}=\bar W_{IJ}=\frac{1}{2}\varepsilon_{IJKL}W^{KL}.
    \label{1.1}
    \end{equation}
   The gauge connections and the superfield strengths in (\ref{1.0}) and  (\ref{1.1}) are defined up to gauge transformations
    \begin{equation}
    \mathcal{A'}^{I}_\alpha  =-ie^{i\tau}(\nabla^{I}_\alpha e^{-i\tau}) ,\qquad {W'}^{IJ}  = e^{i\tau}W^{IJ}e^{-i\tau},
    \label{1.6}
    \end{equation}
    where $\tau$ is a real $\mathcal{N}=4$ superfield parameter. Note that the condition (\ref{1.1}) breaks the $U(4)$ R-symmetry of the ``flat'' $\N=4$ superspace down to $SU(4)$.

    Next we rewrite the constraints (\ref{1.0}), (\ref{1.1}) in terms of indices $i, \ A$ according to the rule ($\ref{rule}$).
    Using the antisymmetry of the superfield strength $ W ^ {IJ} $ and the reality condition  (\ref{1.1}) we express  it in terms of few independent
    components:
    \begin{eqnarray}
    W^{ij}=\epsilon^{ij}W,&\qquad& \bar W_{ij}=-\epsilon_{ij}\bar W,\nn
    W^{AB}=\epsilon^{AB}\bar W,&\qquad& \bar W_{AB}=-\epsilon_{AB} W,\nn
    W^{iA}=-i\phi^{iA},&\qquad& \bar W_{iA}=
    i\phi_{iA}=i\epsilon_{ij}\epsilon_{AB}\phi^{jB}.
    \end{eqnarray}
    Then we plug these expressions back into the constraints (\ref{1.0}) and obtain
    \begin{eqnarray}
     \{\nabla_{\alpha }^{i}, \nabla_{\beta }^j\} = -2i\epsilon_{\alpha\beta}\epsilon^{ij} W\,,&  \quad&
    \{\bar\nabla_{\dot\alpha i}, \bar\nabla_{\dot\beta j}\} = -2i\epsilon_{\dot\alpha\dot\beta}\epsilon_{ij} \bar W\,, \nn
    \{\hat\nabla_{\alpha }^A, \hat \nabla_{\beta }^B\} = -2i\epsilon_{\alpha\beta}\epsilon^{AB}\bar W\,,  &\quad&
    \{\bar{\hat\nabla}_{\dot\alpha A}, \bar{\hat\nabla}_{\dot\beta B}\} = -2i\epsilon_{\dot\alpha\dot\beta}\epsilon_{AB}  W\,,\nn
    \{\nabla_{\alpha}^i, \bar\nabla_{\dot\beta j}\} = -2i\delta^i_j\nabla_{\alpha\dot\beta}\,,  &\quad&
    \{\hat\nabla_{\alpha }^A, \bar{\hat\nabla}_{\dot\beta B}\} = -2i\delta^A_B\nabla_{\alpha\dot\beta} \,, \nn
    \{\nabla_{\alpha}^i, \hat\nabla_{\beta }^B\} = -2\epsilon_{\alpha\beta} \phi^{iB}\,,  &\quad& \{\bar\nabla_{\dot\alpha i}, \bar{\hat\nabla}_{\dot\beta B}\} =- 2\epsilon_{\dot\alpha\dot\beta} \phi_{iB}\,,  \nn
    \{\nabla_{\alpha i}, \bar{\hat\nabla}_{\dot\beta B}\} &=& \{\bar\nabla_{\dot\alpha i}, \hat \nabla_{\beta B}\} = 0\,,
    \label{3.8}
    \end{eqnarray}
    where  the gauge connections are assumed to be rearranged in accord with the rule (\ref{rule}):
     \begin{eqnarray}
    &\nabla^i_\alpha=D^i_\alpha+i\mathcal{A}^i_\alpha,  \qquad
    \hat\nabla^A_\alpha=\hat D^A_\alpha+i\mathcal{\hat A}^A_\alpha,\nn
    &   \bar\nabla_{\dot \alpha j}=\bar D_{\dot\alpha j}+i\mathcal{\bar A}_{\dot\alpha j},  \quad
    \bar {\hat\nabla}_{\dot\alpha B}=\bar{\hat D}_{\dot\alpha B}+i \mathcal{\bar{\hat A}}_{\dot\alpha B},  \nn
    &\nabla_{\alpha \dot \alpha}=\partial_{\alpha \dot \alpha}+i\mathcal{V}_{\alpha \dot \alpha}. \label{GConn}
    \end{eqnarray}

    The constraints (\ref{3.8}) imply some important consequences following from the Bianchi identities. E.g., for the mixed-index superfield
    strength  $\phi_{iB}$ the Bianchi identity implies
    \begin{equation}
    \nabla_{\alpha (i} \phi_{j)B}=0, \quad \hat\nabla_{\alpha (A}\phi_{iB)}=0.
    \label{1.3}
    \end{equation}
    Indeed, let us write the Bianchi for $\nabla^i_\a$
    \begin{eqnarray}
    \{\nabla^j_\gamma   \{\nabla_{\alpha}^i, \hat\nabla_{\beta }^B\}\}+ \{\nabla^i_{\a} \{\hat\nabla_{\b}^B, \nabla_{\gamma }^j\}\}+\{\hat\nabla^B_{\b} \{\nabla_{\gamma}^j, \nabla_{\a }^i\}\}=0.
    \end{eqnarray}
    Substituting the constraints (\ref{3.8}) into it and symmetrizing over indices $i,\ j$, we obtain
    \begin{eqnarray}
    \epsilon_{\alpha\beta}\nabla^{(j}_\gamma \phi^{i)B}+\epsilon_{\gamma\beta}\nabla^{(i}_\alpha \phi^{j)B}=0 \Rightarrow  \nabla_{\alpha (i} \phi_{j)B}=0.
    \end{eqnarray}
    The second equation in (\ref{1.3}) is derived in a similar way.

    As the next step, we define the harmonic projections of the quantities appearing in (\ref{3.8}), (\ref{GConn}),
\bea
\nabla_{\alpha, \dot\alpha}^{\pm} = \nabla_{\alpha, \dot\alpha}^i u^{\pm}_i\,, \quad \nabla_{\alpha, \dot\alpha}^{\hat\pm} =
\hat\nabla_{\alpha, \dot\alpha}^A v^{\hat\pm}_A\,, \quad \phi^{\pm\hat\pm }=\phi^{iA}u^{\pm}_iv^{\hat\pm}_A\,, \;\quad \phi^{\pm\hat\mp }=\phi^{iA}u^{\pm}_iv^{\hat\mp}_A\,, \lb{ProjBi}
\eea
in terms of which the constraints (\ref{3.8}) can be equivalently rewritten as an extended set:
\begin{eqnarray}
    &(\rm{a})&  \;\{\nabla^+_\alpha,\nabla^+_\beta\}=\{\bar\nabla^+_{\dot\alpha},\bar\nabla^+_{\dot\beta}\}=\{\nabla^+_\alpha,\bar\nabla^+_{\dot\beta}\}=0,\nn
    &(\rm{b})& \;\{\nabla^{\hat +}_\alpha,\nabla^{\hat +}_\beta\}=\{\bar\nabla^{\hat +}_{\dot\alpha},\bar\nabla^{\hat +}_{\dot\beta}\}=\{\nabla^{\hat +}_\alpha,\bar\nabla^{\hat +}_{\dot\beta}\}=0,\nn
    &(\rm{c})& \;\{\nabla_\alpha^+,\nabla_\beta^{\hat +}\}=-2\epsilon_{\alpha \beta}\phi^{+\hat+},\qquad\{\bar \nabla_{\dot\alpha}^+,\bar\nabla_{\dot\beta}^{\hat +}\}=-2\epsilon_{\dot\alpha \dot \beta}\phi^{+\hat+},
    \label{3.11}\nn
    &(\rm{d})& \; \{\nabla_{\alpha}^+, \bar\nabla_{\dot\beta }^{\hat +}\}
    = \{\bar\nabla_{\dot\alpha }^{+}, \nabla_{\beta }^{\hat +}\} = 0,\nn
    &(\rm{e})& \;[\partial^{++},\nabla^{+}_{\alpha}]=[\partial^{\hat + \hat +},\nabla^{+}_{\alpha}]=[\partial^{++},\nabla^{\hat +}_{\alpha}]
    =[\partial^{\hat + \hat+},\nabla^{\hat +}_{\alpha}]=0,\nn
    &(\rm{f})& \;[\partial^{++},\bar \nabla^{+}_{\dot \alpha}]=[\partial^{\hat + \hat +},\bar \nabla^{+}_{\dot \alpha}]
    =[\partial^{++},\bar \nabla^{\hat +}_{\dot \alpha}]=[\partial^{\hat+ \hat+},\bar \nabla^{\hat +}_{\dot \alpha}]=0,\nn
    &(\rm{g})& \;[\partial^{++},\partial^{\hat{+} \hat +} ]=0\,.\lb{CBconstr}
    \end{eqnarray}

The equivalency can be shown in the following way which is quite common for the harmonic superspace formulations of the extended supersymmetric gauge theories (see \cite{litlink GIOS}).
 First, from (\ref{CBconstr}e) and (\ref{CBconstr}f) it follows that $\nabla_{\alpha, \dot\alpha}^{\pm}$ and $\nabla_{\alpha, \dot\alpha}^{\hat\pm}$
are linear in the harmonics $u^+_i$ and $v^{\hat +}_A$, $\nabla_{\alpha, \dot\alpha}^{+} = \nabla_{\alpha, \dot\alpha}^i u^+_i$ and
$\nabla_{\alpha, \dot\alpha}^{\hat +} = \hat\nabla_{\alpha, \dot\alpha}^A v^{\hat +}_A$. Then, from (\ref{CBconstr}a) and (\ref{CBconstr}b),
the first three lines in the constraints (\ref{3.8}) follow (e.g., (\ref{CBconstr}a) implies $\{\nabla_\alpha^{(i}, \nabla^{j)}_\beta\} = 0$, etc.) From (\ref{CBconstr}c) and the proper Bianchi identity (see below) the fourth line in
(\ref{3.8}) follows.  At last, (\ref{CBconstr}d) implies the fifth line. The negatively charged objects can be obtained from the positively charged ones by the action
of the harmonic derivatives $\partial^{--}, \partial^{\hat- \hat-}$.

The Bianchi identity mentioned above is obtained by commuting the proper spinor derivatives
with both sides of eq. (\ref{CBconstr}c). It implies
\begin{equation}
\nabla^+_\alpha\phi^{+\hat+ }=\nabla^{\hat{+}}_\alpha\phi^{+\hat+ }=\bar\nabla^+_{\dot\alpha}  \phi^{+\hat+ }=\bar\nabla^{\hat{+}}_{\dot\alpha} \phi^{+\hat+ }=\partial^{++}\phi^{+\hat+ }=\partial^{\hat{+}\hat{+}}\phi^{+\hat+ }=0.
    \label{1.4}
    \end{equation}
    These relations are equivalent to the identities (\ref{1.3}). Indeed, given a real superfield $\phi^{+\hp}$ satisfying (\ref{1.4}),
    it can be written as $\phi^{iA}u^+_iv^{\hp}_A$ with $\phi^{iA}$ satisfying (\ref{1.3}). In particular, the last two relations in (\ref{1.4})
    just imply that $\phi^{+\hp} =\phi^{iA}u^+_iv^{\hp}_A$.

Note that the constraints (\ref{CBconstr}) are written in the central basis of $\N=4$ bi-harmonic superspace, with ``short'' harmonic derivatives $D^{++} = \partial^{++}$ and
$D^{\hat +\hat +} = \partial^{\hat + \hat +}$. However, their form cannot depend on the choice of the basis, so in what follows we will use the general notation $D^{\pm\pm}$ and
$D^{\hat\pm\hat\pm}$ for the harmonic derivatives.

    \subsection{The analytic frame  \label{s2}}

    Following the generalities of the harmonic superspace approach, the crucial step now is passing to the analytic frame  where it will become possible
    to solve the constraints (\ref{CBconstr}) in terms of the appropriate analytic gauge superfields and to express the superfield strengths $ \phi ^ {+ \hat +} $, $W$, $\bar W$
    in terms of these fundamental objects. If some of the harmonic projections of the gauge-covariant spinor derivatives form an anti-commutative subset, the relevant spinor connections are pure gauge and one can always
    choose a frame where these derivatives coincide with the ``flat'' ones, {\it i.e.} involve no gauge superconnections. Clearly, such an anticommuting  subset of spinor derivatives
    is in a one-to-one correspondence
    with the existence of some analytic subspace in the given harmonic superspace.

    In our case, because of the constraint  $\{\nabla_\alpha^+,\nabla_\beta^{\hat +}\}=-2\epsilon_{\alpha \beta}\phi^{+\hat+}$, it is impossible
    to simultaneously make ``flat'' (having no gauge superconnections)
    all the positively charged spinor derivatives. Maximum what one can reach is to remove the gauge connections either
    from $\nabla^{+}_{\alpha, \dot\alpha}$ or from $\nabla^{\hat +}_{\alpha, \dot\alpha}$. Without loss of generality, we will chose the frame in which
    the derivatives $\nabla^{\hat +}_\alpha$ and $\bar\nabla^{\hat +}_{\dot\alpha}$ coincide with the flat ones, so that the  $\zeta_{II}$ analyticity from the sets (\ref{BiAnal}) can be made manifest.

    Thus, consider the constraints $\{\nabla^{\hat +}_\alpha,\nabla^{\hat +}_\beta\}=\{\bar\nabla^{\hat +}_{\dot\alpha},\bar\nabla^{\hat +}_{\dot\beta}\}=\{\nabla^{\hat +}_\alpha,\bar\nabla^{\hat +}_{\dot\beta}\}=0$.
    Their general solution reads
    \begin{eqnarray}
    \nabla^{\hat +}_\alpha=e^{iV}D^{\hat + }_\alpha e^{-iV},\qquad  \bar \nabla^{\hat +}_{\dot \alpha}=e^{iV}\bar D^{\hat + }_{\dot \alpha}
    e^{-iV} \; \Longrightarrow \;\mathcal{A}_{\alpha, \dot\alpha}^{\hat +} = -ie^{iV}(D^{\hat + }_{\alpha, \dot\alpha} e^{-iV})\,,
    \end{eqnarray}
    where  $V$ is a real ``bridge'' superfield ($V=\widetilde{V}$) with the following gauge transformation law
    \begin{equation}
    e^{iV'}=e^{i\tau}e^{iV}e^{i\Lambda},
    \label{1.5}
    \end{equation}
    where $\Lambda$ is  ``$\zeta_{II}$''  analytic superfield, $\Lambda=\Lambda(\zeta_{II})$,
   and $\tau$ is a general real, harmonic-independent ($\partial^{++}\tau = \partial^{\hat + \hat +}\tau =0$ in the central basis),  $\N=4$ superfield.
    Now we perform the similarity transformation
    \begin{eqnarray}
    &\nabla^{\hat +}_{\alpha}\rightarrow e^{-iV}\nabla^{\hat +}_\alpha e^{iV} =  D^{\hat +}_{\alpha}, \qquad \bar \nabla^{\hat +}_{\dot \alpha}\rightarrow e^{-iV}\bar\nabla^{\hat +}_{\dot\alpha} e^{iV}
    = \bar D^{\hat +}_{\dot \alpha}, \nn
    &\nabla^+_{\alpha}\rightarrow e^{-iV}\nabla^{+}_{\alpha}e^{iV}, \qquad \bar\nabla^+_{\dot\alpha}\rightarrow e^{-iV}\bar \nabla^{+}_{\alpha}e^{iV},
    \quad\phi^{+\hat +} \rightarrow e^{-iV}\phi^{+\hat +}e^{iV} \lb{AnalPass1}
    \end{eqnarray}
    and
    \begin{eqnarray}
    && D^{++} \rightarrow \nabla^{+ +}=e^{-iV}D^{+ +}e^{iV} :=D^{+ +}+i {V}^{+ +}, \nn
    && D^{\hat+\hat+} \rightarrow \nabla^{\hat+\hat +}
    =e^{-iV}D^{\hat+\hat +}e^{iV}:=D^{\hat+\hat +}+i V^{\hat +\hat +},\lb{DD} \\
    && {V}^{++}=-ie^{-iV}\left(D^{+ +}e^{iV}\right), \quad V^{\hat +\hat +}=-ie^{-iV}\left(D^{\hat+\hat +}e^{iV}\right),\lb{VV}
    \end{eqnarray}
    where ${V}^{++},\ V^{\hat + \hat +}$ are real bi-harmonic superfields. The transformed spinor and harmonic  derivatives satisfy
    the same algebra (\ref{CBconstr})
 \begin{eqnarray}
    &\{\nabla^+_\alpha,\nabla^+_\beta\}=\{\bar\nabla^+_{\dot\alpha},\bar\nabla^+_{\dot\beta}\}=\{\nabla^+_\alpha,\bar\nabla^+_{\dot\beta}\}=0,
    \label{1}\\
    &\{D^{\hat +}_\alpha,D^{\hat +}_\beta\}=\{\bar D^{\hat +}_{\dot\alpha},\bar D^{\hat +}_{\dot\beta}\}=\{ D^{\hat +}_\alpha,\bar D^{\hat +}_{\dot\beta}\}=0, \label{2}\\
    &\{\nabla_\alpha^+,D_\beta^{\hat +}\}=-2\epsilon_{\alpha \beta} \phi^{+\hat+},\qquad\{\bar \nabla_{\dot\alpha}^+,\bar D_{\dot\beta}^{\hat +}\}=
    -2\epsilon_{\dot\alpha \dot \beta} \phi^{+\hat+},  \label{3}\\
    &\{\nabla_{\alpha}^+, \bar D_{\dot\beta }^{\hat +}\} = \{\bar\nabla_{\dot\alpha }^{+}, D_{\beta }^{\hat +}\} = 0,
    \label{4}\\
    &[\nabla^{++},\nabla^{+}_{\alpha}]=[\nabla^{\hat + \hat +},\nabla^{+}_{\alpha}]=[ \nabla^{++},D^{\hat +}_{\alpha}]=[\nabla^{\hat + \hat+},D^{\hat +}_{\alpha}]=0,\label{5}\\
    &[\nabla^{++},\bar \nabla^{+}_{\dot \alpha}]=[\nabla^{\hat + \hat +},\bar \nabla^{+}_{\dot \alpha}]=[\nabla^{++},\bar D^{\hat +}_{\dot \alpha}]
    =[\nabla^{\hat+ \hat+},\bar D^{\hat +}_{\dot \alpha}]=0, \label{6}\\
    &[\nabla^{++},\nabla^{\hat{+} \hat +} ]=0. \label{7}
    \end{eqnarray}
    This is the final form of the $\N=4$ SYM constraints we will deal with in what follows. It involves two harmonic connections $V^{++}, V^{\hat + \hat +}$ defined in (\ref{VV}) and
    the spinorial connections $\mathcal{A}^+_\alpha\,, \bar{\mathcal{A}}^+_{\dot\alpha}$ entering the gauge-covariant spinor derivatives
    \bea
    \nabla_\alpha^+ = D^+_\alpha + i\mathcal{A}^+_\alpha\,, \quad \bar\nabla_{\dot\alpha}^+ = \bar D^+_{\dot\alpha} + i\bar{\mathcal{A}}^+_{\dot\alpha}\,.
    \eea
    It will be convenient to choose the analytic basis in $\N=4$ bi-harmonic superspace, where  $D^{\hat +}_\alpha=\partial/\partial\theta^{\hat -\alpha}$,
    $\bar D^{ \hp}_{\dot\alpha}=\partial/\partial\bar\theta^{ \hat-\dot\alpha}$ and the $\zeta_{II}$ analyticity is manifest\footnote{In
    the analytic basis, the spinor derivatives $D^+_\alpha, \bar D^+_{\dot\alpha}$ are also "short".}.

    The main advantage of the analytic frame and basis is that, in virtue of the relations (\ref{5}), (\ref{6}) (the last two in both chains), the harmonic connections
    ${V}^{++}$ and $V^{\hat +\hat +}$ live on the reduced subspace $\zeta_{II}$,
    \bea
    && D^{\hat +}_{\alpha}{V}^{++} = \bar D^{\hat +}_{\dot \alpha}{V}^{++} = 0\,, \quad D^{\hat +}_{\alpha}{V}^{\hat + \hat +} = \bar D^{\hat +}_{\dot \alpha}{V}^{\hat +\hat +} = 0\,, \Rightarrow \nn
    && {V}^{++} = {V}^{++}(\zeta_{II}), \;\; {V}^{\hat +\hat +} = {V}^{\hat + \hat +}(\zeta_{II}), \lb{AnalitHat}
    \eea
    {\it i.e.} they {\it do not depend} on the Grassmann coordinates $\theta^{\hat -}_\alpha, \bar\theta_{\dot\alpha}^{\hat -}\,$ in the analytic basis.

    The equations (\ref{1.4}) in the analytic frame are rewritten as
    \begin{equation}
    \nabla^+_\alpha\phi^{+\hat+ }=D^{\hat{+}}_\alpha\phi^{+\hat+ }=\bar\nabla^+_\da\phi^{+\hat+ }=\dD^{\hat{+}}_\da\phi^{+\hat+ }=\nabla^{++}\phi^{+\hat+ }
    =\nabla^{\hat{+}\hat{+}}\phi^{+\hat+ }=0,
    \label{1000}
    \end{equation}
    which  are also Bianchi identities for the constraints (\ref{1})-(\ref{7}). We see that
    $$
    \phi^{+\hat+ } = \phi^{+\hat+ }(\zeta_{II})\,,
    $$
    like the harmonic connections ${V}^{++}$ and ${V}^{\hat + \hat +}$.
    In the next section we will solve the equations (\ref{1000}) and the constraints  (\ref{1})-(\ref{7}).

    Using the gauge transformations (\ref {1.6}) and (\ref {1.5}), one can find the transformation laws of the analytic-frame harmonic and spinor connections, as well as of the superfield strength $\phi^{+\hat +}$
    \begin{eqnarray}
    && \delta  {V}^{++}=\nabla^{++}\Lambda (\zeta_{II}), \quad \delta V^{\hat + \hat +}=\nabla^{\hat + \hat +}\Lambda (\zeta_{II}), \label{1.7} \\
    && \delta \phi^{+\hat +}=-i[\Lambda (\zeta_{II}),\phi^{+\hat +}], \quad
    \delta \mathcal{A}^+_{\alpha, \dot\alpha} =\nabla^{+}_{\alpha, \dot\alpha}\Lambda (\zeta_{II})\,. \;\;\label{TranAanal}
    \end{eqnarray}

    \subsection{Gauge fixings \label{s3.1}}
    Before solving the constraints (\ref {1})-(\ref {7}), some preliminary steps are needed.
    At this stage the harmonic connections $  V ^ {++} $ and $ V^{\hat + \hat +} $ are arbitrary functions of the ``hat''-analytic coordinates
    $ \theta ^ {\hat +}_\alpha, \bar \theta ^ {\hat +} _ {\dot \alpha}$ and harmonics $ v ^ {\hat \pm} _A $ (along with the dependence on other coordinates of the
    analytic subspace $\zeta_{II}$, see (\ref{BiAnal})).

    Now we show that the dependence of $ V ^ {\hat + \hat +} $ on
    $ \theta ^ {\hat +} _ \alpha, \bar \theta ^ {\hat +} _ {\dot \alpha} $ and $ v^{\hat \pm} _A $ can
    be  reduced by choosing a Wess-Zumino gauge with respect to the transformations (\ref{1.7}).

    It is straightforward to see that  the gauge freedom associated with the superfield
    transformation parameter  $\Lambda(\zeta_{II})$ can be partially fixed by casting $V^{\hat +\hat +}$ in the short form
    \begin{eqnarray}
    V^{\hat+ \hat+}&=&-2i\theta^{\hat +}_{\alpha}\bar\theta^{\hat +}_{\dot \alpha}\mathcal{\hat A}^{\alpha \dot \alpha}+(\theta^{\hat+})^2\W+ (\bar \theta^{\hat+})^2{\bW}+
    2(\bar \theta^{\hat +})^2\theta^{\hat +\alpha}\psi^{\hat -}_\alpha+2( \theta^{\hat +})^2\bar \theta^{\hat +}_{\dot\alpha}\widetilde \psi^{\hat -\dot\alpha}\nn
    &&+\,3(\theta^{\hat+})^2(\bar \theta^{\hat+})^2\mathcal{D}^{\hat-2},
    \label{8} \\
    &&\psi^{\hat -}_\alpha=\psi^{A}_\alpha v^{\hat -}_A, \quad
    \widetilde \psi^{\hat -\dot\alpha}=\widetilde \psi^{\dot\alpha}_A v^{\hat -A}=-\widetilde \psi^{A\dot\alpha} v^{\hat -}_A, \quad
    \mathcal{D}^{\hat-2}=\mathcal{D}^{(AB)}v^{\hat -}_Av^{\hat -}_B, \quad \bW=\widetilde \W. \label{vDependWZ}
    \end{eqnarray}
    Here the superfields $\mathcal{\hat A}^{\alpha\dot{\alpha}}$, $\psi^{A}_\da$, $\W$ and $\mathcal{D}^{(AB)}$
    are defined on the coordinate set $(x_{\rm an}^m, \theta^{\pm}_\alpha, \bar\theta^{\pm}_{\dot \alpha}, u_i^{\pm})$. While passing to
(\ref{8}),
    the dependence of $\Lambda (\zeta_{II})$ on $(\theta^{\hat +}_\alpha, \bar \theta^{\hat +}_{\dot{\alpha}},v^{\hat \pm}_A)$
    has been fully spent, so the residual gauge freedom is connected with the gauge function
    $\Lambda_{\text{int}}(x_{\rm an}^m, \theta^{\pm}_\alpha, \bar\theta^{\pm}_{\dot \alpha}, u_i^{\pm})$, $\Lambda (\zeta_{II}) \rightarrow \Lambda_{\text{int}}$.
    Below we show that the dependence of $\Lambda_{\text{int}}$ on $\theta^{-}_\alpha, \bar\theta^{ -}_{\dot{\alpha}}$ can also be fully spent for a proper gauge choice.

    To this end, we need to inspect the structure of the spinor derivative. First, let us examine the spinor part of the relations (\ref{1000}), namely
    \begin{equation}
    (\text{a})\;D^{\hat{+}}_\alpha\phi^{+\hat+ }=\bar D^{\hat{+}}_{\dot\alpha} \phi^{+\hat+ }=0, \qquad (\text{b})\;\nabla^+_\alpha\phi^{+\hat+ }=\bar\nabla^+_{\dot\alpha}  \phi^{+\hat+ }=0.
    \label{1.8}
    \end{equation}
    In this subsection we focus on eq. (\ref{1.8}a), the consequences of (\ref{1.8}b) will be discussed later (in subsection \ref{s3.2.2}).
    As was mentioned earlier,  it follows from (\ref{1.8}a) that $\phi^{ +\hat +}$ does not depend on $\theta^{\hat -}_\alpha$ and $\bar \theta^{\hat -}_{\dot{\alpha}}$.
     In addition, using this property in the constraints (\ref3) and (\ref4) implies that
    \begin{equation}
    \mathcal{A}^+_\alpha= A^+_\alpha+2i\phi^{+\hat{+}}\theta^{\hat -}_\alpha,
    \qquad
    \mathcal{\bar A}^+_{\dot \alpha}=\bar{ {A}}^+_{\dot \alpha}
    +2 i\phi^{+\hat{+}}\bar \theta^{\hat -}_{\dot\alpha},
    \label{9}
    \end{equation}
    where $ A^+_\alpha,\ \bar{ {A}}^+_{\dot \alpha}$ do not depend on $\theta^{\hat -}_\alpha$ and $\bar \theta^{\hat -}_{\dot \alpha}$
    and so can be represented as
    \begin{eqnarray}
    A^{+}_\beta &=& f^{+}_\beta+
    \theta^{\hat + \alpha}f^{+\hat -}_{\alpha\beta}+
    \bar \theta^{\hat +}_{\dot \alpha} g^{+\hat - \dot\alpha}_{\beta}+
    ( \theta^{\hat +})^2f^{+\hat{-}2}_{\beta}+
    (\bar \theta^{\hat +})^2g^{+\hat{-}2}_{\beta}+\theta^{\hat+}_{\alpha}\bar\theta^{\hat+}_{\dot\alpha}
    f^{+\hat-\hat -\alpha\dot{\alpha}}_{\beta} \nn
    && +\,(\bar \theta^{\hat +})^2\theta^{\hat +\alpha}f_{\alpha\beta}^{+\hat{-}3}
    +( \theta^{\hat +})^2 \bar \theta^{\hat +}_{\dot\alpha}
    g^{+\hat{-}3\dot \alpha}_{\beta}
    +
    ( \theta^{\hat +})^2(\bar \theta^{\hat +})^2f^{+\hat -4}_{\beta}.
    \label{788}
    \end{eqnarray}
    Note, that $\bar{ {A}}^+_{\dot \alpha}=-\widetilde{ {A}^+_{\dot
    \alpha}}\,$. All the coefficients in the expansion (\ref{788}) at this stage are arbitrary functions of the remaining coordinates
    $(x_{\rm an}^m, \theta^{\pm}_\alpha, \bar\theta^{\pm}_{\dot \alpha}, u_i^{\pm}, v_A^{\hat\pm})$. Below we will show that all terms except
    the second one can be eliminated either by the constraints or by choosing an additional gauge.

    This additional gauge-fixing will be imposed right now and it goes as follows. First note that in the zeroth order in $\theta^{\hat \pm}$ the
    constraint $[\nabla^{\hat + \hat +},\nabla^{+}_{\alpha, \dot\alpha}] = 0$ in eqs. (\ref{5}), (\ref{6}) implies
    \be
    \partial^{\hat + \hat +}f_{\alpha, \dot\alpha}^+ = 0\,, \lb{1stConseq}
    \ee
which means that $f_{\alpha, \dot\alpha}^+ $ do not depend on the harmonics $v^{\hat \pm}_A$, {\it i.e.} these objects  ``live'' on
the coordinate set $(x_{\rm an}^m, \theta^{\pm}_\alpha, \bar\theta^{\pm}_{\dot \alpha}, u_i^{\pm})$.
    On the other hand, after substituting (\ref{9}) and  (\ref{788}) into the constraint (\ref{1}), we obtain
    \begin{equation}
    D^+_\alpha f^+_\beta+D^+_\beta f^+_\alpha+i\{f^+_\alpha,f^+_\beta\}=0\,,\quad   \bar D^+_{\dot\alpha} f^+_\beta + D^+_\beta \bar f^+_{\dot\alpha}
    +i\{f^+_\alpha, \bar f^+_{\dot\beta}\}=0 \quad ({\rm and \; c.c.}).
    \label{10}
    \end{equation}
    It stems from (\ref{10}) that $f^+_{\alpha, \dot\alpha} = -ie^{i\tilde v}(D^+_{\alpha, \dot\alpha}  e^{-i\tilde v})$, where $\tilde v$ is
    an additional bridge living on the same coordinate set as
    the superfields $f^+_{\alpha, \dot\alpha}$. It transforms as $e^{i\tilde{v}'} = e^{-i\Lambda_{\rm int}}e^{i\tilde{v}} e^{i\Lambda(\zeta)}$, where $\Lambda_{\rm int}$
    was defined after eq. (\ref{vDependWZ}), while the pregauge freedom parameter $\Lambda(\zeta)$ satisfies  the conditions
    $D^{+}_\alpha\Lambda=\bar D^{+}_{\dot \alpha}\Lambda=0\,$ and so can be identified with the $\N=2$ harmonic analytic gauge group parameter.
     Using the newly introduced  bridge,  one can pass  to the frame where
     \be
     f^+_{\alpha, \dot\alpha}=0 \label{2ndGauge}
     \ee
    and the residual gauge group is reduced to the standard $\mathcal{N}=2$ SYM analytic gauge group,  $\Lambda_{\rm int} \rightarrow \Lambda(\zeta)$.
    Actually, this passing can be equivalently interpreted as the gauge choice $\tilde{v} = 0\, \Rightarrow  \, \Lambda_{\rm int} = \Lambda(\zeta)$.

Hereafter we use the spinor connections $ \mathcal{A}^{+}_{\alpha, \dot\alpha} $ in the form (\ref{9}), (\ref{788}) with the condition
    $f^+_{\alpha, \dot\alpha}=0$ and the following $\theta^{\hat +}_\alpha,\bar \theta^{\hat{+}}_{\dot \alpha}$ expansions for  the  ``hat''-analytic
    superfields $\phi^{+\hat{+}}$ and $V^{++}$:
    \begin{eqnarray}
    \phi^{+\hat +}&=& \sqrt{2}q^{+\hat+}+
    \theta^{\hat + \alpha}\W^{+}_{\alpha}+
    \bar \theta^{\hat +}_{\dot \alpha}\widetilde \W^{+\dot\alpha}+
    ( \theta^{\hat +})^2H^{+\hat{-}}+
    (\bar \theta^{\hat +})^2\widetilde H^{+\hat -} -i\theta^{\hat+}_{\alpha}\bar\theta^{\hat+}_{\dot\alpha}
    \beta^{+\hat-\alpha\dot{\alpha}} \nn
    && +\,(\bar \theta^{\hat +})^2\theta^{\hat +\alpha}G_{\alpha}^{+\hat{-}\hat{-}}
    +( \theta^{\hat +})^2 \bar \theta^{\hat +}_{\dot\alpha}
    \widetilde G^{+\hat{-}\hat{-}\dot \alpha}
    +
    ( \theta^{\hat +})^2(\bar \theta^{\hat +})^2G^{+\hat -3},
    \label{11}\\
    V^{+ +}&=&{\V}^{++}+
    \theta^{\hat + \alpha}w^{++\hat-}_{\alpha}+
    \bar \theta^{\hat +}_{\dot \alpha}\widetilde w^{++\hat -\dot\alpha}+
    ( \theta^{\hat +})^2w^{++\hat{-}\hat -}+
    (\bar \theta^{\hat +})^2\widetilde w^{++\hat -\hat{-}} +\theta^{\hat+}_{\alpha}\bar\theta^{\hat+}_{\dot\alpha}
    w^{++\hat-\hat -\alpha\dot{\alpha}}\nn
    && +\,(\bar \theta^{\hat +})^2\theta^{\hat +\alpha}w_{\alpha}^{++\hat{-}3}
    +( \theta^{\hat +})^2 \bar \theta^{\hat +}_{\dot\alpha}
    \widetilde w^{++\hat{-}3\dot \alpha}
    +
    ( \theta^{\hat +})^2(\bar \theta^{\hat +})^2w^{++\hat -4}.
    \label{12}
    \end{eqnarray}
    The superfield coefficients in these expansions will be shown to be severely constrained.  At the moment,
    they are just  $\mathcal {N} = 2 $ harmonic superfields with an extra dependence on the harmonics $ v ^ {\hat \pm} _A $,
    {\it i.e.} defined on the set $(x_{\rm an}^m, \theta^{\pm}_\alpha, \bar\theta^{\pm}_{\dot \alpha}, u_i^{\pm}, v_A^{\hat\pm})$.

    \section{Solving  $\mathcal{N}=4$ SYM constraints in terms of $\mathcal{N}=2$ superfields  \label{s3}}
    In this section we will finish solving the constraints (\ref{1})-(\ref{1000}).
    \subsection{Harmonic equations \label{s3.2}}

    \subsubsection{Constraint $[\nabla^{\hat +\hat +}, \nabla^{++}] = 0$  \label{s3.2.1}}
    We start by showing that  $ V^{++}$ (\ref{12}) in fact does not depend on
    the coordinates $\theta^{\hat +}_\alpha,\bar \theta^{\hat{+}}_{\dot \alpha}, v^{\hat \pm}_A$.
    This follows from the constraint (\ref{7}) which in a more detailed form reads
    \begin{equation}
    D^{++}V^{\hat{+}\hat +}-D^{\hat + \hat +} V^{++}+i[ V^{++},V^{\hat + \hat +}]=0.
    \label{13}
    \end{equation}
        Substituting the expansions (\ref{12}) and  (\ref 8) in (\ref{13}) and equating to zero the coefficients of the
        $\theta^{\hat +\alpha},\ \bar\theta^{\hat +}_{\dot \alpha}$ monomials in the resulting expression, we obtain the set of equations
    \begin{eqnarray}
   && \partial^{\hat{+}\hat +} \V^{++}=0\,, \qquad \partial^{\hat{+}\hat +}w^{++\hat -}_\alpha=0\,, \qquad \partial^{\hat{+}\hat +}\widetilde w^{++\hat -\dot \alpha}=0\,,
    \label{14}\\
   && \partial^{\hat+ \hat +}w^{++\hat -\hat -}- D^{++}\W-i[ \V^{++},\W]=0\,,\quad ({\rm and\;\,c.c.}),
    \label{15}\\
  && \partial^{\hat + \hat +}w^{++\hat{-}\hat -\alpha\dot{\alpha}} + 2iD^{++}\mathcal{\hat A}^{\alpha \dot \alpha}-2[\V^{++},\mathcal{\hat A}^{\alpha \dot \alpha}]
    -2i\partial^{\alpha \dot \alpha}{\V}^{++}=0\,,
    \label{16}\\
  && \partial^{\hat + \hat +}w^{++\hat{-}3}_\alpha - 2D^{++}\psi^{\hat -}_\alpha -2i[\V^{++},\psi^{\hat -}_\alpha]=0\,,\quad ({\rm and\;\,c.c.}),
    \label{17}\\
   && \partial^{\hat +\hat+}w^{++\hat-4} -3D^{++}\mathcal{D}^{\hat - 2}-3i[\V^{++},\mathcal{D}^{\hat - 2}]-
    i[w^{++\hat{-}\hat{-}},\bW]-i[\widetilde w^{++\hat{-}\hat{-}},\W]=0.
    \label{18}
    \end{eqnarray}

    The last two equations in (\ref{14}) imply
    \begin{equation}
    w^{++\hat-}_{\alpha}=\widetilde w^{++\hat-}_{\dot\alpha}=0.
    \label{20}
    \end{equation}
        In addition, the first equation implies $ \V^{++}$ to bear no dependence on $v^{\hat \pm}_A$.
        Inspecting the $\theta^{\hat \pm\alpha}, \bar\theta^{\hat \pm}_{\dot \alpha}$-independent parts of the first constraints in the chains (\ref{5}) and (\ref{6}),
        one also observers that the superfield $\V^{++}$ is ${\cal N}=2$ analytic,  $D^{+}_\alpha \V^{++}=\bar D^{+}_{\dot \alpha} \V^{++}=0$, and in fact
        already at this stage can be identified with the analytic harmonic gauge connection of $\N=2$ SYM theory.

    Eqs. (\ref{15})--(\ref{18}) further imply
    \begin{equation}
    w^{++\hat{-}\hat{-}}=\widetilde w^{++\hat{-}\hat{-}}= w^{++\hat{-}\hat{-}\alpha\dot{\alpha}}=w^{++\hat{-}3 \alpha}=w^{++\hat{-}4}=0.
    \label{21}
    \end{equation}

     Thus, we found
    \begin{equation}
    V^{++}\equiv{\V}^{++}, \quad \Rightarrow \quad \nabla^{++} = D^{++} + i \mathcal{V}^{++}
    \label{26}
    \end{equation}

    Eqs.  (\ref{15})-(\ref{18}) also encode some other consequences appearing in the zeroth order in $v^{\hat \pm}_A$
    \begin{eqnarray}
    & \nabla^{++}\W=  \nabla^{++}{\bW}=0,
    \label{22}\\
    &\nabla^{++}\mathcal{\hat A}^{\alpha \dot \alpha}=\partial^{\alpha \dot \alpha}{\V}^{++},
    \label{24}\\
    &\nabla^{+ +}\psi^{A}_{\alpha} = \nabla^{+ +}\tilde{\psi}^{A}_{\dot\alpha}=0,\qquad {\nabla}^{++}\mathcal{D}^{(AB)}=0.
    \label{25}
    \end{eqnarray}
Note, that (\ref {24}) is equivalent to the vanishing of the commutator
   \begin{equation}
   [\nabla^{++},\hat{\nabla}^{\alpha\dot{\alpha}}]=0, \qquad \hat{\nabla}^{\alpha\dot{\alpha}}=\partial^{\alpha \dot \alpha}+
    i\mathcal{\hat A}^{\alpha\dot\alpha}.
    \label{19}
    \end{equation}

    Thus, the constraint (\ref{7})  has been fully resolved. The validity of the conditions (\ref{22}) - (\ref{25}) on the final solution of all constraints
    will become clear later, in the end of subsection \ref{s4}.
    \subsubsection{Constraints $\nabla^{\hat+\hat+}\phi^{+\hat+}=0$  and $\nabla^{++}\phi^{+\hat+}=0$ \label{s3.2.2}}
    Our task here is to further specify the structure of  spinor connection (\ref{9}). It involves the superfield  $\phi^{+\hat +}$. Consider it in more details.
    Besides the analyticity conditions (\ref{1.8}), it satisfies the harmonic equations
    \begin{equation}
    (\text{a})\;\nabla^{\hat+\hat+}\phi^{+\hat+}=0, \qquad(\text{b})\;\nabla^{++}\phi^{+\hat+}=0.
    \label{27}
    \end{equation}

    We start with (\ref{27}a). Substituting  the expansions of $\phi^{+\hat{+}}$ and $V^{\hat{+}\hat{+}}$ from eqs. (\ref{11}) and (\ref{8}), one obtains the set of equations and their solutions
    \begin{eqnarray}
    && \partial^{\hat +\hat +} q^{+\hat+}=0 \;\Longrightarrow\; q^{+\hat +}=q^{+A}v^{\hat{+}}_A,
    \label{28}\\
   &&  \partial^{\hat +\hat +} \W^{+}_{\alpha}=\partial^{\hat +\hat +}\widetilde \W^{+ \dot\alpha}=0,
    \label{29}\\
    && \partial^{\hat +\hat +}H^{+\hat -}+\sqrt 2 i[\W,q^{+A}]v^{\hat +}_A=0  \;\Longrightarrow\; H^{+\hat -}=-\sqrt 2 i[\W,\,q^{+A}]v_A^{\hat -}\,,
    \label{30}\\
   && \partial^{\hat + \hat +}\beta^{+\hat- \alpha \dot\alpha}
    +2\sqrt{2}\hat\nabla^{\alpha\dot{\alpha}}q^{+ A}v^{\hat+}_A=0\;
    \Longrightarrow\; \beta^{+\hat -\alpha \dot \alpha}=-2\sqrt{2}\hat\nabla^{\alpha\dot{\alpha}}q^{+A}v_A^{\hat-},
     \label{31}\\
   && \partial^{\hat+\hat+}G^{+\hat -\hat -}_\alpha+i\hat{\nabla}_{\alpha\dot{\alpha}}
    \widetilde \W^{+\dot{\alpha}}+2\sqrt{2}i[\psi^{\hat -}_{\alpha},q^{+A}]v_A^{\hat{+}}+i[\bW,\,\W^{+}_{\alpha}]=0\,,
    \label{32}\\
    && 2\partial^{\hat +\hat +}G^{+\hat -3}-\hat\nabla^{\alpha \dot \alpha}\beta^{+\hat -}_{\a\da}-2i\{\psi^{\hat -\alpha},\W^{+}_{\alpha}\}
    -2i\{\widetilde \psi^{\hat -}_{\dot \alpha},\widetilde \W^{+ \dot \alpha}\}\nn
    &&+\,6i\sqrt{2}[\mathcal{D}^{AB}, q^{+C}]v^{\hat -}_Av^{\hat -}_Bv^{\hat +}_C +2\sqrt{2}[ \bW,[\W,q^{+A}]] v^{\hat -}_A
    +2\sqrt{2}[\W,[ \bW,q^{+A}]]v^{\hat -}_A =0\,.
    \label{3444}
    \end{eqnarray}
    Eqs. (\ref {32}), (\ref {3444}) constrain the superfields $G^{+\hat-\hat-}_\alpha$ and $G^{+\hat-3}$ as
    \begin{eqnarray}
    G^{+\hat{-}\hat -}_{\alpha}=G^{+(AB)}_{\alpha}v^{\hat -}_Av^{\hat -}_B, \qquad G^{+(AB)}_{\alpha}=-\sqrt{2}i[\psi^{(A}_{\alpha},q^{+B)}],
    \label{35}\\
    G^{+\hat-3}=G^{+(ABC)}v^{\hat -}_Av^{\hat -}_Bv^{\hat -}_C, \qquad G^{+(ABC)}=-i\sqrt{2}[\mathcal{D}^{(AB},q^{+C)}].
    \label{37}
    \end{eqnarray}
    While finding these solutions, we used the relations
    \begin{equation}
        v^{\hat +}_Av^{\hat -}_B=v^{\hat +}_{(A}v^{\hat -}_{B)}+\frac{1}{2}\epsilon_{AB}, \qquad v^{\hat -}_A v^{\hat -}_Bv^{\hat +}_C=v^{\hat -}_{(A}v^{\hat -}_B v^{\hat +}_{C)}
        +\frac{1}{3}\left(\epsilon_{CA}v^{\hat -}_B+\epsilon_{CB}v^{\hat -}_A\right).
    \end{equation}
    Eq. (\ref {29}) implies the independence of $\W ^ {+} _ \alpha, \tilde{\W} ^{+} _{\dot\alpha }$ from $ v^{\hat \pm}_A$. The constraint (\ref{32}) and  (\ref{3444}),
    in the zeroth and first orders in $v^{\hat \pm}_A$, also imply some self-consistency conditions which are quoted in the appendix \ref{s7.2}.
    These conditions do not bring any new information, but must be satisfied on the final solution of all constraints (like eqs. (\ref{22}) - (\ref{25})), and so they provide
    a good self-consistency check.

    Thus, we have completely fixed the  $ v ^ {\hat \pm}_A $ dependence of the coefficients in the $ \theta ^ {\hat +} _ \alpha, \bar \theta ^ {\hat +} _ {\dot {\alpha}} $ expansion
    (\ref {11}) for $ \phi ^ {+ \hat +} $. The full expression for $\phi^{+\hat +}$ at this step reads
    \begin{eqnarray}
    &&\phi^{+\hat +}=\sqrt{2}q^{+A}v^{\hat +}_A+
    \theta^{\hat + \alpha}\W^{+}_{\alpha}+
    \bar \theta^{\hat +}_{\dot \alpha}\widetilde \W^{+\dot\alpha}-\sqrt{2}i(\bar \theta^{\hat +})^2\theta^{\hat +\alpha}[\psi^{(A}_{\alpha},q^{+B)}]v^{\hat -}_Av^{\hat -}_B\nn
    &&- \sqrt 2 i( \theta^{\hat +})^2[\W,q^{+ A}]v_A^{\hat -}-
    \sqrt 2 i(\bar \theta^{\hat +})^2[ \bW,\, q^{+A}]v_A^{\hat -}
    +\sqrt{2}i( \theta^{\hat +})^2 \bar \theta^{\hat +}_{\dot\alpha}
    [\widetilde\psi^{(A\dot\alpha},q^{+B)}]v^{\hat -}_Av^{\hat -}_B\nn
    &&+2i\sqrt{2}\theta^{\hat+}_{\alpha}\bar\theta^{\hat+}_{\dot\alpha}
    \hat\nabla^{\alpha\dot{\alpha}}q^{+A}v^{\hat -}_A
    -i( \theta^{\hat +})^2(\bar \theta^{\hat +})^2[\mathcal{D}^{AB}, q^{+C}]v_A^{\hat -}v_B^{\hat -}v_C^{\hat -}.
    \label{41}
    \end{eqnarray}
    All the  coefficients in this expansion are harmonic $\N=2$ superfields.

    Now we are ready to display the constraints imposed by the second harmonic equation (\ref {27}b). In the zeroth order in the ``hat''-variables it
    entails just the equation of  motion for the hypermultiplet $q^{+}_A $
    \begin{equation}
    \nabla^{++}q^{+}_{A}=0\,.
    \label{42}
    \end{equation}
    In higher orders, there again appear some extra self-consistency relations to be automatically satisfied on the complete solution of the constraints.
     Note that the reality of the superfield $\phi^{+\hat +}$ implies the reality of $q^+_A$. Indeed,

    \begin{equation}
    \phi^{+\hat +}=\widetilde{\phi^{+\hat +}}\Rightarrow q^{+\hat +}=\widetilde{q^{+\hat +}} \Rightarrow q^{+A}v^{\hat +}_A
    =\widetilde{q^{+A}v^{\hat +}_A}=\widetilde q^+_{A}v^{\hat+A}=-\widetilde q^{+A}v^{\hat+}_A,
    \end{equation}
    or, equivalently,
    \begin{equation}
    {\widetilde q}^A=-q^A \Leftrightarrow \widetilde{ q_A}=q^A.
    \end{equation}

    \subsubsection{Constraints $[\nabla^{\hat + \hat+},\nabla^{ +}_{\alpha}]=0$ and $[\nabla^{\hat + \hat+},\bar \nabla^{ +}_{\dot\alpha}]=0$ \label{s3.2.3}}
Now we can return to the problem of fully fixing the spinor connections $ \mathcal {A}^{+}_\alpha $ and $\mathcal {\bar A} ^ {+} _ {\dot \alpha} $.
The key role in achieving this is played by the constraints
    \begin{equation}
    (\text a)\;[\nabla^{\hat + \hat+},\nabla^{ +}_{\alpha}]=0, \qquad (\text b)\;[\nabla^{\hat + \hat+},\bar \nabla^{ +}_{\dot\alpha}]=0.
    \label{1.9}
    \end{equation}

Like the constraint (\ref{7}) for $V^{++}$, the constraint (\ref{1.9}a) eliminates all the negatively charged components in the expansion (\ref{788}) of $A^+_\alpha$, except for the component
$f^{+\hat -}_{\alpha\beta}\,$,
    \begin{equation}
    g^{+\hat -\dot\alpha}_{\beta}=f^{+\hat{-}2}_{\beta}=g^{+\hat{-}2}_{\beta}=
     f^{+\hat -\hat{-}\alpha\dot\alpha}_\beta=f^{+\hat{-}3}_{\alpha{\beta}}=
     g^{+\hat{-}3\dot\alpha}_{\beta}=f^{+\hat -4}_{\beta}=0.\lb{gfetc}
    \end{equation}
    For $f^{+\hat -}_{\alpha\beta}$  we obtain from (\ref{1.9}a) the harmonic equation
    \begin{equation}
    \partial^{\hat +\hat +}f^{+\hat -}_{\alpha\beta} + 2i\sqrt 2\epsilon_{\beta\alpha} q^{+\hat +} = 0\; \Longrightarrow \;
    f^{+\hat{-}}_{\alpha\beta}=-2\sqrt{2}i\epsilon_{\beta\alpha}q^{+A}v_A^{\hat-}.
    \label{54}
    \end{equation}

    We also obtain the set of self-consistency conditions which are listed in appendix \ref{s7.2}. Here we quote only one important
    condition which will be needed for the subsequent analysis,
        \begin{eqnarray}
    D^{+}_{\beta}\mathcal{\hat A}^{\alpha\dot{\alpha}}+\delta^{\alpha}_{\beta}\widetilde \W^{+\dot{\alpha}}=0 \quad {\rm (and \,\,c.c.)}.
    \label{48}
    \end{eqnarray}

    Since (\ref {1.9}b) is a complex conjugate of eq. (\ref {1.9}a), the restrictions associated with $\bar{A}^+_{\dot \alpha}$ correspond
    just to conjugating the relations (\ref{gfetc}) - (\ref{48}).

    The final form of the spinor connections is obtained by  substituting the solution (\ref {54}) into (\ref {9}):
    \begin{eqnarray}
    \mathcal{A}^+_\alpha =-2\sqrt{2}i\theta^{\hat+}_\alpha q^{+ A}v_A^{\hat -}+2i\theta^{\hat -}_\alpha\phi^{+\hat+}, \quad
    \mathcal{\bar A}^+_{\dot \alpha} = -2\sqrt{2}i\bar \theta^{\hat+}_{\dot \alpha} q^{+ A}v_A^{\hat -}+2i\bar \theta^{\hat -}_{\dot \alpha}\phi^{+\hat+}.
    \label{53}
    \end{eqnarray}

    It is the proper place to come back to the analyticity condition (\ref {1.8}b). Using the exact expressions (\ref {53}) for spinor connections, we draw
    the following consequences of it
    \begin{eqnarray}
    D^+_\alpha q^{+A} =0, \qquad    D^+_\alpha \W^+_\beta=-2\epsilon_{\alpha \beta}[q^{+A},q^{+}_A], \qquad D^+_\alpha \widetilde \W^{+\dot{\alpha}}=0.
    \label{74}
    \end{eqnarray}
    The first relation and its conjugate, $\bar D_{\dot\alpha}^+ q^{+ A} = 0\,$, are just the $\N=2$ Grassmann analyticity conditions for $q^{+A}$. As it will become clear later,
    the other two relations encode the equations of motion for $\N=2$ gauge superfield and the $\N=2$ chirality conditions for the $\N=2$ superfield strengths. We also note that, taking
    into account  (\ref{74}) and the constraint $\nabla^+_{\alpha, \dot\alpha}\phi^{+ \hat +} = 0$ in (\ref{1000}) (to be discussed later),  one can check that
    the short connections (\ref{53}) solve the constraints (\ref{1}). The
    gauge transformation law (\ref{TranAanal}), with $\Lambda(\zeta_{II}) \rightarrow \Lambda(\zeta)$, is reduced to the homogeneous law
    $\delta\,\mathcal{A}^+_{\alpha, \dot\alpha} = i[\mathcal{A}^+_{\alpha, \dot\alpha}, \Lambda(\zeta)]$, taking into account the analyticity of
    $\Lambda(\zeta)$, that is $D^+_{\alpha \dot\alpha}\Lambda(\zeta) = 0\,$.

    \subsection{Supersymmetry transformations \label{s3.2.5}}

    In this subsection we discuss the implementation of hidden supersymmetry. In the analytic basis its transformations on the superspace coordinates  are as follows
  \begin{equation}
  \delta\, x^m_{\rm an}=-2i(\epsilon^{\hat -}\sigma^m\bar \theta^{\hat +}+\theta^{\hat+}\sigma^m\bar \epsilon^{\hat -}),\qquad
  \delta\,\theta^{\hat \pm}_\alpha=\epsilon^{\hat \pm}_\alpha=\epsilon^{A}_\alpha v^{\hat \pm}_A, \qquad
  \delta\,\bar\theta^{\hat \pm}_{\dot\alpha}= \bar\epsilon^{\hat \pm}_{\dot\alpha} = \bar\epsilon^{A}_{\dot\alpha} v^{\hat \pm}_A.  \label{supershifts}
  \end{equation}

    In order to preserve Wess-Zumino gauge of the superfield $V^{\hat +\hat+}$ (\ref{18}), as well as
the ``short'' form of the spinor connections $\mathcal{ A}^+_{\alpha}$ and $ \mathcal{\bar A}^+_{\dot\alpha}$, eqs. (\ref{53}), one
 needs to add the compensating gauge transformation. So the second supersymmetry transformations (in the ``active'' form, {\it i.e.} taken at the fixed ``superpoint'')
should be
\bea
\delta\, V^{\hat +\hat +} = \delta'\, V^{\hat +\hat +} +\nabla^{\hat +\hat +}\Lambda^{({\rm comp})}\,, \quad  \delta\, \mathcal{ A}^+_{\alpha, \dot\alpha}=\delta'\,
\mathcal{ A}^+_{\alpha, \dot\alpha} + \nabla_{\alpha, \dot\alpha}\Lambda^{({\rm comp})}\,,
\eea
where $\delta'$ means the variation under the shifts  (\ref{supershifts}), e.g., $\delta'\, V^{\hat +\hat +} = -\delta x^m_{\rm an} \partial_m V^{\hat +\hat +} + \ldots $, and
\begin{equation}
\Lambda^{(\text{comp})}=\Lambda^{(\text{comp})}_1+\Lambda^{(\text{comp})}_2,
\end{equation}
where $\Lambda^{(\text{comp})}_1$, $\Lambda^{(\text{comp})}_2$ are chosen, respectively,  to preserve (\ref{18}) and (\ref{53}). These composite gauge parameters are
easily found to be
\begin{eqnarray}
    &&\Lambda^{(\text{comp})}_1=\bar \theta^{\hat +}_{\dot{\alpha}}\left(2i\epsilon^{\hat -}_{\alpha}\mathcal{\hat A}^{\alpha\dot{\alpha}}+ 2 \bar \epsilon^{\hat-\dot \alpha}\bW\right)
    -\theta^{\hat +\alpha}\left(2i\bar\epsilon^{\hat -\dot\alpha}\mathcal{\hat A}_{\alpha\dot{\alpha}}- 2\epsilon^{\hat -}_\alpha\W\right)
    \nn
    &&-(\theta^{\hat +})^{2}\bar \epsilon^{(A}_{\dot{\alpha}}\widetilde \psi^{B)\dot\alpha} v^{\hat -}_Av^{\hat -}_B+(\bar \theta^{\hat +})^{2}\epsilon^{(A\alpha} \psi^{B)}_{\alpha} v^{\hat -}_Av^{\hat -}_B
    -2\theta^{\hat +}_\alpha\bar \theta^{\hat +}_{\dot\alpha}
    \left(\epsilon^{(A\alpha} \widetilde \psi^{B)\dot \alpha}+\bar \epsilon^{(A\dot\alpha} \psi^{B)\alpha}\right) v^{\hat -}_Av^{\hat -}_B\nn
    &&+\theta^{\hat +\alpha}(\bar \theta^{\hat +})^{2}\epsilon^{(A}_\alpha\mathcal{D}^{BC)}v^{\hat -}_Av^{\hat -}_Bv^{\hat -}_C
+\bar \theta^{\hat +}_{\dot\alpha}( \theta^{\hat +})^{2}\bar\epsilon^{(A\dot\alpha}\mathcal{D}^{BC)}v^{\hat -}_Av^{\hat -}_Bv^{\hat -}_C\,, \label{Lambda1}
    \end{eqnarray}
    \begin{equation}
  \Lambda^{(\text{comp})}_2=2\sqrt{2}i\theta^{ -\alpha}q^{+A}\epsilon_{A\alpha}+2\sqrt{2}i\bar\theta^{ -\dot\alpha}q^{+A}\bar\epsilon_{A\dot\alpha}.\label{Lambda2}
    \end{equation}

    For the variation of the superfield $\V ^ {++} $ we obtain
    \begin{equation}
    \delta\, \V^{ ++}=2i(\epsilon^{\hat -\alpha}\bar \theta^{\hat +\dot{\alpha}}+\theta^{\hat+\alpha}\bar \epsilon^{\hat -\dot{\alpha}})\partial_{\alpha \dot \alpha}\V^{ + +}+\nabla^{ + +}\Lambda^{(\text{comp})}.
    \label{1.11}
    \end{equation}
    Let us inspect $\nabla^{ + +}\Lambda^{(\text{comp})}$. Using the relations (\ref{22}) - (\ref{25}) we find
    \begin{eqnarray}
    &&\nabla^{++}\Lambda^{(\text{comp})}=-2i(\epsilon^{\hat -\alpha}\bar \theta^{\hat +\dot{\alpha}}
    +\theta^{\hat+\alpha}\bar \epsilon^{\hat -\dot{\alpha}})\partial_{\alpha \dot \alpha}\V^{ + +}\nn
    &&+\,\nabla^{++}\left(2\sqrt{2}i\theta^{ -\alpha}q^{+A}\epsilon_{A\alpha}+2\sqrt{2}i\bar\theta^{ -\dot\alpha}q^{+A}\bar\epsilon_{A\dot\alpha}\right).
    \label{1.12}
    \end{eqnarray}
    The first term precisely cancels the unwanted term in (\ref{1.11}) involving $\theta^{\hat + \alpha, \dot\alpha}$, while the second term, with taking into account the on-shell condition  $\nabla^{++}q^+_A=0$, yields the already known transformation
    \begin{equation}
    \delta\,\V^{++}=-\left[2\sqrt{2}i\epsilon^{A\alpha}\theta^{ +}_{\alpha}-2\sqrt{2}i\bar\epsilon^A_{\dot\alpha}\bar\theta^{ +\dot\alpha}\right]q^{+}_{A}.
    \end{equation}

    Similarly, considering the transformations of the superfield $ \phi ^ {+ \hat +} $ and using the equations of motion that will be obtained below (eq. (\ref{261})),
    one obtains the transformation law of the hypermultiplet $ q ^ {+} _ A$
    \begin{eqnarray}
    \sqrt{2}\delta\, q^{+\hat +}&=&-\epsilon^{\hat +\alpha}\W^{+}_{\alpha}+2\theta^{-\alpha}\epsilon^{\hat +}_\alpha[q^{+A},q^{+}_A]+ {\rm c.c.} \nn
    &=&\frac{i}{8}\epsilon^{\hat +\alpha}\left(2D^+_{\alpha}(\bar D^+)^2\V^{--}+\theta^{-}_\alpha(\bar D^+)^2\V^{--}\right)+ {\rm c.c.} \nn
    &=&\frac{i}{8}\epsilon^{\hat +\alpha}(D^{+})^2(\bar D^{+})^2(\theta^{-}_\alpha \V^{--})-\frac{i}{8}\bar\epsilon^{\hat +}_{\dot \alpha}(D^{+})^2(\bar D^{+})^2(\bar\theta^{-\dot\alpha} \V^{--}),
    \end{eqnarray}
    or, equivalently,
    \begin{equation}
    \delta\, q^{+}_A=\frac{1}{16\sqrt{2}}(D^{+})^2(\bar D^{+})^2\left[2i\epsilon^{\alpha}_A\theta^{-}_\alpha \V^{--}-2i\bar\epsilon^{}_{A\dot \alpha}\bar\theta^{-\dot\alpha} \V^{--}\right].
    \end{equation}

    Rescaling the parameters $\epsilon$ as
    \begin{equation}
     \epsilon_{A\alpha}=\frac{i}{2\sqrt 2}\epsilon_{A\alpha}',
    \label{5.48}
    \end{equation}
    we recover the already known realization of the hidden supersymmetry (\ref{00.5})
    \begin{eqnarray}
    \delta\, \V^{++}=\left[\epsilon^{A\alpha}\theta^{ +}_{\alpha}-\bar\epsilon^A_{\dot\alpha}\bar\theta^{ +\dot\alpha}\right]q^{+}_{A}, \quad
    \delta\, q^{+}_A=-\frac{1}{32}(D^{+})^2(\bar D^{+})^2\left[\epsilon^{\alpha}_A\theta^{-}_\alpha \V^{--}+\bar\epsilon^{}_{A\dot \alpha}\bar\theta^{-\dot\alpha} \V^{--}\right].
    \end{eqnarray}

    Now it is quite legitimate to identify $\N=2$ superfields $q^{+}_A$ and $\V^{++}$ with the hypermultiplet and gauge multiplet superfields from section \ref{s2.0}.
    At this stage, we  have expressed all the geometric quantities of $ \mathcal {N} = 4 $ SYM theory in terms of $ \mathcal {N} = 2 $ superfields.
    It remains to relate the superfield coefficients  appearing in (\ref {8}) to the basic $ \mathcal {N} = 2 $ superfields $\V^{++}, q^{+ A}$. This can be done  in an algebraic way,
    without solving any differential equations, by requiring that the vector connections and the superfield strengths obtained from the relations
     with and without ``hats'' coincide with each other.

        \subsection{Identifying vector connections \label{s4}}
    So we are led to explore the superfield vector connections in the sectors with and without ``hat''.  First, we will consider the sector including
    derivatives with respect to the ordinary coordinates (without ``hats''). We define  $\bar\nabla^{-}_{\dot\alpha}$ in the standard way
    \begin{eqnarray}
    &\bar\nabla^{-}_{\dot\alpha}:=\bar D^{-}_{\dot\alpha}+i\mathcal{\bar A}^{-}_{\dot \alpha}=[\nabla^{--},\bar \nabla^{+}_{\dot\alpha}],  \qquad
     \mathcal{\bar A}^{-}_{\dot\alpha}=\mathcal{\bar A}^{-(0)}_{\dot \alpha}-2i\sqrt 2\bar\theta^{\hat+}_{\dot\alpha} q^{-A}v^{\hat -}_A +2i\bar\theta^{\hat -}_{\dot\alpha}\nabla^{--}\phi^{+\hat+},
    \label{83}\\
     &\mathcal{\bar A}^{-(0)}_{\dot\alpha}=-\bar D^{+}_{\dot\alpha} \V^{--},
    \end{eqnarray}
    where
    \begin{equation}
    \nabla^{--}=D^{--}+i\V^{--}, \quad q^{- A} := \nabla^{--} q^{+ A}\,,
    \label{81}
    \end{equation}
    and $\V^{--}$ is related to $\V^{++}$ via the harmonic zero curvature condition
    \begin{equation}
    D^{++}\V^{--}-D^{--}\V^{++}+i[\V^{++},\V^{--}]=0.
    \label{82}
    \end{equation}

        Accordingly, vector connection is defined in the standard way,
    \begin{equation}
    \{\nabla^{+}_\alpha,\bar\nabla^{-}_{\dot\beta}\}=-2i(\partial_{\alpha\dot \beta}+i\mathcal{V}_{\alpha \dot\beta}),\qquad
    \mathcal{V}_{\alpha\dot \beta}=-\frac{1}{2i}(\nabla^{+}_\alpha\mathcal{\bar A}^{-}_{\dot \beta}+\bar D^{-}_{\dot \beta}\mathcal{A}^{+}_\alpha).
    \label{84}
    \end{equation}
    Using the expressions (\ref{83}) for $\mathcal{\bar A}^{-}_{\dot \beta}$ and (\ref{53}) for $\mathcal{A}^{+}_\alpha$, we find the expression for $\mathcal{V}_{\a \db}$
    \begin{eqnarray}
    &&\mathcal{V}_{\alpha\dot{\beta}}=\mathcal{A}_{\alpha\dot{\beta}}- \sqrt{2}\,\bar\theta^{\hat+}_{\dot\beta}D^{+}_{\alpha} q^{-A}v^{\hat -}_A + \bar\theta^{\hat -}_{\dot\beta}
    D^{+}_{\alpha}\nabla^{--}\phi^{+\hat+}-\sqrt{2} \theta^{\hat+}_{ \alpha}\bar \nabla^{-(0)}_{\dot{\beta}} q^{+A}v^{\hat -}_A \nn
    &&+ \theta^{\hat -}_{ \alpha}\bar \nabla^{-(0)}_{\dot{\beta}}\phi^{+\hat+}
    +4\theta^{\hat+}_{\alpha}\bar \theta^{\hat+}_{\dot{\beta}}[q^{+A},q^{-B}]v^{\hat{-}}_A v^{\hat{-}}_B - 2\sqrt{2}\theta^{\hat+}_{\alpha}\bar \theta^{\hat-}_{\dot{\beta}}
    [q^{+A},\nabla^{--}\phi^{+\hat +}]v^{\hat -}_A\nn
    && -2\sqrt{2}i\theta^{\hat-}_{\alpha}\bar \theta^{\hat+}_{\dot{\beta}}[\phi^{+\hat{+}},q^{-A}]v^{\hat -}_A
    + 2\theta^{\hat-}_{\alpha}\bar \theta^{\hat-}_{\dot{\beta}}[\phi^{+\hat +},\nabla^{--}\phi^{+\hat+}]\,,  \quad
\label{85}
    \end{eqnarray}
    where
      \be
    \bar \nabla^{-(0)}_{\dot \beta}=\bar D^{-}_{\dot \beta}+i \mathcal{\bar A}^{-(0)}_{\dot \beta}\,, \quad
    \mathcal{A}_{\alpha\dot{\beta}}=-\dfrac{1}{2i}D^{+}_\alpha\mathcal{\bar A}^{-(0)}_{\dot{\beta}}.\lb{defnabl0}
    \ee
   The $\N=4$ vector connection  (\ref{85}) displays a restricted dependence on $\theta^{\hat-}_\alpha,\bar\theta^{\hat-}_{\dot \alpha}$
     (only monomials of the first and second orders appear),
    but includes all $\theta^{\hat +}_\alpha,\bar\theta^{\hat +}_{\dot \alpha}$ monomials. For what follows it will be useful to quote the opposite chirality
    counterpart of the $\N=2$ spinor covariant derivative (\ref{defnabl0})
  \be
    \nabla^{-(0)}_{\beta}= D^{-}_{\beta}+i \mathcal{A}^{-(0)}_{\beta}\,, \quad \mathcal{A}^{-(0)}_{\beta}= - D^{+}_{\beta} \V^{--}\,.\lb{defnabl00}
    \ee

    One can perform an analogous construction for the derivatives with ``hats''. We define the relevant second harmonic connection $V^{\hat{-}\hat -}$
    by the ``hat'' flatness condition
    \begin{equation}
    D^{\hat+\hat+}V^{\hat-\hat-}-D^{\hat-\hat-}V^{\hat+\hat+}+i[V^{\hat+\hat+},V^{\hat-\hat-}]=0,
    \label{87}
    \end{equation}
    and then define the relevant spinor and vector connections
    \begin{eqnarray}
    \bar\nabla^{\hat-}_{\dot\alpha}:=\bar D^{\hat-}_{\dot\alpha}+i\mathcal{\bar A}^{\hat-}_{\dot \alpha}=[\nabla^{\hat-\hat-},\bar D^{\hat+}_{\dot\alpha}], &\qquad& \mathcal{\bar A}^{\hat-}_{\dot\alpha}=-\dfrac{\partial}{\partial \bar\theta^{\hat-\dot{\alpha}}}V^{\hat-\hat-},
    \label{88}\\
    \{D^{\hat+}_\alpha,\bar\nabla^{\hat-}_{\dot\beta}\}=-2i(\partial_{\alpha\dot \beta}+i\mathcal{\hat V}_{\alpha \dot\beta}),
    &\qquad& \mathcal{\hat V}_{\alpha\dot \beta}
    =-\dfrac{i}{2}\dfrac{\partial}{\partial \theta^{\hat-{\alpha}}} \dfrac{\partial}{\partial \bar\theta^{\hat-\dot{\beta}}}V^{\hat-\hat-},
    \label{89}
    \end{eqnarray}
    where $\nabla^{\hat -\hat-}=D^{\hat -\hat-}+iV^{\hat -\hat-}$.

        In order to perform further calculations we need the expression for $V^{\hat-\hat{-}}$.
        We parametrize the $\theta^{\hat-}_\alpha,\bar\theta^{\hat-}_{\dot \alpha}$ expansion  of $V^{\hat-\hat{-}}$ in the following way
    \begin{equation}
    V^{\hat- \hat-}=-2i\theta^{\hat -}_{\alpha}\bar\theta^{\hat -}_{\dot \alpha}w^{\alpha \dot \alpha}+ (\theta^{\hat-})^2w+(\bar \theta^{\hat-})^2{\widetilde w}
    +(\bar \theta^{\hat -})^2\theta^{\hat -\alpha}w^{\hat +}_\alpha+( \theta^{\hat -})^2\bar \theta^{\hat -}_{\dot\alpha}\widetilde w^{\hat +\dot\alpha}+(\theta^{\hat-})^2(\bar \theta^{\hat-})^2w^{\hat+2}.
    \label{90}
    \end{equation}
    The $(\theta^{\hat+}_\alpha,\bar\theta^{\hat+}_{\dot \alpha}, v^{\hat +}_A)$-dependence of the coefficients in this expression will be determined from eq. (\ref{87}), with
    $V^{\hat +\hat+}$ taken in the form (\ref{8}). Possible coefficients of the monomials of first and zeroth orders in $\theta^{\hat -}_\alpha, \bar\theta^{\hat -}_{\dot \alpha}$
    can be shown to vanish as a consequence of
    equations like   $\partial^{\hat+\hat +}\omega^{\hat -}=0\Rightarrow \omega^{\hm}=0$.
    The process of solving the equation (\ref{87}) is rather tiresome and the solution looks rather bulky.
    We give it in the appendix \ref{s4.1}. Here we collect only the information that is needed for further steps, viz., the expressions for the
    coefficients $w^{\beta\dot{\beta}}$ and $w$,
        \begin{eqnarray}
   & w^{\beta\dot{\beta}}=\mathcal{\hat A}^{\beta\dot{\beta}}
    -i\theta^{\hat + \beta}\widetilde \psi^{A\dot{\beta}}v_A^{\hat -}
    -i\bar \theta^{\hat +\dot \beta}\psi^{ A\beta}v_A^{\hat -}
    +2i\theta^{\hat+\beta}\bar\theta^{\hat+\dot\beta}
    \mathcal{D}^{AB}v^{\hat -}_A v^{\hat -}_B,\label{4.56}\\
  & w= \W
    -\bar \theta^{\hat +}_{\dot \beta}\widetilde\psi^{A\dot{\beta}}v^{\hat -}_A+
    (\bar \theta^{\hat +})^2 \mathcal{D}^{AB}v^{\hat -}_A v^{\hat -}_B,
    \label{4.57}
    \end{eqnarray}
    where we made use of the relations (\ref{vDependWZ}). Note that one cannot calculate these coefficients directly from eq. (\ref{87}).
    The procedure of finding them
    requires the knowledge of all the coefficients obeying a set of coupled harmonic equations.

    Now one can determine the vector connection $\mathcal{\hat V}_{\alpha\dot \beta}$. Substituting (\ref{90}) into (\ref{89}) and using eq. (\ref{4.56})
    we obtain, in zeroth order in $\theta^{\hat-}_\alpha,\bar\theta^{\hat-}_{\dot \alpha}$,
    \begin{equation}
    \mathcal{\hat V}^{\b\dot \beta}|_{\theta^{\hat-}_\alpha,\bar\theta^{\hat-}_{\dot \alpha}=0}=\mathcal{\hat A}^{\beta\dot{\beta}}
    -i\theta^{\hat + \beta}\widetilde \psi^{A \dot{\beta}}v^{\hat -}_A
    -i\bar \theta^{\hat +\dot \beta}\psi^{A \beta}v^{\hat -}_A
    +2i\theta^{\hat+\beta}\bar\theta^{\hat+\dot\beta}
    \mathcal{D}^{AB} v^{\hat -}_A v^{\hat -}_B\,.
    \label{4.58}
    \end{equation}

    Identifying vector connections from both sectors,
    \begin{equation}
    \mathcal{\hat V}_{\alpha\dot \beta}=\mathcal{ V}_{\alpha\dot \beta},
    \end{equation}
    and,  in particular, the two expansions (\ref {4.58}) and (\ref {85}), in zeroth order in $ \theta ^ {\hat -} _ \alpha, \bar \theta ^ {\hat -} _ {\dot { \alpha}} $
    we get
    \begin{equation}
    \mathcal{\hat A}_{\alpha\dot \beta}=\mathcal{A}_{\alpha\dot \beta}, \quad
    \mathcal{D}^{AB}=-2i[q^{+(A},q^{-B)}],\quad
    \widetilde \psi^{A}_{\dot \beta}= -i\sqrt 2\bar\nabla^{-(0)}_{\dot{\beta}}q^{+A}\,,
    \quad
    \psi^{A}_\beta=i\sqrt 2\nabla^{-(0)}_{{\beta}}q^{+ A}.
    \label{242}
    \end{equation}

        Thus, we have succeeded to express a part of $\N=2$ superfields in (\ref{8}) and (\ref{41}) in terms of  the hypermultiplet $q^{+}_A$ and
        the gauge superfield $\V^{++}$. However, the superfields $\W^+_\alpha,\ \widetilde \W^+ _ {\dot \alpha} $ and $ \W,\ \bW $
        still remain unspecified.

        It is easy to obtain the expression for $ \W ^ + _ \alpha,\ \widetilde \W ^ + _ {\dot \alpha}\, $. To this end,
        we substitute $ \mathcal {\hat A} _ {\alpha \dot {\alpha}} $ from (\ref {242}) into (\ref {48}). As a result, we find
        \begin{equation}
    \W^{+}_\alpha=-\frac{i}{4}D^{+}_{\alpha}\left(\bar D^{+}\right)^2\V^{--},
    \qquad
    \widetilde  \W^{+}_{\dot\alpha}=-\frac{i}{4}\bar D^{+}_{\dot\alpha}\left( D^{+}\right)^2\V^{--}.
    \label{211}
    \end{equation}
The rest of the constraint (\ref {48}) is reduced to $D^+_\alpha \mathcal{A}_{\beta\dot\beta} +D^+_\beta \mathcal{A}_{\alpha\dot\beta} = 0\; {\rm (and\; c.c.)},$ which is satisfied identically
since $\mathcal{A}_{\alpha\dot\beta} \sim D^+_\alpha \bar{\mathcal{A}}^{-(0)}_{\dot\beta}$.

More effort is required to determine the superfields $\W,\ \bW$.
    One needs to take into account that the superfield strength $W$, like the vector connection, can de expressed
    in two ways. First, we can use the relation
    \begin{equation}
    \{D^{\hat +}_\alpha,\nabla^{\hat -}_\beta\}=2i\epsilon_{\alpha\beta}\bar W \quad {\rm (and\; c.c.)}\,,
    \label{244}
    \end{equation}
    where
    \begin{equation}
    \nabla^{\hat-}_\alpha=[\nabla^{\hat -\hat -},D^{\hat +}_\alpha].
    \label{245}
    \end{equation}
    It follows from the second line of (\ref{3.8}) by contracting its both sides with $v^{\hp A}v^{\hm B}$ and then passing
    to the analytical frame.  Substituting  (\ref{245}) into (\ref{244}),  we obtain
    \begin{equation}
    \bar W=-\frac{1}{4}\big( D^{\hat+}\big)^2V^{\hat-\hat-}, \qquad
    W=-\frac{1}{4}\big( \bar D^{\hat+}\big)^2V^{\hat-\hat-}.
    \label{212}
    \end{equation}
Note that the definition (\ref{244})  implies, through Bianchi identities, the covariant harmonic independence of $W$, $\nabla^{\pm\pm} W = \nabla^{\hat \pm \hat \pm} W = 0$,
as well as the reality condition $\big(D^{\hat +}\big)^2\bar W=    \big(\bar D^{\hat +}\big)^2 W$.

    Alternatively, we can use the relation
    \begin{equation}
    \{\nabla^{ +}_\alpha,\nabla^{ -}_\beta\}=2i\epsilon_{\alpha\beta}W \quad {\rm (and\; c.c.)}\,,
    \label{246}
    \end{equation}
    with $\nabla^{-}_\alpha$ being defined in the standard way,
    $\nabla^{-}_\alpha=[\nabla^{ - -},\nabla^{+}_\alpha]$.  Eq.  (\ref{246}) amounts to
    \begin{equation}
    D^{+}_\alpha\mathcal{A}^{-}_\beta+D^{-}_\beta\mathcal{A}^{+}_\alpha+i\{\mathcal{A}^{+}_\alpha,\mathcal{A}^{-}_\beta\}=2\epsilon_{\alpha \beta}W \quad {\rm (and\; c.c.)}\,.
    \end{equation}
    Then, substituting the definition of  $\mathcal{A}^{+}_\beta$ from (\ref{53}), we obtain another expression for the superfield strength $W$
    (in zeroth order in $\theta^{\hat\pm}_\a,\bar \theta^{\hat\pm}_\da$)
    \begin{equation}
    W|_{\theta^{\hat\pm}_\a=\bar \theta^{\hat\pm}_\da=0}=-\frac{1}{4}\left(D^{+}\right)^2\V^{--}\quad {\rm (and\; c.c.)}\,,
    \label{214}
    \end{equation}

    Substituting the expansion (\ref{90}) for $ V ^ {\hat - \hat -} $ in (\ref{212}) and using that $w|_{\theta^{\hat +}_\a=\bar \theta^{\hat +}_\da=0} = \mathcal{W}$
    as follows from (\ref{4.57}), we equate (\ref{214}) to (\ref {212}) and obtain the sought expressions for $ \W,\  \bW$
    \begin{equation}
    \W=-\frac{1}{4}\left(\bar D^{+}\right)^2\V^{--}, \qquad   \bW=-\frac{1}{4} \left( D^{+}\right)^2\V^{--}.
    \label{254}
    \end{equation}
    These expressions coincide with those defined in subsection \ref {s2.2}. Eqs. (\ref{211}) can be rewritten as
    \begin{equation}
    \W^{+}_\alpha= iD^+_\alpha \W\,, \qquad \widetilde  \W^{+}_{\dot\alpha}=i \bar D^+_{\dot\alpha}\bW.
    \label{2111}
    \end{equation}

    Thus, we have expressed all superfield components of $V^{\hat -\hat -}$ in terms of the hypermultiplet and $\N=2$ gauge superfields $q^{+}_A$ and $\V^{++}$.
    Now one can be convinced that the previously deduced conditions (\ref{22}) - (\ref{19}) are indeed satisfied.

    To summarize, all the bi-harmonic $\N=4$ SYM superfields we started with proved to be expressed in terms of the two basic analytic  $\N=2$ superfields involved
    in the $\N=2$ harmonic superspace action principle for $\N=4$ SYM theory considered in section 2.

    \subsection{Guide to Section 4 \label{s6}}

        For reader's convenience, in this subsection we quote the expressions of the basic involved $\N=4$ superfields in terms
        of the hypermultiplet and gauge superfields $q^{+}_A$ and $\V^{++}$, as well
        as the equations of motion for the latter, as the result of solving the $\N=4$ SYM constraints (\ref{1}) - (\ref{7}).

    The expressions for the $\N=4$ spinor connections were obtained in subsection \ref{s3.2.3}
    \begin{eqnarray}
    \mathcal{A}^+_\alpha=-2\sqrt{2}i\theta^{\hat+}_\alpha q^{+\hat -}+2i\theta^{\hat -}_\alpha\phi^{+\hat+}, \quad
    \mathcal{\bar A}^+_{\dot \alpha}=-2\sqrt{2}i\bar \theta^{\hat+}_{\dot \alpha} q^{+\hat -}+2i\bar \theta^{\hat -}_{\dot \alpha}\phi^{+\hat+}.
    \end{eqnarray}

        The $\N=2$ vector and spinor connection  were found in subsection \ref{s4}
    \begin{equation}
    \mathcal{A}_{\alpha\dot{\alpha}}=\frac{1}{2i}D^{+}_\alpha\bar D^{+}_{\dot{\alpha}}\V^{--}, \qquad \mathcal{ A}^{-(0)}_{\alpha}=- D^{+}_{\alpha} \V^{--}, \qquad
   \mathcal{\bar A}^{-(0)}_{\dot\alpha}=-\bar D^{+}_{\dot\alpha} \V^{--}.
    \label{connections}
    \end{equation}

        The expressions for $\N=2$ superfields entering  $ V^{\hat + \hat +} $ in WZ gauge (\ref {8}) were obtained
        in the sections \ref{s4} in eqs. (\ref {242}) and (\ref {254}):
    \begin{eqnarray}
    &\mathcal{\hat A}_{\alpha\dot{\alpha}} = \mathcal{A}_{\alpha\dot{\alpha}}\,, \qquad \W=-\dfrac{1}{4}\left(\bar D^{+}\right)^2\V^{--}, \qquad   \bW=-\dfrac{1}{4}\left( D^{+}\right)^2\V^{--},
    \label{258}\\
    &\widetilde \psi^{ A}_{\dot \beta}=-i\sqrt 2\bar\nabla^{-(0)}_{\dot{\beta}}q^{+A},
    \quad
    \psi^{ A}_\beta=i\sqrt 2\nabla^{-(0)}_\beta q^{+A},
    \quad
    \mathcal{D}^{AB}=-2i[q^{+(A},q^{-B)}].
    \label{259}
    \end{eqnarray}

    Analyticity of the hypermultiplet  $q^{+}_A$ was shown in subsection \ref{s3.2}, eq. (\ref{74}).

    One of the equations of motion, namely,
    \begin{equation}
    \nabla^{++}q^{+}_A=0,
    \end{equation}
    was found in subsection \ref{s3.2.2}, eq. (\ref {42}). To obtain another equation, one uses the relation (\ref {74})
    from subsection \ref{s3.2.3}, as well as the expression (\ref {211}) for $ \W ^ {+}_\alpha $. After substituting one into another, the second
    equation of motion follows
    \begin{equation}
    F^{++} = -i[q^{+A},q^{+}_A], \qquad F^{++}=\frac{1}{16}\left(D^{+}\right)^{2}\left(\bar D^{+}\right)^{2}\V^{--}
    \label{261}.
    \end{equation}
    Note that the rest of eq. (\ref {74}), $D^+_\alpha \W^+_\beta  + D^+_\beta \W^+_\alpha = 0$ (and c.c),  is satisfied identically.

    The definition of the superfield strength $ \phi^{+\hat +} $ is given by eq. (\ref {41}) of subsection \ref {s3.2.2}. Substituting
    the expressions (\ref {2111}) and (\ref {259}) into (\ref {41}), we deduce the final expression for $ \phi^{+\hat +} $
        \begin{eqnarray}
    &&\phi^{+\hat +}=\sqrt{2}q^{+A}v^{\hat +}_A+
    i\theta^{\hat + \alpha}D^{+}_\a\W+
    i\bar \theta^{\hat +}_{\dot \alpha}\dD^{+\da} \bW+2(\bar \theta^{\hat +})^2\theta^{\hat +\alpha}[\nabla^{-(0)}_\alpha q^{+(A},q^{+B)}]v^{\hat -}_Av^{\hat -}_B\nn
    &&- \sqrt 2 i( \theta^{\hat +})^2[\W,q^{+A}] v^{\hat -}_A
     -\sqrt 2 i(\bar \theta^{\hat +})^2[ \bW,q^{+ A}]v^{\hat -}_A
    +2( \theta^{\hat +})^2 \bar \theta^{\hat +}_{\dot\alpha}
    [\bar \nabla^{-(0)\dot\alpha} q^{+(A},q^{+B)}]v^{\hat -}_Av^{\hat -}_B\nn
    &&+2\sqrt{2}i\theta^{\hat+}_{\alpha}\bar\theta^{\hat+}_{\dot\alpha}
    \hat\nabla^{\alpha\dot{\alpha}}q^{+A}v^{\hat -}_A
    -2\sqrt{2}( \theta^{\hat +})^2(\bar \theta^{\hat +})^2[[q^{+A},q^{-B}],q^{+C}]v^{\hat -}_A v^{\hat -}_B v^{\hat -}_C,
    \end{eqnarray}
where the spinor covariant derivatives were defined after eq. (\ref{85}), and the vector covariant derivative was defined in (\ref{19}).
They are derivatives with connections (\ref{connections}):
\bea
\hat\nabla_{\alpha\dot{\alpha}} =  \partial_{\alpha\dot{\alpha}} + \frac{1}{2}D^{+}_\alpha\bar D^{+}_{\dot{\alpha}}\V^{--}\,, \;
\nabla^{-(0)}_\alpha = D^-_\alpha -iD^{+}_{\alpha} \V^{--}\,,\;\bar\nabla^{-(0)}_{\dot\alpha} = \bar D^-_{\dot\alpha} -i\bar D^{+}_{\dot\alpha} \V^{--}\,.\lb{0Deriv}
\eea
Now we have all the necessary ingredients to directly check the on-shell validity of the constraints (\ref{1000}) and, hence, of (\ref{1}) (recall the discussion
in the end of subsection \ref{s3.2.3}). When doing so, one should take into account that on shell,
with $\nabla^{++}q^{+ A} =0$, the following condition is valid
$$
\nabla^{++} q^{- A} = \nabla^{++} \nabla^{--} q^{+ A} = q^{+ A}\,.
$$

    The expression for the superfield strength $ W $ was found in subsection \ref{s4}
        \begin{equation}
    \bar W=-\frac{1}{4}\big( D^{\hat+}\big)^2V^{\hat-\hat-}, \qquad
    W=-\frac{1}{4}\big( \bar D^{\hat+}\big)^2V^{\hat-\hat-}.
    \end{equation}
    Bianchi identity for the superfield strength $ W $ follows from its definition (eqs. (\ref{244}), (\ref{246}))
    \begin{eqnarray}
   && \big(D^{\hat +}\big)^2\bar W= \big(\bar D^{\hat +}\big)^2 W\,, \quad  \nabla^{+ +} W = \nabla^{\hat +\hat +} W = 0 \quad ({\rm and\;\; c.c.})\,, \nn
   &&  D^{\hat +}_\alpha W = -i\nabla^-_\alpha \phi^{+\hat +}, \quad \nabla^+_\alpha \bar W= i \nabla^{\hat -}_\alpha\phi^{+\hat +} \quad ({\rm and\;\; c.c.})\,. \label{AdditBianchi}
     \end{eqnarray}
    The chirality conditions are also obvious
    \begin{equation}
    D^{\hat +}_\alpha\bar W=\nabla^{\hat -}_\alpha\bar W=0 \quad  ({\rm and } \;{\rm c.c.}).
    \end{equation}
   The explicit expressions for $W$ and $\bar W$ are rather cumbersome, so we will prefer to give them only for abelian case.

        \subsubsection{Abelian case  \label{s8.4}}
    This subsection presents some important consequences of constraints in abelian case.  In this case everything becomes simpler
    as all commutators vanish.

    The expressions for $\N=2$ superfields appearing in the definition of $ V^{\hat + \hat +} $ (\ref {8}) become
    \begin{eqnarray}
    & \W=-\dfrac{1}{4}\left(\bar D^{+}\right)^2\V^{--}, \qquad    \bW=\dfrac{1}{4}\left( D^{+}\right)^2\V^{--},
    \label{404}\\
    &\widetilde \psi^{ A}_{\dot \beta}=-i\sqrt 2\bar D^-_{\dot{\beta}}q^{+A},
    \quad
    \psi^{ A}_\beta=i\sqrt 2 D^-_\beta q^{+A},
    \quad
    \mathcal{D}^{AB}=0.
    \label{405}
    \end{eqnarray}

    The equations of motions from the previous subsection read
    \begin{equation}
    D^{++}q^{+}_A=0,\qquad  \left(D^{+}\right)^{2}\left(D^{-}\right)^{2}\V^{--}=0.
    \label{407}
    \end{equation}

    The definition of the superfield strength $ \phi ^ {+ \hat +} $ is given by eq. (\ref {41}). Substituting   the expressions (\ref {404}) and (\ref {405}) in it, we obtain
    \begin{equation}
    \phi^{+\hat +}=\sqrt{2}q^{+A}v^{\hat +}_A+
    i\theta^{\hat + \alpha}D^{+}_\alpha \mathcal{W}+
    i\bar \theta^{\hat +}_{\dot \alpha}\bar D^{+\dot \alpha} \mathcal{ \bar W}+2\sqrt{2}i\theta^{\hat +}_\alpha\bar\theta^{\hat +}_{\dot \alpha}\partial^{{\alpha}\dot \alpha}
    q^{+ A}v^{\hat -}_A.
    \label{4.67}
    \end{equation}

    The expression for the superfield strengths $ W $  and $\bar W$ were found in subsection \ref{s4}:
   \begin{equation}
    \bar W=-\frac{1}{4}\big( D^{\hat+}\big)^2V^{\hat-\hat-}, \qquad
    W=-\frac{1}{4}\big( \bar D^{\hat+}\big)^2V^{\hat-\hat-}.
    \label{4.68}
    \end{equation}
     Taking into account the equations of motion, the $ \theta ^ {\hat \pm}_\a, \bar \theta ^ {\hat \pm}_\da $ expansion of these quantities reads
    \begin{eqnarray}
   W&=&\mathcal{\bar W}+i\sqrt{2}\theta^{\hat +\beta}D^{-}_\beta q^{+A}v^{\hat -}_A-i\sqrt{2}\theta^{\hat -\beta}D^{-}_\beta q^{+ A}v^{\hat +}_A
   +2i\theta^{\hat -\beta}\bar\theta^{\hat +\dot \alpha}\partial_{\beta\dot{\alpha}}\mathcal{\bar W}\nn
   &&-\theta^{\hat -\alpha}\theta^{\hat +\beta}D^{-}_{\beta}D^{+}_{\alpha}\mathcal{W}
   +2\sqrt{2}\theta^{\hat +}_\alpha\bar\theta^{\hat +}_{\dot \alpha}\partial^{\alpha\dot{\alpha}}\theta^{\hat -\beta}D^{-}_\beta q^{+A}v^{\hat -}_A,
   \label{4.69}\\
   \bar W&=&\mathcal{ W}+i\sqrt{2}\bar\theta^{\hat +}_{\db}\bar D^{-\dot\beta} q^{+ A}v^{\hat -}_A
   -i\sqrt{2}\bar \theta^{\hat -}_{\dot\beta}\bar D^{-\dot\beta} q^{+A}v^{\hat +}_A-2i\bar\theta^{\hat -}_{\dot\beta}\theta^{\hat +}_{ \alpha}\partial^{\a\db}\mathcal{ W}\nn
   &&-\bar\theta^{\hat -}_{\dot\alpha}\bar\theta^{\hat +}_{\dot\beta}\bar D^{-\dot\beta}\bar D^{+\dot\alpha}\mathcal{\bar W}+
   2\sqrt{2}\theta^{\hat +}_\alpha\bar\theta^{\hat +}_{\dot \alpha}\partial^{\alpha\dot{\alpha}}\bar\theta^{\hat -}_{\dot\beta}\bar D^{-\dot\beta} q^{+A}v^{\hat -}_A.
   \label{4.70}
   \end{eqnarray}

    It is instructive to list here  a few further properties of the superfield strengths that will be used later.
    The  zero curvature condition (\ref {87}) and the definitions of $ W $ and $ \bar W $ (\ref {4.68}) imply
    \begin{equation}
    D    ^{\hat +\hat+}W=D^{\hat-\hat-}W=D^{\hat+\hat+}\bar W=D^{\hat-\hat-}\bar W.
    \end{equation}
    The definitions of $W$ and $\bar W$ (\ref{4.68})  and the conditions listed previously imply
   the  chirality and antichirality of $\bar W$ and $ W$ in the ``hat''-sector
    \begin{eqnarray}
    D^{\hat \pm}_{\alpha}\bar W=\bar D^{\hat \pm}_{\dot \alpha} W=0.
    \label{501}
    \end{eqnarray}
    In addition, the expansions (\ref{4.69}) and (\ref{4.70}) imply
    chirality and antichirality of $W$ and $\bar W$ in the sector without  "hat"
    \begin{eqnarray}
    D^{ \pm}_{\alpha} W=\bar D^{ \pm}_{\dot \alpha}\bar W=0.
    \label{502}
    \end{eqnarray}
    The relations (\ref{244}) and (\ref{246}) entail Bianchi identities relating the superfields $W,\ \bar W,\ \phi^{+\hat +}$
    \begin{eqnarray}
    D^{\hat+}_\alpha W =-iD^{-}_\alpha \phi^{+\hat +}\,, &\quad& D^{+}_\alpha\bar W = iD^{\hat -}_\alpha \phi^{+\hat +}\,,\nn
    \label{500}
    \bar D^{\hat+}_{\dot\alpha}\bar W = i\bar D^{-}_{\dot\alpha} \phi^{+\hat +}\,, &\quad& \bar D^{+}_{\dot\alpha} W = -i\bar D^{\hat -}_{\dot\alpha} \phi^{+\hat +}\,.
    \end{eqnarray}
   The superfield $\phi^{+\hat +}$ also satisfies the conditions
    \begin{eqnarray}
    D^{\hat +}_{\alpha}\phi^{+\hat -}=-D^{\hat -}_{\alpha}\phi^{+\hat +}, &\quad&
    D^{ +}_{\alpha}\phi^{-\hat +}=-D^{ -}_{\alpha}\phi^{+\hat +},\nn
    D^{\hat -}_\alpha\phi^{\pm\hat -}=0,&\quad&
    D^{ -}_\alpha\phi^{-\hat \pm}=0, \label{FurtherBianchi}
    \end{eqnarray}
    where $\phi^{+\hat -}=D^{\hat -\hat -}\phi^{+\hat +}$, $\phi^{-\hat+ }=D^{ - -}\phi^{+\hat +}$ and $\phi^{-\hat -}=D^{ - -}D^{\hat -\hat -}\phi^{+\hat +}$.

    The expansions of  $W$ and $\bar W$ (\ref{4.69}), (\ref{4.70}) also imply the well-known on-shell relations
    \begin{equation}
    (D^{+})^2W=(D^{\hat +})^2W=(D^{+})^2\bar W=(D^{\hat+})^2\bar W=0.
    \label{510}
    \end{equation}
    All these relations will be employed in section  \ref {s6.0}, while constructing the expression
    for invariant effective action.

     \section{$ \mathcal {N} = 4 $ supersymmetric invariants \label{s6.0}}
     In this section we apply the bi-harmonic superspace technique to construct examples of the
     $ \mathcal {N} = 4 $ supersymmetric invariants which, being rewritten through $\N=2$ superfields, possess an extra on-shell
     hidden $\N=2$ supersymmetry. In particular, we will show how the
     low-energy effective action (\ref {2.50}) can be written in terms  of $\N=4$ bi-harmonic
     superfields, so as to secure, from the very beginning, the hidden second $\N=2$ supersymmetry. For sake of simplicity,  we will focus on the case of Abelian gauge group. As
     was earlier mentioned, just this case corresponds to the Coulomb branch of $\N=4$ SYM theory.

    \subsection{General structure of invariants}
     We begin with describing a general structure of invariants in bi-harmonic superspace.
     The simplest expression is
    \begin{eqnarray}
    \int \ du\ d v\ d^{20}z\mathcal L,
    \label{5.1}
    \end{eqnarray}
    where $\mathcal L$ is some $\mathcal N=4$ superfield and integral goes over the full bi-harmonic superspace. It is invariant under $\N=4$ supersymmetry
    due to the presence of integration over all $\theta$'s the number of which is twice as bigger than in $\N=2$ harmonic superspace.
    The invariant (\ref{5.1}) can be rewritten, in an obvious way,  as an integral over $\N=2$ harmonic superspace
     \begin{eqnarray}
   \int \ du\ d v\ d^{20}z\ \mathcal L=\int  du \ d^{12}z\left( \int d v
   \ (D^{\hp})^4(D^{\hm})^4\ \mathcal L\right),
   \label{5555}
     \end{eqnarray}
 where the measure $du d^{12}z$ was defined in (\ref{measHSS}). The relation (\ref{5555}) allows one to transform the invariants originally
 written in bi-harmonic superspace to the invariants ``living'' in the standard $\N=2$ harmonic superspace.

    There are other types of the $\N=4$ invariants which can be constructed as integrals over various invariant analytic subspaces (\ref{BiAnal})
    in bi-harmonic superspace:
    \begin{eqnarray}
   && \int du\int d v\int d^4 x \ d^4\theta^{+} \ d^4\theta^{\hm} \ d^4\theta^{\hp} \mathcal L^{+4}(\zeta_I) =\int du\int d v\int d^4 x  (D^{-})^4(D^{\hp})^4(D^{\hat -})^4\mathcal L^{+4}(\zeta_I),\nn
   &&  \int du\int d v\int d^4 x \ d^4\theta^{+}\ d^4\theta^{-} \ d^4\theta^{\hp} \mathcal L^{\hp4}(\zeta_{II})=
   \int du\int d v\int d^4 x  (D^{+})^4(D^{-})^4(D^{\hat -})^4\mathcal L^{\hp4}(\zeta_{II}), \nn
   && \int du\int d v\int d^4 x \ d^4\theta^{+} \ d^4\theta^{\hp} \mathcal L^{+4\hp4}(\zeta_A)= \int du\int d v\int d^4 x  (D^{-})^4(D^{\hat -})^4 \mathcal L^{+4\hp4}(\zeta_A),\
   \label{5.2}
   \end{eqnarray}
    where $\mathcal L^{+4}, \ \mathcal L^{\hat+4}, \mathcal L^{+4\hp 4}$ are (half)analytic $\mathcal{ N}=4$ superfields defined by the constraints
        \begin{eqnarray}
 && D^+_\a \mathcal L^{+4}=\bar D^+_{\da} \mathcal L^{+4}=0,\nn
    &&    D^{\hp}_{\a} \mathcal L^{\hp4}=\bar D^{\hp}_{\da}\mathcal L^{\hp4}=0, \nn
    &&  D^+_\a \mathcal L^{+4\hp 4}=\bar D^+_{\da} \mathcal L^{+4\hp 4}=D^{\hp}_\a\mathcal L^{+4\hp4}=\bar D^{\hp}_{\da} \mathcal L^{+4\hp4}=0.
    \end{eqnarray}
    The superfield Lagrangian densities  in (\ref{5.2}) are integrated over all those $\theta$'s on which they depend. Hence,
    the expressions  (\ref{5.2}) are invariants of $\N=4$  supersymmetry. These invariants can also be rewritten as integrals over
    $\N=2$ harmonic superspace:
    \begin{eqnarray}
   && \int du\int d v\int d^4 x \ d^4\theta^{+} \ d^4\theta^{\hm} \ d^4\theta^{\hp} \mathcal L^{+4} =\int d\zeta^{-4}
   \left(\int d v \ (D^{\hp})^4(D^{\hat -})^4\mathcal L^{+4}\right),\nn
   &&  \int du \int d v\int d^4 x \ d^4\theta^{+}\ d^4\theta^{-} \ d^4\theta^{\hp} \mathcal L^{\hp4}=\int du \ d^{12} z \left( \int d v \ (D^{\hat -})^4\mathcal L^{\hp4}\right), \nn
   && \int du \int d v\int d^4 x \ d^4\theta^{+} \ d^4\theta^{\hp} \mathcal L^{+4\hp4}= \int d\zeta^{-4}\left(\int d v \  (D^{\hat -})^4 \mathcal L^{+4\hp4}\right),\
   \label{5.5}
   \end{eqnarray}
 where the eventual integrals go over $\N=2$ harmonic superspace or its analytic subspace and the measure $d\zeta^{-4}$ was defined in (\ref{measHSS}).

In a similar manner, one can construct $\N=4$  invariants as integrals over some other invariant subspaces of $\N=4$ bi-harmonic superspace, e.g.,
over chiral subspaces.

    \subsection{From bi-harmonic $\N=4$  superinvariants to $\N=2$ superfields \label{s5.10}}
     In this subsection we will consider three examples of the higher-derivative invariants admitting a formulation in bi-harmonic
     superspace where the whole on-shell $\N=4$ invariance is manifest. They will be transformed to some
     invariants in $\N=2$ harmonic superspace, where only $\N=2$ supersymmetry is manifest, while the
     invariance under the second, hidden $\N=2$ supersymmetry requires a non-trivial check. We will deal with the abelian $U(1)$ gauge group as a remnant
     of $SU(2)$ gauge group in the Coulomb branch of the corresponding $\N=4$ SYM theory\footnote{The consideration can easily be extended to Cartan subalgebra of
     any gauge group.}.

    Let us start with the expression
    \begin{equation}
    I=\int d {u} \ d  v \ d^{20} {z}\, (W\bar W)^2,
    \label{6.33}
    \end{equation}
    where the integration goes over the total bi-harmonic superspace. The functional (\ref{6.33}) is manifestly
    on-shell $\N=4$ supersymmetric by construction. To transform the expression
    (\ref{6.33}) to $\N=2$ harmonic superspace, we substitute
    the expressions for the superfield strengths $ W $, $ \bar W $ (\ref {4.69}), (\ref {4.70}),
    then do the integral over Grassmann and harmonic variables with ``hat'' and finally obtain the expression
    in terms of harmonic $ \mathcal {N} = 2 $ superfields
    \begin{equation}
    I=\int d {u} \ d^{12} {z} \ \mathcal{L},
    \label{6.34}
    \end{equation}
    where
    \begin{eqnarray}
    &&\mathcal{L}= (\partial \mathcal W)^2(\partial \mathcal{\bar W})^2-2i(\partial^{\alpha\dot{\alpha}}D^{-\beta} q^{+}_A
    \partial_{\beta\dot{\alpha}}\bar \W)(\bar D^{-\dot \beta}q^{+A} \partial_{\alpha\dot{\beta}} \W)
    \nn
    &&+i(\partial^{\alpha\dot{\alpha}}D^{-}_\beta q^{+}_A
    D^{-}_{\alpha} D^{+\beta} \W)(\bar D^{-}_{\dot \beta}q^{+A} \bar D^{-}_{\dot \alpha}\bar D^{+\dot \beta}\bar \W)-\partial^{\beta\dot{\alpha}}\bar \W\bar D^{-}_{\dot \alpha}\bar D^{+}_{\dot \beta}\bar \W\partial^{\a \db} \W D^-_{ \alpha} D^{+}_{ \beta} \W\nn
    &&-
    \dfrac{2}{3}\partial^2(D^{-\beta} q^{+}_AD^{-}_{\beta} q^{+}_B)(\bar D^{-}_{\dot \beta}q^{+A}
    \bar D^{-\dot \beta}q^{+B}).
    \label{invariant1}
    \end{eqnarray}
    Here $\mathcal W, \bar{\mathcal W}$ are the
    $\N=2$ superfield strengths, $(\partial \mathcal W)^2=(\partial_{\a\da} \W)(\partial^{\da\a}
    \W)$, and $\partial^2=\partial_{m}\partial^{m}$. The expression (\ref{invariant1}) is manifestly $\N=2$ supersymmetric,
    while its  hidden on-shell $\N=2$ supersymmetry is not evident in advance and requires a rather non-trivial check. However, we are guaranteed to have it
    since we started from the manifestly $\N=4$ supersymmetric expression.

Consider next an invariant of the same dimension containing both the superfields
$\phi^{+\hp}$ and $\phi^{-\hm}$. Since the  total harmonic charge of such an
expression has to be zero, it should simultaneously include $\phi^{+\hp}$ and
$\phi^{-\hm}$. For example, let us write the $\N=4$
invariant of the form
    \begin{equation}
I'=\int d {u} \ d  v \ d^{20} {z} \,(\phi^{+\hp}\phi^{-\hm})^2.
\label{6.35}
\end{equation}
After the procedure similar to what we used to derive
(\ref{invariant1}), we arrive at the action in $\N=2$ superspace
\begin{equation}
I'=\int d {u} \ d^{12} {z} \ \mathcal{L}',
\end{equation}
where
\begin{eqnarray}
&&\mathcal{L}'=-\frac{16}{3}q^{+A}q^{+B}\partial^4\left(q^-_Aq^-_B\right)-
D^{+\a}\W \bar D^{+\da}\bar\W\partial^2\left(D^{-}_{\a}\W \bar D^{-}_{\da}\bar\W\right)\nn
&&-4iD^{+\a}\W \bar
D^{+\da}\bar\W\partial^2\left(q^{-A}\partial_{{\alpha}\dot
\alpha}q^-_A\right).
\label{invariant2}
\end{eqnarray}
This expression is manifestly $\N=2$ supersymmetric, while its
invariance under hidden $\N=2$ supersymmetry is not immediately seen.
Note that (\ref{6.35}) can be evidently rewritten in the central basis, where some additional  harmonic projections of $\phi^{iA}$ can be defined,
viz., $\phi^{+\hat -}, \phi^{- \hat +}$. Using the evident relations like $ \phi^{+\hat -} = \partial^{++}\phi^{-\hat -}, \phi^{- \hat +}
= \partial^{\hat + \hat +}\phi^{- \hat -}\,$, etc, one can check that any neutral product of four such projections is reduced to (\ref{6.35}) via integrating by parts
with respect to the harmonic derivatives.

As the last example of invariants of the same dimension, we consider the $\N=4$ invariant including both
$W$ and $\phi^{+\hp}$ superfields
    \begin{equation}
I''=\int d {u} \ d  v \ d^{20} {z}\,
(\phi^{+\hp}\phi^{-\hm})(W\bar W).
\label{I''}
\end{equation}
After descending to $\N=2$ superspace, we obtain
\begin{equation}
I''=\int d {u} \ d^{12} {z} \ \mathcal{L}'',
\label{6.36}
\end{equation}
where
\begin{eqnarray}
&&\mathcal{L}''=-\dfrac{8i}{3}\partial^2(q^+_A\partial_{\a\da} q^-_B)(D^{-\a}q^{+A}\bar D^{-\da}q^{+B})\nn
&&+\dfrac{1}{4}(D^{+\b}\W\partial_{\a\da}D^{-}_{\b}\W)
\partial^{\a\dot \gamma}\bar \W \bar D^{-}_{\dot\gamma}\bar D^{+\da}\bW
+\dfrac{1}{4}(\bar D^{+\db}\bar\W\partial_{\a\da}\bar D^{-}_{\db}\bar\W)
\partial^{\da \gamma} \W  D^{-}_{\gamma} D^{+\a}\W
\nn
&&-\dfrac{i}{4}(D^{+\b}\W\partial_{\a\da}\bar D^{-}_{\db}\bar\W) D^{-}_{\b} D^{+\a}\W \bar D^{-\db} \bar D^{+\da}\bW
-i(D^{+}_{\b}\W\partial_{\a\da}\bar D^{-}_{\db}\bar\W)
\partial^{ \b\da} \W \partial^{\a \db} \bW \nn
&&+(D^{+}_\b\W \partial_{\a\da}q^{-A})
\partial^{\b\dg} D^{-\a}q^{+}_{A}\bar D^{-}_{\dg}\bar D^{+\da}\bW
+i(D^{+\b}\W \partial_{\a\da}q^{-A})
\partial_{\b\dg} \bar D^{-\da}q^{+}_{A} \partial^{\a\dg}\bW\nn
&&+(\bar D^{+}_{\db}\bW \partial_{\a\da}q^{-A})\partial^{\db\g} \bar D^{-\da}q^{+}_{A} D^{-}_{\g} D^{+\a}\W
-i(\bar D^{+\db}\bW \partial_{\a\da}q^{-A})
\partial_{\db\g}  D^{-\a}q^{+}_{A} \partial^{\da\g}\W.
\label{5.21}
\end{eqnarray}
This expression is written in terms of $\N =2$ harmonic
superfields. It is on-shell $\N=4$
supersymmetric since it was derived from the manifestly $\N=4$
supersymmetric invariant (\ref{I''}). If we would forget about the $\N=4$ superfield origin of (\ref{5.21}),
the proof of its hidden on-shell  $\N=2$ supersymmetry is a rather involved procedure (though it could be performed, of course).
Note that (\ref{I''}) is unique among the invariants of this type: the possible invariant $\sim \phi^{+\hat -} \phi^{-\hat +}\, W\bar W $ is reduced
to (\ref{I''})  after integrating by parts with respect to harmonic derivatives and taking into account the harmonic independence of $W, \bar W$.

Thus we have given three examples of superinvariants in bi-harmonic superspace. All of them are on-shell $\N=4$
supersymmetric by construction. We have shown how they can be
equivalently rewritten in $\N=2$ harmonic superspace, where
only $\N=2$ supersymmetry remains manifest, while the  proof of invariance under
additional hidden $\N=2$ supersymmetry is a non-trivial  job.

These three examples demonstrate a power  of bi-harmonic
superspace approach for constructing $\N =4$ supersymmetric
invariants. The manifestly $\N=4$ supersymmetric invariants look
simple when written in terms of bi-harmonic superspace, however, are converted
into the rather complicated expressions  after passing to their $\N =2$ harmonic superspace form.
Moreover, the inverse problem of promoting these $\N =2$ harmonic superfield densities  to their $\N=4$ bi-harmonic prototypes
cannot be accomplished in a simple way.

As we saw, the above on-shell $\N=4$ superinvariants admit a unique
representation in terms of $\N=2$ harmonic superfields. Since the
technique of deriving the component structures of the local
functionals defined on $\N=2$ harmonic superspace is well developed,
we can in principle calculate the component structure of above
superinvariants. All of these component Lagrangians  contain the higher
derivatives. We suppose that such invariants could arise as some
sub-leading contributions to $\N=4$ SYM low-energy
effective action. As a simple exercise, we calculated the terms depending on Maxwell field strength.
Making in (\ref{invariant1}), (\ref{invariant2}) and  (\ref{5.21}) the substitutions
\bea
\W \; \Rightarrow \; 2\theta^{+}_{\a}\theta^{-}_{\b} F^{\a\b}\,, \quad \bW  \; \Rightarrow \; 2\bar\theta^{-}_{\db}\bar\theta^{+}_{\da}\dF^{\da\db}\,, \quad
q^{\pm A}\; \Rightarrow \; 0\,,
\eea
and integrating over Grassmann and harmonic variables, we deduce (modulo terms vanishing on the free equations of motion)
\bea
&& I \quad \Longrightarrow \quad 2 \int \ d^4x\left[\partial^2(\dF^2)\partial^2(F^2)+ 2\partial^2(\dF_{\dg\da}F_{\g\a})\partial^2(\dF^{\dg\da}F^{\g\a})\right], \nn
&& I'\quad \Longrightarrow \quad 2 \int \ d^4x\left[\partial^2(\dF^2)\partial^2(F^2)+ 2\partial^2(\dF_{\dg\da}F_{\g\a})\partial^2(\dF^{\dg\da}F^{\g\a})\right], \nn
&&I''\quad \Longrightarrow \quad \int \ d^4x\left[\partial^2(\dF^2)\partial^2(F^2)+ 2\partial^2(\dF_{\dg\da}F_{\g\a})\partial^2(\dF^{\dg\da}F^{\g\a})\right]. \label{Maxwell}
\eea

One of the reasons why these expressions proved to be the same is the relation
\bea
\int du\ dv\ d^{20}z\ \left[(W\bar W)^2-4 (\phi^{+\hp})(\phi^{-\hm})(W\bar W)+(\phi^{+\hp}\phi^{-\hm})^2\right]=0,
\eea
which is a consequence of the condition
$$
D^{\hat+}_{\a, \da}\big[(W\bar W)^2-4 (\phi^{+\hp})(\phi^{-\hm})(W\bar W)+(\phi^{+\hp}\phi^{-\hm})^2\big] \sim
D^{-}_{\a, \da}\big[W\bar W^2\phi^{+\hp}-\bar W(\phi^{+\hp})^2\phi^{-\hm}\big],
$$
following  from Bianchi identities (\ref{501}) - (\ref{FurtherBianchi}). For the time being  it remains unclear to us why {\it all three} $\N=4$ superfield actions
give rise to the same Maxwell higher-derivative action. Perhaps, these invariants are related by the $\N=4$ $R$-symmetry group $SU(4)$ \footnote{The superfield realization of the $SU(4)$
R-symmetry in $\N=2$ harmonic superspace was given in \cite{BII1}. We plan to discuss its implications in bi-harmonic $\N=4$ superspace elsewhere.}.

\subsection{Low-energy effective action in bi-harmonic superspace \label{s5.3}}
Our aim here is to recast the invariant (\ref{2.50}) in the bi-harmonic superspace.

Let us consider the following functional
\begin{eqnarray}
\Gamma=c \ \int du\int d v\int d^4 x \ d^4\theta^{+} \
d^4\theta^{\hp}\, \mathcal{L}^{+4 \hat +4}_{\rm eff}\,, \quad  \mathcal{L}^{+4 \hat +4}_{\rm eff} = \frac{(\phi^{+\hat+})^4} {(W\bar W)^2}\mathcal{
M}(Z), \label{6.29}
\end{eqnarray}
where
\begin{equation}
\mathcal{ M}(Z)=\sum\limits_{n=0}^\infty
\dfrac{(-1)^n(n+1)!^2}{n!(n+4)!}Z^n
\label{M}
\end{equation}
and
\begin{equation}
Z=\frac{\phi^{+\hat+}\phi^{-\hat -}}{W\bar W}.
 \label{Z}
\end{equation}
The series in (\ref{M}) is summed up into the following expression
\begin{eqnarray}
\mathcal{ M}(Z)= \dfrac{ \left(6 + 8 Z + 2Z^2\right)\text{ln}(1 + Z)-6Z - 5 Z^2}{4 Z^4} = \frac{1}{24} + \mathcal{O}(Z)\,. \label{M1}
\end{eqnarray}
Below we check consistency of the integral (\ref{6.29}) and prove
that its part leading in derivatives  actually coincides with (\ref{2.50}).

The expression (\ref{6.29}) is written as an integral over full
analytic subspace. In order to show that it is $\N=4$ supersymmetric,  one needs, first of all, to check that the integrand is
analytic or at least analytic up to derivative.
We should act by the derivatives $D^{\hp}_\a, \bar D^{\hp}_\da, D^+_\a, \bar D^+_\da$ on a generic term in series  (\ref{6.29}). To have a feeling
what happens  we consider in some detail the action of $D^{\hp}_\a$. Using the identities (\ref{501})-(\ref{510}) and the result of acting on them various harmonic
derivatives, we are able to show that
\begin{eqnarray}
D^{\hp}_\a\left[\frac{(\phi^{+\hat +})^{n+4}
(\phi^{-\hm})^{n}}{(W\bar W)^{n+2}}\right]&=&
-i \frac{n}{n+1}D^{-}_\a\left[\frac{(\phi^{+\hat +})^{n+4}
(\phi^{-\hm})^{n-1}}{W^{n+2}\bar W^{n+1}}\right] -\frac{(\phi^{+\hat +})^{n+4} (\phi^{-\hm})^{n}}{W^{n+3}\bar W^{n+2}}(n+2)D^{\hp}_\a W \nn
&&-\,\frac{n(n+4)}{n+1}\frac{(\phi^{+\hat +})^{n+3}
(\phi^{-\hm})^{n-1}D^{\hp}_\a W}{W^{n+2}\bar W^{n+1}}. \lb{Sum1}
\end{eqnarray}
From this generic relation one can deduce that the second and third terms in (\ref{Sum1}) are canceled by the contributions from the adjacent
terms in the sum (\ref{6.29}) and that all such unwanted terms are mutually canceled when acting by  $D^{\hp}_\a$ on the whole series (\ref{6.29}).
So finally we obtain
\begin{eqnarray}
&& D^{\hp}_\a\,\mathcal{L}^{+4 \hat +4}_{\rm eff} =D^{-}_\a G^{+5\hp5}_{(1)}\,, \qquad \dD^{\hp}_\da\,\mathcal{L}^{+4 \hat +4}_{\rm eff}
=\dD^{-}_\da \widetilde G^{+5\hp5}_{(1)},\lb{D-first} \\
&& G^{+5\hp5}_{(1)} = i \sum\limits_{n=0}^\infty
\frac{(-1)^{n+1} n(n+1)!}{(n+4)!} \frac{(\phi^{+\hat
+})^{n+4} (\phi^{-\hm})^{n-1}}{W^{n+2}\bar W^{n+1}}\,.
\end{eqnarray}
Analogously, one can check that a similar result holds as well for $+$ derivatives
\begin{eqnarray}
D^{+}_\a\,\mathcal{L}^{+4 \hat +4}_{\rm eff} =  D^{\hm}_\a G^{+5\hp5}_{(2)}, \qquad
\dD^{+}_\da\,\mathcal{L}^{+4 \hat +4}_{\rm eff}=\dD^{\hm}_\da \widetilde G^{+5\hp5}_{(2)}\,,\lb{D-second}
\end{eqnarray}
where $G^{+5\hp5}_{(2)}$ is some expression depending on $W, \ \bar W,\ \phi^{+\hp}, \phi^{-\hm}$. Thus,
the integrand in  (\ref{6.29})  is analytic up to derivative,
\bea
&& \mathcal{L}^{+4 \hat +4}_{\rm eff} =\mathcal{L}^{+4 \hat +4}_{\rm eff (0)}(\zeta_A) - \theta^{-\alpha}D^{\hm}_\a G^{+5\hp5}_{(2)}
- \bar\theta^{-\dot\alpha}\bar D^{\hm}_\da \widetilde G^{+5\hp5}_{(2)} \nn
&&-\, \theta^{\hat -\alpha}D^{-}_\a G^{+5\hp5}_{(1)}-
\bar\theta^{\hat -\dot\alpha}\bar D^{-}_\da \widetilde G^{+5\hp5}_{(1)} + \ldots \,, \lb{ExpLeff}
\eea
where ``dots'' stand for terms of higher orders in $\theta^{-}_{\alpha, \dot\alpha}$ and $\theta^{\hat -}_{\alpha, \dot\alpha}$, each involving negatively charged
spinor derivatives of the appropriate superfield expressions. Because
of the presence of the  operators $(D^-)^4(D^{\hat -})^4$ in the analytic integration measure in (\ref{6.29}), all terms in (\ref{ExpLeff}), except for the first one,
do not contribute,
\bea
\mathcal{L}^{+4 \hat +4}_{\rm eff} \; \Longrightarrow \; \mathcal{L}^{+4 \hat +4}_{\rm eff (0)}(\zeta_A)\,, \lb{ToAnalL}
\eea
and so (\ref{ExpLeff}) is indeed an on-shell $\N=4$  superinvariant (the Bianchi identities (\ref{501})-(\ref{510}) which were used in deriving (\ref{Sum1}) are valid
on shell).

It remains to prove that  (\ref{6.29}) coincides with (\ref{2.50}). To this end, one first needs to rewrite (\ref{6.29}) as an integral over $\N=2$ harmonic superspace.
Taking into account (\ref{ToAnalL}), one can put, from the very beginning, $\theta^{\hat -}_\alpha = \bar\theta^{\hat -}_\da = 0$ in all objects entering (\ref{2.50}),
\begin{eqnarray}
&&W \,\Rightarrow \; \mathcal{\bar W}+i\sqrt{2}\theta^{\hat +\beta}D^{-}_\beta
q^{+\hat -}, \qquad
\bar W \,\Rightarrow \;  \mathcal{ W}+i\sqrt{2}\bar\theta^{\hat +}_{\beta}\bar D^{-\dot\beta} q^{+\hat -}, \nn
&&\phi^{+\hat +} \,\Rightarrow \; \sqrt{2}q^{+\hp}+ i\theta^{\hat +
\alpha}D^{+}_\alpha \mathcal{W}+
i\bar \theta^{\hat +}_{\dot \alpha}\bar D^{+\dot \alpha} \mathcal{ \bar W}+2\sqrt{2}i\theta^{\hat +}_\alpha\bar\theta^{\hat +}_{\dot \alpha}\partial^{{\alpha}\dot \alpha}q^{+\hat -}, \nn
&&\phi^{-\hat -} \,\Rightarrow \; \sqrt{2}q^{-\hm}. \label{5.40}
\end{eqnarray}

Next we consider some special cases, because they can clarify why
the expressions  (\ref{6.29}) and (\ref{2.50}) coincide. The general
case is rather involved but it can also be worked out in a similar fashion.

First, consider the case when hypermultiplet $q^{+\hp}$ equals zero. Due
to (\ref{5.40}) $\phi^{-\hat -  }$ also equals zero. Hence, $Z=0$ and
$\mathcal M(Z) \Rightarrow 1/24$. Thus, we have
\begin{eqnarray}
&& \Gamma =\frac{c}{24} \int du\ d v\ d^4 x \ d^4\theta^{+} \ d^4\theta^{\hp} \frac{(\phi^{+\hat+})^4}{(W\bar W)^2}
=\frac{c}{16} \int  d\zeta^{-4} \frac{(D^{+}\W)^2(\dD^{+}\bW)^2}{\W^2\bW^2}\nn
&&=c\  \int du \ d^{12}z \ \text{ln}\left(\frac{
    \W}{\Lambda}\right)\text{ln}\left(\frac{{\bW}}{\Lambda}\right)=\int du\ d^{12} z \ \mathcal{H}(\W,\bW),
\end{eqnarray}
where we made use of the equations of motion (\ref{equations}), when  passing to the
last line. Hence, if $q^{+\hp}$ equals zero, (\ref{6.29}) coincides  with (\ref{2.50}).

As the next step, consider the case when hypermultiplet $q^{+}_A$ does not depend
on  $x_m$ and $\theta$'s, i. e. all derivatives of $q^{+}_A$ are equal
to zero. This is just the standard requirement to distinguish the leading term in the effective action.
Then all terms with $\theta^{\hp}_\a$ and  $\btheta^{\hp}_\da$ prove to be
located in $\phi^{+\hp}$. Hence, in this case the expression
(\ref{6.29}) equals
\begin{eqnarray}
&&\Gamma = c\ \int du\ d v\ d^4 x \ d^4\theta^{+} \ d^4\theta^{\hp} \frac{(\phi^{+\hat+})^4}{(\mathcal{W}\mathcal{\bar W})^2}\mathcal{ M}(Z) \nn
&& =\frac{c}{16} \ \int d\zeta^{-4} \sum\limits_{n=0}^{\infty}\frac{(-1)^n(n+1)(2q^{+A}q^{-}_A)^n(D^{+}\mathcal{ W})^2(\bar D^{+}\mathcal{ \bar W})^2}{(\mathcal W\mathcal{\bar W})^{n+2}}\nn
&&=c\  \int du \ d^{12}z  \left[\text{ln}\left(\frac{
    \W}{\Lambda}\right)\text{ln}\left(\frac{{\bW}}{\Lambda}\right)+\sum\limits_{n=1}^{\infty}
\frac{(-1)^n(2q^{+A}q^{-}_A)^n}{n^2(n+1)(\W\bW)^n}\right]\nn
&&=\  \int du \ d^{12}z\left[
\mathcal{H}(\W,\bW)+\mathcal{L}\left(-2\frac{q^{+A}q^{-}_A}{\W\bW}\right)\right],
\end{eqnarray}
where we used equations of motion (\ref{equations})  when passing to the
next-to-last line. Hence, if $q^{+\hp}$ is constant, expression
(\ref{6.29}), once again, coincides with (\ref{2.50}). Note that the approximation just used gives rise
to a term $F^4$ of the fourth order in the Maxwell field strength in the component Lagrangian. The effective action
(\ref{2.50}) also encodes the Wess-Zumino term \cite{WeZu} which was derived in \cite{litlink 22} by applying to another limit of (\ref{2.50}), such that $x$-derivatives of the scalar fields
are retained while all other components of $\N=4$ SYM multiplet are put equal to zero. Using the same background in (\ref{6.29}) we arrive
at the same WZ term in components.

Thus, we expressed the effective actions in bi-harmonic superspace in
terms of $\N=4$ superfields (\ref{6.29}). It is given as an integral
over the full analytic subspace of $\N=4$ bi-harmonic superspace. A significant difference of (\ref
{6.29}) from the $\N=2$ superspace effective action (\ref {2.50}) is that (\ref {6.29}) is
$\N=4$ supersymmetric only on shell, while in (\ref {2.50}) the equations
of motion are required only to prove the invariance under hidden $\mathcal
{N} = 2 $ supersymmetry. This peculiarity seems to be not too essential since after passing to
$\N=2$ superfield form of the effective action one can ``forget'' about its
$\N=4$ superfield origin and stop to worry on its manifest $\N=2$ supersymmetry. Anyway,
one {\it needs} to assume the equations of motion, once the hidden supersymmetry is concerned.
The same is true for $\N=4$ invariants considered in subsection \ref{s5.10}. To rephrase this argument, in $\N=4$ formulation both $\N=2$
supersymmetries enter on equal footing and so both are on shell, while after passing to $\N=2$ formulation one $\N=2$
becomes formally manifest, while another one remains hidden and on-shell.

\section{Conclusions}
Let us summarize the results. We have developed a new superfield
method of constructing the on-shell $\N=4$ supersymmetric
invariants in $4D\,, \N=4$ SYM theory. To know the precise structure of
such superinvariants is of high necessity  when calculating the effective action in
$\N=4$ SYM quantum field theory formulated in $\N=2$ harmonic superspace and when studying the
low-energy limit of string/brane theory. The method is based on the
concept of bi-harmonic $\N=4$ superspace, which properly generalizes the notion of $\N=2$
harmonic superspace \cite{litlink GIOS} to an extension of the latter with the double sets of the Grassmann and harmonic
coordinates, so that the automorphism group  $SU(2)\times SU(2) \times U(1) \subset SU(4)$ remains manifest. Using the formulation of $4D\,, \N=4$ SYM theory
in this bi-harmonic superspace it becomes possible to construct the on-shell $\N=4$ superinvariants
in a manifestly $\N=4$ supersymmetric fashion and then pass to their equivalent $\N=2$ superfield form by a simple general recipe.

The basic merit of the new formulation is that, within its framework,  the defining constraints of $ \mathcal {N} = 4 $ SYM theory in $\N=4$ superspace
can be resolved in terms of the basic objects of $ \mathcal {N} = 2 $ harmonic superfield description of this theory, the gauge superfield $\V^{++}$ and the hypermultiplet
superfield $q^{+ A}$. The relevant $\N=2$ superfield
equations of motion directly follow from the bi-harmonic form of the defining  $ \mathcal {N} = 4 $ SYM constraints. Thus, there was established the precise correspondence
between the on-shell bi-harmonic $\N=4$ superfields and the superfields underlying  the $\N=2$ harmonic superspace formulation
of $\N=4$ SYM theory, in which two supersymmetries are manifest and the other two are on-shell and hidden.

As an illustration of how the proposed method works, we
have constructed three abelian  manifestly on-shell
$\N=4$ supersymmetric higher-derivative invariants and rewrote them in
terms of $\N=2$ harmonic superfields.
In the $\N=2$ superfield formulation, the second $\N=2$ supersymmetry of these invariants looks highly implicit and it would be very hard to guess the structure of these
invariants in advance. Also, we showed how the $\N=4$ SYM leading low-energy
effective action (\ref{2.50}) can be recast in the manifestly
$\N=4$ supersymmetric form.

The proposed method of constructing the manifestly $\N=4$
supersymmetric on-shell invariants seems general enough to apply it
for constructing and analyzing  various other invariants in $\N=4$
SYM theory, for instance those which are $\N=4$
completions of the $F^6$, $F^8\,, \ldots $ invariants from the gauge
field sector by adding the proper hypermultiplet terms. Such invariants can
correspond to possible contributions to quantum effective actions in the Coulomb
phase from higher loops, generalizing the one-loop $F^4$
effective action described by the invariant (\ref{2.50}) (see, e.g.,
\cite{Ch-T}, \cite{BKT}, \cite{BPT}, \cite{Kuze}). Such terms also arise as the next-to-leading corrections in the one-loop effective action
(see, e.g., \cite{BBP}). Among other interesting problems it is worth to mention a possible stringy extension of
$\N=4$ SYM constraints in the form (\ref{1}) - (\ref{7}) and
construction of $\N=4$ supergravity in $\N=4$ bi-harmonic superspace.

\section*{Acknowledgements}
The authors thank Pierre Fayet for valuable comments.
The work of I.L.B. and E.A.I. was partially supported by Ministry of
Education of Russian Federation, project No FEWF-2020-003.

    \appendix

    \section{Self-consistency conditions \label{s7.2} }

    Self-consistency conditions arise as additional relations on superfields, when solving the constraints. They are to be identically satisfied on the final solution of all constraints.
    In this appendix we list the self-consistency conditions which come out from the equations in subsection \ref{s3.2}.

    The conditions following from eq. (\ref{27}a):
    \begin{eqnarray}
    &&\hat{\nabla}_{\alpha\dot{\alpha}}
    \widetilde \W^{+\dot{\alpha}}+\sqrt{2}[\psi_{A\alpha},q^{+A}]+[\bW,\W^{+}_{\alpha}]=0,
    \label{38}\\
    &&\hat{\nabla}^{\alpha\dot{\alpha}}
    \W^{+}_{{\alpha}}+\sqrt{2}[\widetilde \psi^{\dot \alpha}_A,q^{+A}]-[\W,\widetilde \W^{+}_{\dot \alpha}]=0,
    \label{39}\\
    &&\hat\nabla^{\alpha \dot \alpha}\hat\nabla_{\alpha\dot{\alpha}}q^{+A}-i\{\psi^{ A\alpha},\W^{+}_{\alpha}\}+i\{\widetilde \psi^{A}_{\dot \alpha},\widetilde \W^{+ \dot \alpha}\}\nn
    &&-2i[\mathcal{D}^{AB},q^{+}_B]+\sqrt{2}[ \bW,[\W,q^{+A}] ]+\sqrt{2}[\W,[ \bW,q^{+A}]]=0.
    \label{40}
    \end{eqnarray}

    The conditions following from eqs. (\ref{27}b):
    \begin{eqnarray}
    &\nabla^{++}\W^+_{\alpha}=0,\quad  \nabla^{++}\hat\nabla^{\alpha\dot{\alpha}}q^{+A}=0, \quad \nabla^{++}[\widetilde\psi^{(A\dot\alpha},q^{+B)}]=0,\nn  &\nabla^{++}[\mathcal{D}^{(AB},q^{+C)}]=0,\quad
    \nabla^{++}[ \bW,q^{+A}]=0,\qquad \nabla^{++}[\W,q^{+A}]=0.
    \end{eqnarray}

    The conditions following from eqs. (\ref{1.9}):
    \begin{eqnarray}
   && D^{+}_\beta\W+i\W^{+}_\beta=0,
    \label{55}\\
   && D^{+}_\beta{\bW}=0,
    \label{56}\\
    &&D^{+}_{\beta}\mathcal{\hat A}^{\alpha\dot{\alpha}}+\delta^{\alpha}_{\beta}\widetilde \W^{+\dot{\alpha}}=0,
    \label{57}\\
    &&D^{+}_\beta\psi^{A}_\alpha+2\sqrt 2i\epsilon_{ \beta\alpha}[\bW,q^{+A}]=0,
    \label{58}\\
    &&D^{+}_\beta\widetilde\psi^{A\dot{\alpha}}+2\sqrt{2}\epsilon_{ \beta\alpha}\hat{\nabla}^{\alpha\dot{\alpha}}q^{+A}=0,
    \label{8.12}\\
    &&D^{+}_\beta\mathcal{D}^{(AB)}+\sqrt{2}[\psi^{(A}_\beta,q^{+B)}]=0.
    \label{60}
    \end{eqnarray}

    The conditions following from eqs. (\ref{1.8}b):
    \begin{eqnarray}
    &&D^+_\beta\hat\nabla^{\alpha\dot \alpha}q^{+\hat-}-i\delta^{\alpha}_{\beta}[q^{+\hat -},\widetilde \W^{+\dot{\alpha}}]=0,
    \label{75}\\
    &&D^{+}_\beta[\W,q^{+A}]-i[q^{+A},\W_{\beta}]=0,\\
    && D^{+}_\beta[ \bW,q^{+A}]=0,\\
    &&D^+_\beta[\psi^{(A}_\alpha,q^{+B)}]+2\sqrt 2 i\epsilon_{ \beta\alpha}[q^{+(A}, [ \bW,q^{+B)}]]=0,
    \label{76}\\
   && D^+_\beta[\widetilde \psi^{(A\dot\alpha},q^{+B)}]-2\sqrt{2}\epsilon_{ \beta\alpha}[q^{+(A},\hat\nabla^{\alpha \dot \alpha}q^{+B)}]=0,
    \label{77}\\
   && D^+_\beta[\mathcal{D}^{(AB},q^{+C)}]-\sqrt{2}[q^{+(A},[\psi^{B}_\beta,q^{+C)}]=0.
    \label{78}
    \end{eqnarray}

    All these conditions become identities on the final solution of all constraints (subsection \ref {s6}). For example, consider the relation (\ref {8.12})
    \begin{equation}
    D^{+}_\beta\widetilde\psi^{A\dot{\alpha}}+2\sqrt{2}\epsilon_{ \beta\alpha}\hat{\nabla}^{\alpha\dot{\alpha}}q^{+A}=0.
    \end{equation}
    Substituting into it the expression for $ \widetilde \psi ^ {A \dot {\alpha}} $ from (\ref {259})
    \begin{equation}
    \widetilde \psi^{ A}_{\dot \beta}=-i\sqrt 2\bar\nabla^{-(0)}_{\dot{\beta}}q^{+A},
    \end{equation}
    we obtain
    \begin{equation}
    i\sqrt 2D^{+}_\beta\bar\nabla^{-(0)\dot{\alpha}}q^{+A}-2\sqrt{2}\epsilon_{ \beta\alpha}\hat{\nabla}^{\alpha\dot{\alpha}}q^{+A}=0.
    \end{equation}
    Hence, we get the identity,
    \begin{equation}
    \hat{\nabla}^{\alpha\dot{\alpha}}q^{+A}-\hat{\nabla}^{\alpha\dot{\alpha}}q^{+A}=0.
    \end{equation}

    \section{Calculation of $V^{\hat -\hat -}$ \label{s4.1}}
    The non-analytic gauge connection $V^{\hat -\hat{-}}$  is defined as a solution of the  zero curvature condition
    \begin{equation}
    D^{\hat+\hat+}V^{\hat-\hat-}-D^{\hat-\hat-}V^{\hat+\hat+}+i[V^{\hat+\hat+},V^{\hat-\hat-}]=0,
    \label{77777}
    \end{equation}
    where
    \begin{eqnarray}
    V^{\hat+ \hat+}&=& -2i\theta^{\hat +}_{\alpha}\bar\theta^{\hat +}_{\dot \alpha}\mathcal{\hat A}^{\alpha \dot \alpha}+ (\theta^{\hat+})^2\W+ (\bar \theta^{\hat+})^2{\bW}
    +2(\bar \theta^{\hat +})^2\theta^{\hat +\alpha}\psi^{\hat -}_\alpha+2( \theta^{\hat +})^2\bar \theta^{\hat +}_{\dot\alpha}\widetilde \psi^{\hat -\dot\alpha}\nn
    &&+\,3(\theta^{\hat+})^2(\bar \theta^{\hat+})^2\mathcal{D}^{\hat-2}.
    \end{eqnarray}
    We parametrize the $\theta^{\hat-}_\alpha,\bar\theta^{\hat-}_{\dot \alpha}$ expansion of  $V^{\hat-\hat{-}}$ in the following way
    \begin{eqnarray}
    V^{\hat- \hat-}&=&-2i\theta^{\hat -}_{\alpha}\bar\theta^{\hat -}_{\dot \alpha}w^{\alpha \dot \alpha}+ (\theta^{\hat-})^2w+(\bar \theta^{\hat-})^2{\widetilde w}
    +(\bar \theta^{\hat -})^2\theta^{\hat -\alpha}w^{\hat +}_\alpha+( \theta^{\hat -})^2\bar \theta^{\hat -}_{\dot\alpha}\widetilde w^{\hat
    +\dot\alpha}\nn
    &&+\,(\theta^{\hat-})^2(\bar \theta^{\hat-})^2w^{\hat+2}.
    \label{900}
    \end{eqnarray}
    All the coefficient do not depend on $\theta^{\hat-}_\alpha,\bar\theta^{\hat-}_{\dot \alpha}$ and their  $\theta^{\hat+}_\alpha,\bar\theta^{\hat+}_{\dot \alpha}, v^{\hat +}_A$
    dependence will be determined from eq. (\ref{77777}). Possible coefficients of the first and zeroth order monomials in these coordinates  are killed by the
    equations like   $\partial^{\hat+\hat +}\omega^{\hat -}=0\Rightarrow \omega^{\hm}=0$. Here we give only the general scheme of finding the solution and the final answer.

    The first step consists of finding equations on the coefficients in (\ref{900}). The  $\theta^{\hat-}_\alpha,\bar\theta^{\hat-}_{\dot \alpha}$ expansion (\ref{77777}) contains the  monomials of the first, second, third and fourth degrees.  Equating the corresponding coefficients
    to zero, we obtain the  set of equations
   \begin{eqnarray}
 && i\epsilon_{\alpha\beta}\bar\theta^{\hat +}_{\dot \alpha}w^{\beta \dot \alpha}+ \theta^{\hat+}_\alpha w
   =i\epsilon_{\alpha\beta}\bar\theta^{\hat +}_{\dot \alpha}\mathcal{\hat A}^{\beta \dot \alpha}+\theta^{\hat+}_\alpha\W
   +(\bar \theta^{\hat +})^2\psi^{\hat -}_\alpha
   +2\theta^{\hat +}_\alpha\bar \theta^{\hat +}_{\dot\alpha}\widetilde \psi^{\hat -\dot\alpha}+3\theta^{\hat+}_\alpha(\bar \theta^{\hat+})^2\mathcal{D}^{\hat-2},
   \label{91}\\
  && i\theta^{\hat +}_{\alpha}w^{\alpha \dot \alpha}
   +\bar \theta^{\hat+\dot{\alpha}}{\widetilde w}
   =i\theta^{\hat +}_{\alpha}\mathcal{\hat A}^{\alpha \dot \alpha}+ \bar \theta^{\hat+\dot{\alpha}}{\bW}
   +2\bar \theta^{\hat +\dot{\alpha}}\theta^{\hat +\alpha}\psi^{\hat -}_\alpha+( \theta^{\hat +})^2\widetilde \psi^{\hat -\dot\alpha}+6(\theta^{\hat+})^2\bar \theta^{\hat+\dot{\alpha}}\mathcal{D}^{\hat-2},
   \label{92}\\
  && \nabla^{\hat{+}\hat +}w^{\beta\dot{\beta}}-i\bar\theta^{\hat +\dot{\beta}}w^{\hat +\beta}+i\theta^{\hat +\beta}\widetilde w^{\hat +\dot{\beta}}+2i\theta^{\hat +}_{\alpha}\bar\theta^{\hat +}_{\dot \alpha}\partial^{\beta\dot{\beta}}\mathcal{\hat A}^{\alpha \dot \alpha}- (\theta^{\hat+})^2\partial^{\beta\dot{\beta}}\W- (\bar \theta^{\hat+})^2\partial^{\beta\dot{\beta}}{\bW}\nn
 &&  -2(\bar \theta^{\hat +})^2\theta^{\hat +\alpha}\partial^{\beta\dot{\beta}}\psi^{\hat -}_\alpha-2( \theta^{\hat +})^2\bar \theta^{\hat +}_{\dot\alpha}\partial^{\beta\dot{\beta}}\widetilde \psi^{\hat -\dot\alpha}-3(\theta^{\hat+})^2(\bar \theta^{\hat+})^2\partial^{\beta\dot{\beta}}\mathcal{D}^{\hat-2}=0,
   \label{93}\\
   &&\nabla^{\hat{+}\hat +}w+\bar \theta^{\hat +}_{\dot\alpha}\widetilde w^{\hat +\dot\alpha}=0
   \label{94},\\
  && \nabla^{\hat{+}\hat +}\widetilde w+\theta^{\hat +\alpha}w^{\hat +}_\alpha=0
   \label{95},\\
  && \nabla^{\hat{+}\hat +}\widetilde w^{\hat +}_{\dot{\alpha}}+2\bar\theta^{\hat +}_{\dot\alpha}w^{\hat+2}=0
   \label{96},\\
   &&\nabla^{\hat{+}\hat +}w^{\hat{+}}_\alpha+2\theta^{\hat +}_{\alpha}w^{\hat+2}=0,
   \label{97}\\
 &&  \nabla^{\hat{+}\hat +}w^{\hat+2}=0  \label{98}.
   \end{eqnarray}

    To solve eqs.  (\ref{91})-(\ref{98}), we expand the corresponding unknowns in (\ref{90}) over $\theta^{\hat+}_\alpha,\bar\theta^{\hat+}_{\dot \alpha}$ and then fix
    the $v^{\hat \pm}_A$ dependence of the coefficients by these equations.
    We explicitly perform this operation only for eq. (\ref{98}). We write
    \begin{eqnarray}
    &&w^{\hat+2}=r^{\hat+2}+
    \theta^{\hat + \alpha}r^{\hat +}_{\alpha}+
    \bar \theta^{\hat +}_{\dot \alpha}\widetilde r^{\hat +\dot\alpha}_{ }+
    ( \theta^{\hat +})^2r^{}_{ }+
    (\bar \theta^{\hat +})^2\widetilde r^{}_{ }\nn
    &&+(\bar \theta^{\hat +})^2\theta^{\hat +\alpha}r_{\alpha }^{\hat{-}}
    +( \theta^{\hat +})^2 \bar \theta^{\hat +}_{\dot\alpha}
    \widetilde r^{\hat{-}\dot \alpha}_{ }
    -2i\theta^{\hat+}_{\alpha}\bar\theta^{\hat+}_{\dot\alpha}
    r^{\alpha\dot{\alpha}}_{ }+
    ( \theta^{\hat +})^2(\bar \theta^{\hat +})^2r^{\hat -2}_{ },
    \label{99}
    \end{eqnarray}
    and obtain the following set of equations and their solutions
    \begin{eqnarray}
    &&\partial^{\hat +\hat +}r^{\hat{+}2}=0 \quad\Rightarrow\quad r^{\hat{+}2}=r^{AB}v^{\hat +}_Av^{\hat +}_B,
    \label{100}\\
    &&\partial^{\hat +\hat +}r^{\hat{+}}_{\alpha}=0 \quad\Rightarrow\quad r^{\hat{+}}_{\alpha}=r^{A}_{\alpha}v^{\hat +}_A,
    \label{101}\\
    &&\partial^{\hat +\hat +}\widetilde r^{\hat{+}\dot{\alpha}}=0 \quad\Rightarrow\quad \widetilde r^{\hat{+}\dot{\alpha}}=-\widetilde r^{A\dot{\alpha}}v^{\hat +}_A,
    \label{102}\\
    &&\partial^{\hat +\hat +}r^{\alpha \dot \alpha}+\hat{\nabla}^{\alpha \dot \alpha}r^{+2}=0    \quad\Rightarrow\quad r^{\alpha \dot \alpha}=r^{\alpha \dot \alpha}_0-\hat{\nabla}^{\alpha \dot \alpha}r^{(AB)}v^{\hat +}_Av^{\hat -}_B
    \label{103},\\
    &&\partial^{\hat +\hat +}r+i[\W,r^{\hat +2}]=0
    \quad\Rightarrow\quad r=r_0-i[\W,r^{(AB)}v^{\hat +}_Av^{\hat -}_B]
    \label{104},\\
    &&\partial^{\hat +\hat +}\widetilde{r}+i[\bW,r^{\hat +2}]=0
    \quad\Rightarrow\quad \widetilde r=\widetilde r_0-i[\bW, r^{(AB)}v^{\hat +}_Av^{\hat -}_B]
    \label{105},\\
    &&\partial^{\hat +\hat +}r^{\hat -}_\alpha+i\hat{\nabla}_{\alpha \dot \alpha}\widetilde{r}^{\hat +\dot\alpha}+2i[\psi^{\hat -}_{\alpha},r^{\hat +2}]+ [ \bW, r^{\hat +}_\alpha]=0
    \label{106},\\
    &&\partial^{\hat +\hat +}\widetilde r^{\hat -\dot\alpha}-i\hat{\nabla}^{\alpha \dot \alpha} {r}^{\hat +}_{\alpha}+2i[\widetilde \psi^{\hat -}_{\alpha},r^{\hat+2}]
    +i[\W,\widetilde r^{\hat +\dot\alpha}]=0
    \label{107},\\
    &&\partial^{\hat +\hat +}r^{\hat-2}-\hat{\nabla}^{\alpha{\dot \alpha}}r_{\alpha\dot{\alpha}}+3i[\mathcal{D}^{\hat{-}2},r^{\hat+2}]-2i\{\psi^{\hat -\alpha},r^{\hat +}_\alpha\}\nn
    &&-2i\{\widetilde \psi^{\hat -}_{\dot\alpha},\widetilde r^{\hat +\dot\alpha}\}+i[\W,\widetilde r]+i[\bW,{r}]=0.
    \label{108}
    \end{eqnarray}

    Eqs. (\ref {106})-(\ref {108}) are rather cumbersome. Their solutions are given below
    \begin{eqnarray}
    {\rm (\ref{106})} \Rightarrow&&  r^{\hat -}_\alpha=-i [ \bW, r^{A}_\alpha]v^{\hat -}_A+i\hat{\nabla}_{\alpha\dot{\alpha}}\widetilde r^{A\dot{\alpha}}v^{\hat -}_A-\frac{4i}{3}[\psi_{A\alpha},r^{(AB)}]v^{\hat -}_B\nn
   &&-i[\psi^{(A}_\alpha,r^{BC)}]v^{\hat +}_Av^{\hat -}_Bv^{\hat -}_C
    \label{109},\\
    {\rm (\ref{107})} \Rightarrow && \widetilde r^{\hat -\dot\alpha}=i[\W, \widetilde r^{A\dot\alpha}]v^{\hat -}_A+i\hat{\nabla}^{\alpha\dot{\alpha}}r^{A}_{\alpha}v^{\hat -}_A+\frac{4i}{3}[\widetilde \psi_{A}^{\dot\alpha},r^{(AB)}]v^{\hat -}_B\nn
    &&+i[\widetilde\psi^{(A\dot\alpha},r^{BC)}]v^{\hat +}_Av^{\hat -}_Bv^{\hat -}_C
    \label{110},\\
    {\rm (\ref{108})}\Rightarrow&&  r^{\hat -2}=
    \frac{1}{2}\bigg(-\hat{\nabla}^{\alpha{\dot \alpha}}\hat{\nabla}_{\alpha \dot \alpha}r^{(AB)}
    +i\{\psi^{A\alpha},r^{B}_\alpha\}+i\{\widetilde \psi^{A}_{\dot\alpha},\widetilde r^{B\dot\alpha}\}\nn
   && -[\W,[\bW,r^{(AB)}]]
    -[\W,[\bW,r^{(AB)}]]
    \bigg)v^{\hat -}_Av^{\hat -}_B
    -\frac{3i}{2}[\mathcal{D}^{A}_C,r^{CB}]v^{\hat -}_Av^{\hat -}_B\\
    &&-i[\mathcal{D}^{(AB},r^{CD)}]v^{\hat -}_{(A}v^{\hat -}_Bv^{\hat -}_Cv^{\hat +}_{D)}.
    \label{112}
    \end{eqnarray}
    Eqs. (\ref{100})- (\ref{105})  and (\ref{109}), (\ref{110}) and (\ref{112}) include definitions of the coefficients in the expansion of $w^{\hat +2}$.

    The remaining equations can be solved analogously. Instead of giving the analogs of eqs. (\ref{100})-(\ref{108},) we will
    present  at once the full solution for the coefficients in (\ref{900})
    \begin{eqnarray}
    &&w^{\hat+2}=\mathcal{D}^{\hat+2}+
    \theta^{\hat + \alpha} \left(-i\hat \nabla_{{\alpha}\dot \alpha}\widetilde\psi^{A\dot{\alpha}}+  i [\psi^{A}_\alpha,\W]\right)v^{\hat +}_A
    -\bar \theta^{\hat +}_{\dot \alpha}
    \left(i\hat\nabla^{{\alpha}\dot \alpha}\psi^{A}_{{\alpha}}+  i [\widetilde \psi^{A\dot{\alpha}}, \bW]\right)v^{\hat +}_A\nn
    &&+
    \dfrac{1}{4}( \theta^{\hat +})^2\Big(-2\hat{\nabla}^{\alpha\dot \alpha}\hat{\nabla}_{\alpha\dot{\alpha}}\W-2[\W,[\bW,\W]]+i\{\widetilde\psi^{\dot \alpha}_A,\widetilde\psi^A_{\dot\alpha}\}-4i[\W, \mathcal{D}^{\hat+\hat -}] \Big)\nn
    &&+
    \dfrac{1}{4}( \bar \theta^{\hat +})^2\Big(-2\hat{\nabla}^{\alpha\dot
     \alpha}\hat{\nabla}_{\alpha\dot{\alpha}}\bW+2[\bW,[\bW,\W]]+i\{\psi_{ \alpha A},\psi^{A\alpha}\}-4i[ \bW,\mathcal{D}^{\hat+\hat -}] \Big)\nn
    &&+(\bar \theta^{\hat +})^2\theta^{\hat +\alpha}
    \Big(- [ \bW,\hat \nabla_{{\alpha}\dot \alpha}\widetilde\psi^{A\dot{\alpha}}]v^{\hat -}_A
    + [ \bW, [\psi^{A}_\alpha,\W]]v^{\hat -}_A-2\hat{\nabla}_{\alpha\dot{\alpha}}\hat \nabla^{\beta\dot{\alpha}}\psi^A_{{\beta}} v^{\hat -}_A\nn
    &&- \hat{\nabla}_{\alpha\dot \alpha}[\widetilde \psi^{A\da},\bW]v^{\hat -}_A
    -\dfrac{4i}{3}[\psi_{A\alpha},\mathcal{D}^{(AB)}]v^{\hat -}_B
    -i[\psi^{(A}_\alpha,\mathcal{D}^{BC)}]v^{\hat +}_Av^{\hat -}_Bv^{\hat -}_C\Big)\nn
    &&+( \theta^{\hat +})^2 \bar \theta^{\hat +}_{\dot\alpha}
    \Big(- [ \W,\hat \nabla^{{\alpha}\dot \alpha}\psi^{A}_{{\alpha}}]v^{\hat -}_A
    - [\W, [\widetilde\psi^{A\dot\alpha},\bW]]v^{\hat -}_A+\hat{\nabla}^{\alpha\dot{\alpha}}\hat \nabla_{\alpha\dot{\beta}}\widetilde\psi^{A\dot{\beta}} v^{\hat -}_A\nn
    &&- \hat{\nabla}^{\alpha\dot \alpha}[ \psi^{A}_{\alpha},\W]v^{\hat -}_A
    +\dfrac{4i}{3}[\widetilde\psi_{A}^{\dot\alpha},\mathcal{D}^{(AB)}]v^{\hat -}_B
    +i[\psi^{(A\alpha},\mathcal{D}^{BC)}]v^{\hat +}_Av^{\hat -}_Bv^{\hat -}_C\Big)\nn
    &&-i\theta^{\hat+}_{\alpha}\bar\theta^{\hat+}_{\dot\alpha}
    \Big(i[\hat{\nabla}^{\alpha{\dot{\alpha}}}\bW,\W]-i[\bW,\hat{\nabla}^{\alpha{\dot{\alpha}}}\W]-\{\psi_{A}^{\alpha},\widetilde\psi^{A\dot{\alpha}}\}-2\hat{\nabla}^{\alpha{\dot{\alpha}}}\mathcal{D}^{\hat +\hat -}\nn
    &&+
    \hat\nabla^{\alpha}_{\text{ }\text{ }\dot \beta}\left(2\partial_{\beta}^{\text{ }\text{ }(\dot \beta}\mathcal{ \hat A}^{\beta\dot{\alpha})}+i[\mathcal{\hat A}_{\beta}^{\text{ }\text{ }\dot \beta},\mathcal{ \hat A}^{\beta\dot{\alpha}}]\right)\Big)\nn
    &&+
    \frac{1}{2}( \theta^{\hat +})^2    ( \bar\theta^{\hat +})^2\Big(\big(-\hat{\nabla}^{\alpha{\dot \alpha}}\hat{\nabla}_{\alpha \dot \alpha}\mathcal{D}^{(AB)}
    +\{\psi^{A\alpha},\hat \nabla_{{\alpha}\dot \alpha}\widetilde\psi^{B\dot{\alpha}}\}
    - \{\psi^{A\alpha},
    [\psi^{B}_\alpha,\W]\}\nn
    &&-\{\widetilde \psi^{A}_{\dot\alpha},\hat\nabla^{{\alpha}\dot \alpha}\psi^{B}_{{\alpha}}\}
    -  \{\widetilde \psi^{A}_{\dot\alpha},[\widetilde \psi^{B\dot{\alpha}}, \bW]\}
    -[\W,[\bW,\mathcal{D}^{(AB)}]]-
    [\W,[\bW,\mathcal{D}^{(AB)}]]
    \big)v^{\hat -}_Av^{\hat -}_B\nn
    &&-3i[\mathcal{D}^{A}_C,\mathcal{D}^{CD}]v^{\hat -}_Av^{\hat -}_B\Big),
    \end{eqnarray}
    \begin{eqnarray}
    &&w^{\hat +}_\beta=-\psi^{A}_\beta v^{\hat +}_A+
    \theta^{\hat +}_{\beta}\Big(i[\bW,\W]-2\mathcal{D}^{\hat +\hat -}\Big)+\theta^{\hat +\alpha}\left(2\partial_{(\alpha\dot{\alpha}}\mathcal{  \hat A}_{\beta)}^{\text{ }\text{ }\dot{\alpha}}+i[\mathcal{ \hat A}_{\alpha\dot \alpha},\mathcal{\hat A}_\beta^{\text{ }\text{ }\dot{\alpha}}]\right)\nn
    &&+2i\bar \theta^{\hat +\dot \alpha}\hat{\nabla}_{\beta\dot{\alpha}}\bW
    -i( \theta^{\hat +})^2\hat \nabla_{{\beta}\dot \alpha}\widetilde\psi^{A\dot{\alpha}}v^{\hat -}_A
    +i(\bar \theta^{\hat +})^2[\bW,\psi^A_\beta]v^{\hat -}_A\nn
    &&+(\bar \theta^{\hat +})^2\theta^{\hat +\alpha}
    \left(\sqrt 2i\epsilon_{\beta\alpha}[\bW,\mathcal{D}^{(AB)}]v^{\hat -}_Av^{\hat -}_B+i\{\psi^{(A}_\alpha,\psi^{B)}_{\beta}\}v^{\hat -}_Av^{\hat -}_B\right)\nn
    &&-( \theta^{\hat +})^2 \bar \theta^{\hat +}_{\dot\alpha}
    \left(i\{\widetilde\psi^{(A\dot\alpha},\psi^{B)}_{\beta}\}v^{\hat -}_Av^{\hat -}_B
    +2i\epsilon_{\beta\alpha}\hat{\nabla}^{\alpha\dot{\alpha}}\mathcal{D}^{AB}v^{\hat -}_Av^{\hat -}_B\right)\nn
    &&-2i\theta^{\hat+}_{\alpha}\bar\theta^{\hat+}_{\dot\alpha}
    \Big(\hat{\nabla}^{\alpha\dot{\alpha}}\psi^{A}_\beta v^{\hat -}_A-\delta^{\alpha}_{\beta}\left(\hat\nabla^{{\gamma}\dot \alpha}\psi^{A}_{{\gamma}}+  [\widetilde \psi^{A\dot{\alpha}}, \bW]\right)v^{\hat -}_A\Big)\nn
    &&+
    \dfrac{5i}{3}( \theta^{\hat +})^2(\bar \theta^{\hat +})^2[\mathcal{D}^{(AB},\psi^{C)}_\beta]v^{\hat -}_Av^{\hat -}_Bv^{\hat -}_C,
    \end{eqnarray}

    \begin{eqnarray}
    &&\widetilde w^{\hat +}_{\dot\beta}=\widetilde\psi^{A}_{\dot\beta} v^{\hat +}_A-
    \bar\theta^{\hat +}_{\dot\beta}\Big(i[\bW,\W]+2\mathcal{D}^{\hat +\hat -}\Big)+\bar\theta^{\hat +\dot\alpha}\left(2\partial_{\alpha(\dot{\alpha}}\mathcal{ \hat A}^\alpha_{\text{ }\text{ }\dot{\beta})}+i[\mathcal{\hat A}_{\alpha\dot \alpha},\mathcal{ \hat A}_\beta^{\text{ }\text{ }\dot{\alpha}}]\right)\nn
    &&+2i\theta^{\hat + \alpha}\hat{\nabla}_{\alpha\dot{\beta}}\W
    +i(\bar \theta^{\hat +})^2\hat \nabla_{{\alpha}\dot \beta}\psi^{A{\alpha}}v^{\hat -}_A
    -i( \theta^{\hat +})^2[\bW,\widetilde\psi^A_\beta]v^{\hat -}_A\nn
    &&+(\bar \theta^{\hat +})^2\theta^{\hat +\alpha}
    \left(2i\hat{\nabla}_{\alpha\dot\beta}\mathcal{D}^{AB}v^{\hat{-}}_Av^{\hat -}_B-i\{ \psi^{(A}_\alpha,\widetilde \psi^{B)}_{\dot{\beta}}\}v^{\hat{-}}_Av^{\hat -}_B\right)\nn
    &&-( \theta^{\hat +})^2 \bar \theta^{\hat +}_{\dot\alpha}
    \left(2i\delta^{\dot{\alpha}}_{\dot{\beta}}[\W,\mathcal{D}^{(AB)}]v^{\hat{-}}_Av^{\hat -}_B-i\{\widetilde\psi^{(A\dot{\alpha}},\widetilde \psi^{B)}_{\dot{\beta}} \}v^{\hat{-}}_Av^{\hat -}_B\right)\nn
    &&+2i\theta^{\hat+}_{\alpha}\bar\theta^{\hat+}_{\dot\alpha}
    \Big(\hat{\nabla}^{\alpha\dot{\alpha}}\widetilde\psi^{A}_{\dot\beta} v^{\hat -}_A-\delta^{\dot\alpha}_{\dot\beta}\left(\hat \nabla^{{\alpha}\dot \gamma}\widetilde\psi^{A}_{\dot{\gamma}}+  [\psi^{A\alpha},\W]\right)v^{\hat -}_A\Big)\nn
    &&-
    \dfrac{5i}{3}( \theta^{\hat +})^2(\bar \theta^{\hat +})^2[\mathcal{D}^{(AB},\widetilde\psi^{C)}_\beta]v^{\hat -}_Av^{\hat -}_Bv^{\hat -}_C,
    \end{eqnarray}
    \begin{eqnarray}
    w&=&\W
    -\bar \theta^{\hat +}_{\dot \beta}\widetilde\psi^{A\dot{\beta}}v^{\hat -}_A+
    (\bar \theta^{\hat +})^2 \mathcal{D}^{\hat -2},\\
    \widetilde w&=& \bW
    + \theta^{\hat + \beta}\psi^{A}_{{\beta}}v^{\hat -}_A+
    ( \theta^{\hat +})^2 \mathcal{D}^{\hat -2},\\
    w^{\beta\dot{\beta}}&=&\mathcal{\hat A}^{\beta\dot{\beta}}
    -i\theta^{\hat + \beta}\widetilde \psi^{A\dot{\beta}}v^{\hat -}_A-i
    \bar \theta^{\hat +\dot \beta} \psi^{A\beta}v^{\hat -}_A
    +i\theta^{\hat+\beta}\bar\theta^{\hat+\dot\beta}
    \mathcal{D}^{\hat -2}.
    \end{eqnarray}
    Substituting all these expressions in (\ref {900}), one can obtain the full expression for $ V ^ {\hat - \hat -} $.

    In the Abelian case the expressions for the superfield coefficients are essentially simplified
    \begin{eqnarray}
    &&w^{\hat+2}=\mathcal{D}^{\hat+2}-
    i\theta^{\hat + \alpha}\partial_{{\alpha}\dot \alpha}\widetilde\psi^{A\dot{\alpha}}v^{\hat +}_A
    -i\bar \theta^{\hat +}_{\dot \alpha}
    \partial^{{\alpha}\dot \alpha}\psi^{A}_{{\alpha}}v^{\hat +}_A\nn
    &&-
    \dfrac{1}{2}( \theta^{\hat +})^2\partial^{\alpha\dot \alpha}\partial_{\alpha\dot{\alpha}}\W
    -\dfrac{1}{2}( \bar \theta^{\hat +})^2\partial^{\alpha\dot \alpha}\partial_{\alpha\dot{\alpha}}\bW\nn
    &&-(\bar \theta^{\hat +})^2\theta^{\hat +\alpha}
    \partial_{\alpha\dot{\alpha}}\partial^{\beta\dot{\alpha}}\psi^A_{{\beta}} v^{\hat -}_A
    +( \theta^{\hat +})^2 \bar \theta^{\hat +}_{\dot\alpha}
    \partial^{\alpha\dot{\alpha}}\partial_{\alpha\dot{\beta}}\widetilde\psi^{A\dot{\beta}} v^{\hat -}_A\nn
    &&-2i\theta^{\hat+}_{\alpha}\bar\theta^{\hat+}_{\dot\alpha}
    \Big(-\partial^{\alpha{\dot{\alpha}}}\mathcal{D}^{\hat +\hat -}+
    \partial^{\alpha}_{\text{ }\text{ }\dot \beta}\partial_{\beta}^{\text{ }\text{ }(\dot \beta}\mathcal{\hat A}^{\beta\dot{\alpha})}\Big)
    -\dfrac{1}{2}( \theta^{\hat +})^2   ( \bar\theta^{\hat +})^2\partial^{\alpha{\dot \alpha}}\partial_{\alpha \dot \alpha}\mathcal{D}^{(AB)}v^{\hm}_Av^{\hm}_B,
    \end{eqnarray}
    \begin{eqnarray}
    &&w^{\hat +}_\beta=-\psi^{A}_\beta v^{\hat +}_A
    -2\theta^{\hat +}_{\beta}\mathcal{D}^{\hat +\hat -}+2\theta^{\hat +\alpha}\partial_{(\alpha\dot{\alpha}}\mathcal{ A}_{\beta)}^{\text{ }\text{ }\dot{\alpha}}
    +2i\bar \theta^{\hat +\dot \alpha}\partial_{\beta\dot{\alpha}}\bW
    -i( \theta^{\hat +})^2\partial_{{\beta}\dot \alpha}\widetilde\psi^{A\dot{\alpha}}v^{\hat -}_A\nn
    &&-2i( \theta^{\hat +})^2 \bar \theta^{\hat +}_{\dot\alpha}
    \epsilon_{\beta\alpha}\partial^{\alpha\dot{\alpha}}\mathcal{D}^{AB}v^{\hat -}_Av^{\hat -}_B
    -i\theta^{\hat+}_{\alpha}\bar\theta^{\hat+}_{\dot\alpha}
    \Big(\partial^{\alpha\dot{\alpha}}\psi^{A}_\beta v^{\hat -}_A-\delta^{\alpha}_{\beta}\partial^{{\gamma}\dot \alpha}\psi^{A}_{{\gamma}}v^{\hat -}_A\Big),
    \end{eqnarray}
    \begin{eqnarray}
    &&\widetilde w^{\hat +}_{\dot\beta}=\widetilde\psi^{A}_{\dot\beta} v^{\hat +}_A-
    2\bar\theta^{\hat +}_{\dot\beta}\mathcal{D}^{\hat +\hat -}+2\bar\theta^{\hat +\dot\alpha}\partial_{\alpha(\dot{\alpha}}\mathcal{ A}^\alpha_{\text{ }\text{ }\dot{\beta})}+2i\theta^{\hat + \alpha}\partial_{\alpha\dot{\beta}}\bW +i(\bar \theta^{\hat +})^2\partial_{{\alpha}\dot \beta}\psi^{A{\alpha}}v^{\hat -}_A\nn
    &&+2i(\bar \theta^{\hat +})^2\theta^{\hat +\alpha}
    \partial_{\alpha\dot\beta}\mathcal{D}^{AB}v^{\hat{-}}_Av^{\hat -}_B+2i\theta^{\hat+}_{\alpha}\bar\theta^{\hat+}_{\dot\alpha}
    \Big(\partial^{\alpha\dot{\alpha}}\widetilde\psi^{A}_{\dot\beta} v^{\hat -}_A-\delta^{\dot\alpha}_{\dot\beta}\partial^{{\alpha}\dot \gamma}\widetilde\psi^{A}_{\dot{\gamma}}v^{\hat -}_A\Big),
    \end{eqnarray}
    \begin{eqnarray}
    w&=& \W
    -\bar \theta^{\hat +}_{\dot \beta}\widetilde\psi^{A\dot{\beta}}v^{\hat -}_A+
    (\bar \theta^{\hat +})^2 \mathcal{D}^{\hat -2},\\
    \widetilde w&=& \bW
    + \theta^{\hat + \beta}\psi^{A}_{{\beta}}v^{\hat -}_A+
    ( \theta^{\hat +})^2 \mathcal{D}^{\hat -2},\\
    w^{\beta\dot{\beta}}&=&\mathcal{\hat A}^{\beta\dot{\beta}}
    -i\theta^{\hat + \beta}\widetilde \psi^{A\dot{\beta}}v^{\hat -}_A
    -i\bar \theta^{\hat +\dot \beta} \psi^{A\beta}v^{\hat -}_A
    +2i\theta^{\hat+\beta}\bar\theta^{\hat+\dot\beta}
    \mathcal{D}^{\hat -2}.
    \end{eqnarray}

\end{document}